\documentclass[11pt]{article}
\pdfoutput = 1

\usepackage[utf8]{inputenc}
\usepackage{color,graphicx}
\usepackage{amsmath}
\usepackage{verbatim}
\usepackage{amssymb}
\usepackage{physics}
\usepackage{amsfonts}
\usepackage{array}
\usepackage{setspace}
\usepackage{float}
\usepackage{url}
\usepackage{standalone}
\usepackage{leftindex}
\usepackage{tikz}
\usepackage{fancybox}
\usepackage{tensor}
\usepackage{tikz}
\usepackage{caption}
\usepackage{subcaption}
\usepackage{dsfont}
\usepackage{jheppub}
\usepackage{relsize}
\definecolor{pinkk}{rgb}{1,0,1}

\newcommand{\diff}[1]{\mathop{}\!\mathrm{d}{#1}\mathop{}\!}

\usepackage{hyperref}
\hypersetup{
    colorlinks,
    citecolor=blue,
    filecolor=black,
    linkcolor=blue,
    urlcolor=blue,
    linktocpage=true
}

\newcommand{\be}{\begin{equation}}
\newcommand{\ee}{\end{equation}}
\newcommand{\bea}{\begin{align}}
\newcommand{\eea}{\end{align}}
\newcommand{\bi}{\begin{itemize}}
\newcommand{\ei}{\end{itemize}}

\numberwithin{equation}{section}

\title{De Sitter quantum gravity and the emergence of local algebras}

\author{Molly Kaplan$^1$,} \emailAdd{mekaplan@ucsb.edu}
\author{Donald Marolf$^1$,} \emailAdd{marolf@ucsb.edu}
\author{Xuyang Yu$^1$,} \emailAdd{xuyangyu@ucsb.edu}
\author{Ying Zhao$^{2,3}$} \emailAdd{zhaoying@mit.edu}

\affiliation{$^1$ Department of Physics, University of California, Santa Barbara, CA 93106, USA}
\affiliation{$^2$ Kavli Institute for Theoretical Physics, Santa Barbara, CA 93106, USA}
\affiliation{$^3$ Massachusetts Institute of Technology, Cambridge, MA 02139, USA}

\abstract{Quantum theories of gravity are generally expected to have some degree of non-locality, with familiar local physics emerging only in a particular limit.  Perturbative quantum gravity around backgrounds with isometries and compact Cauchy slices provides an interesting laboratory in which this emergence can be explored.  In this context, the remaining isometries are gauge symmetries and, as a result, gauge-invariant observables cannot be localized.  Instead, local physics can arise only through certain relational constructions.

We explore such issues below for perturbative quantum gravity around de Sitter space.  In particular, we describe a class of
gauge-invariant observables which, under appropriate conditions, provide good approximations to certain algebras of local fields.  Our results suggest that, near any minimal $S^d$ in dS$_{d+1}$,  this approximation can be accurate  only over regions in  which the corresponding global time coordinate $t$ spans an interval $\Delta t \lesssim O(\ln G^{-1})$.  In contrast, however, we find that the approximation can be accurate over arbitrarily large regions of global dS$_{d+1}$ so long as those regions are located far to the future or past of such a minimal $S^d$.
This in particular includes arbitrarily large parts of any static patch.}

\begin{document}

% Title Page
% \input{editionlegend.tex}
 \maketitle

%%%%%%%%%%%%%%%%%%%%%%%%%%%%%%%%%%%
%\pagebreak
%\setcounter{page}{1}
%\tableofcontents

%\newpage

\section{Introduction}\label{sec:intro}

It has long been recognized that the physics of quantum gravity will involve at least some degree of non-locality, with familiar local physics emerging in the perturbative limit $G \rightarrow 0$, where $G$ is Newton's gravitational constant.  While some such effects may stem from topology-changing processes in the gravitational path integral, we will focus here on a form of non-locality that is directly associated with diffeomorphism-invariance (see
e.g. discussions in
\cite{DeWitt:1962,DeWitt:1967yk,Banks:1984cw,Hartle:1986gn,Rovelli:1990jm,Rovelli:1990pi,Kiefer:1993fg,Marolf:1994nz,Giddings:2005id,Marolf:2015jha}), which are expected to arise even when topology-change is absent.

In addition, recent progress on understanding gravitational entropy in this limit  \cite{Witten:2021unn,Chandrasekaran:2022cip,Chandrasekaran:2022eqq,Jensen:2023yxy,Penington:2023dql,Kudler-Flam:2024psh,Chen:2024rpx}  has emphasized the importance of the emergence of an {\it algebra} of local fields.  Our goal here is to perform the next steps in investigating just how such algebras appear as $G \rightarrow 0$ by exploring a construction advocated in \cite{Giddings_2007}  for the interesting-but-tractable context of perturbative gravity around global de Sitter (dS$_D$) space, with metric
\begin{equation}
ds^2 = -dt^2 + \ell^2 \cosh^2(t/\ell) d\Omega_d^2,
\end{equation}
where $d=D-1$, $\ell$ is the de Sitter scale, and  $d\Omega_d^2$ is the round metric on the unit sphere $S^d$.

As emphasized in \cite{Marolf:2015jha}, perturbative quantum gravity is manifestly local when formulated around a background that completely breaks diffeomorphism-invariance in a local manner\footnote{I.e., when the background fields can be used to build a set of local scalars rich enough that, given any two distinct points $p,q$ in the background, we have $\phi(p)\neq\phi(q)$ for at least one such scalar field $\phi$. Such scalars may include the Ricci scalar or other scalars built from derivatves of the metric.}.  In particular, in that context it can be described by gauge-invariant operators that satisfy exact microcausality.  But this is not the case when the background leaves a subgroup of gauge diffeomorphisms unbroken; i.e., when the Cauchy surfaces of the background are compact (so that all diffeomorphisms are gauge) and when there is an isometry that also leaves invariant any matter fields that may be present.  In this more subtle context, even at the perturbative level any gauge-invariant observable must be invariant under the unbroken isometries.\footnote{This is, of course, just the gravitational version of a general fact about gauge symmetries and perturbation expansions.  Given a gauge transformation $g$ that acts on fields $\phi$ via $\phi \mapsto \phi_g$, we may choose a classical background $\bar \phi$ and define the perturbative field
$\delta \phi:= \phi - \bar \phi$ and the perturbative gauge transformation $(\delta \phi)_g : = \phi_g - \bar\phi$.  When $\bar \phi_g \neq \bar \phi$, the space of small perturbations is preserved only when $g-1$ is of order $\delta \phi$, so that dropping terms of order $(\delta \phi)^2$ yields $\phi_g \approx \delta \phi + \bar\phi_g$ and thus $\delta \phi_g \approx \delta \phi +(\bar\phi_g - \bar\phi)$.  On the other hand, for a family of transformations with $\bar \phi_g = \bar \phi$, we may take $g$ arbitrarily large.  Furthermore, the action of $g$ on $\delta \phi$ is then essentially the same as the action of $g$ on $\phi$.  In particular, in the gravitational case the unbroken diffeomorphisms act as finite diffeomorphisms on the perturbative fields $\delta \phi$.}
As a result, a gauge-invariant observable in perturbative gravity can be supported in a small region of spacetime localized near a single point $p$ only if $p$ is a fixed point of every unbroken isometry.

When expanding around global de Sitter, such observables must be invariant under the full connected component of the de Sitter group.  Since this group acts transitively on dS$_d$,  any observable will be maximally delocalized. 
Nevertheless, we expect to recover a notion of local physics by making use of relational constructions; see e.g. \cite{DeWitt:1962,DeWitt:1967yk,Banks:1984cw,Hartle:1986gn,Rovelli:1990jm,Rovelli:1990pi,Kiefer:1993fg,Marolf:1994nz,Giddings:2005id,Giddings_2007}.
Indeed, in an appropriate limit we should obtain the usual local algebra of quantum fields on a fixed spacetime background.

Related issues were recently addressed in \cite{Chen:2024rpx}, which explored how a rolling inflaton field could replace the clock used for the construction described in \cite{Chandrasekaran:2022cip} of a type II von Neumann algebra for the static patch of dS. However, our treatment differs from that of \cite{Chen:2024rpx} in three important ways.  The first is that \cite{Chen:2024rpx} assumed that a definition of a preferred static patch $P$ of their de Sitter space had already been given in a gauge-invariant manner.  This then left only the isometry associated with time translations within $P$ to be treated explicitly.  One might thus say that they took locality in space as a given and focussed instead on issues associated with the emergence of locality in time.   In contrast, we treat all de Sitter isometries on an equal footing and explicitly study the emergence of locality in both space and time.  A second difference is that,
in addition to understanding the limiting algebra, we will also characterize the departures from the $G=0$ limit that arise at small-but-finite values of $G$.  Finally, a third difference is that we consider perturbations around a {\it stable} de Sitter space, and in particular one in which all matter fields (including any field that might be called an `inflaton') has a stable vacuum.  We expect this to be a good starting point for discussion of more interesting scenarios that involve eternal inflation with a small probability of ending inflation in each Hubble volume; see e.g. \cite{Kachru:2003aw,McAllister:2024lnt} for progress on embedding such constructions in string theory.

Our focus on stable (or nearly-stable) dS$_{d+1}$ vacua has important implications for our construction of gauge-invariant observables.  To explain the details, it will be useful to refer to the theory at order $G^0$ as quantum field theory on a fixed de Sitter background (dS QFT), where we take this to include the theory of linearlized gravitons.  To construct perturbative observables, it may then seem natural to follow \cite{DeWitt:1967yk} and consider observables of the form
\begin{equation}
\label{eq:ObsInt}
{\cal O} = \int dx\sqrt{-g} A(x)
\end{equation}
for some local scalar field $A(x)$ in our dS QFT.  This approach has been shown to be successful in certain simple models of quantum gravity \cite{Marolf:1994wh,Marolf:1994ss,Marolf:1994nz,Giddings:2005id}.  The analogous construction was also used in \cite{Chen:2024rpx} (where the integral was only over static patch time translations since, as noted above, that work assumed that a preferred notion of a static patch had already been given), and in \cite{Chandrasekaran:2022cip,Jensen:2023yxy,Kudler-Flam:2024psh} (though with an ad hoc observer clock instead of just local quantum fields).  However, since local correlators in any state\footnote{Due to our introduction of the group averaging inner product in section \ref{sec:review}, we use square brackets to denote bra states $\bigl[\Psi\big|$ and ket states $\big|\Psi\bigr]$ of dS QFT. } $\big|\Psi \bigr]$ are well-approximated by correlators in the vacuum $\big|0\bigr]$ at late times, the integral in \eqref{eq:ObsInt} will diverge when acting on {\it any} state $\big|\Psi\bigr]$ in the dS QFT Hilbert space $\mathcal H_{QFT}$ \cite{Giddings:2005id,Giddings_2007}.   

In particular, for any $\big|\Psi\bigr]$ and for ${\cal O}$ as in \eqref{eq:ObsInt}, in a computation of the norm-squared \begin{equation}
\label{eq:normsquared}
\Big| {\cal O} \big|\Psi\bigr] \Big|^2
= 
\bigl[\Psi\big|
{\cal O}
{\cal O} \big|\Psi\bigr] 
= \int dx dy \sqrt{-g(x)}\sqrt{-g(y)}
\bigl[\Psi\big|
A(x) A(y) \big|\Psi\bigr],
\end{equation}
the leading term at large separations between $x$ and $y$
is given by the norm-squared of the state
\begin{equation}
\label{eq:actonvac}
\int dx \sqrt{-g} A(x) \big|0\bigr].
\end{equation}
 But by dividing the  integral over dS$_{d+1}$ in \eqref{eq:actonvac} into an infinite number of large-but-finite regions, and using the decay of dS correlators at large separations, we may write \eqref{eq:actonvac} as an infinite sum over approximately-orthogonal states.  This representation thus makes manifest the divergent nature of its norm.  An equivalent observation was also mentioned in \cite{Chandrasekaran:2022cip,Chen:2024rpx} using the static patch language that every state in dS will thermalize at late times.  As noted in \cite{Giddings_2007}, the issue may be considered to be an operator-realization of the so-called `Boltzmann brain' problem since, no matter how complicated we make the operator $A(x)$ (perhaps in an attempt to make the operator respond only to large and complicated excitations of $\big|0\bigr]$), our $A(x)$ will still fail to annihilate the vacuum $\big|0\bigr]$ and will thus respond to virtual (or, in the thermal static patch description, Boltzmann) versions of such excitations with at least some small probability $p$ per unit spacetime volume.  For any $p>0$, integrating over the infinite volume of dS$_{d+1}$ then gives the divergence described above.

The success of using \eqref{eq:ObsInt} in \cite{Chen:2024rpx} was thus directly tied to the assumption of a rapidly decaying inflaton field made in that work.
Since we take all matter fields to be stable, we will require a different approach.  In particular, we choose to follow \cite{Giddings_2007} in replacing the local observable $A(x)$ with a distinctly {\it non-local} operator $A$ that does in fact annihilate the dS vacuum (though the actual form of the operators we will use is rather different from that described in \cite{Giddings_2007}).    Since $A$ is not a local field, there is then no meaning to $A(x)$, and thus no direct analogue of \eqref{eq:ObsInt}.  However, we can still apply a de Sitter transformation $g$ to the operator $A$ by 
computing $U(g) A U(g^{-1})$, where $U(g)$ is the unitary representation of $g$ on the 
Hilbert space of the associated quantum field theory on a fixed de Sitter background (dS QFT). We may then again follow \cite{Giddings_2007} in constructing de Sitter-invariant observables by writing 
\begin{equation}
\label{eq:ObsdS}
{\cal O} = \int_{g\in \text{SO}_0(D,1)} dg U(g) A U(g^{-1}).
\end{equation}
 The expression \eqref{eq:ObsdS} uses the Haar measure $dg$ to integrate over all elements $g$ of the subggroup SO$_0(D,1)\subset$SO$(D,1)$ of isometries of dS$_D=dS_{d+1}$ that are connected to the identity (i.e., over the orthochronous Lorentz group). 

In the main text below we will consider a context with two independent fields, $\phi$ and $\psi$, so that our dS QFT Hilbert space takes the form $\mathcal H_{QFT} = \mathcal H_{QFT}^\phi \otimes H_{QFT}^\psi$. Here $\phi$ and $\psi$ need not be scalars and, in particular, we can include linearized gravitons in our dS QFT by taking them to be part of the field $\phi$.    We then choose a local operator $\tilde A(x)$ on $\mathcal H_{QFT}^\phi$ and a vaccum-orthogonal state $\big|\psi_0\bigr] \in \mathcal H_{QFT}^\psi$ (so that $[ 0 | \psi_0 ]=0$).  Taking $A: = \tilde A(x) \otimes \big|\psi_0\bigr] \bigl[\psi_0\big|$ for any fixed $x$ will then define a finite ${\cal O}$ with the desired properties for appropriate choices of $\big|\psi_0\bigr]$.  In effect, as will be made manifest in section \ref{sec:GAobs}, we will use the state $\big|\psi_0\bigr]$ to define a quantum version of a reference frame with respect to which positions and directions in de Sitter space can then be specified\footnote{There is a vast literature on so-called quantum reference frames; see e.g. \cite{Aharonov:1967zza,Aharonov:1984zz} for foundational works, \cite{Loveridge:2017pcv,Hoehn:2019fsy} for recent works including broad reviews,  and \cite{Fewster:2024pur,DeVuyst:2024pop,AliAhmad:2024wja} for relations to \cite{Chandrasekaran:2022cip,Jensen:2023yxy,Chen:2024rpx,Kudler-Flam:2024psh}. This literature addresses themes that strongly overlap with our current discussion, though often with a slightly different emphasis and formalism. Related issues are also discussed in e.g. \cite{Klinger:2023auu}.} .  In particular, despite the integral over de Sitter transformations in \eqref{eq:ObsdS}, we will see explicitly below that the definition of the observable ${\cal O}$ depends on the choice of the point $x$. 
 Note that what is really needed is just an origin for this reference frame, together with a way to specify directions emanating from that origin, as one can then use the background de Sitter metric to construct e.g. a set of Riemann normal coordinates (or any other coordinate system on $dS_D$) with respect to which the point $x$ can be specified.  Below, we will thus refer to $\big|\psi_0\bigr]$ as a {\it reference state}.  We emphasize that the associated reference field $\psi$ is a part of the dS QFT and, in particular, that it will backreact on the geometry at higher orders in perturbation theory.

For the above non-local operators $A$,
we will see explicitly in section \ref{sec:GAobs} that the operators \eqref{eq:ObsdS} act like local quantum fields in any limit where $\bigl[\psi_0\big|U(g)\big|\psi_0\bigr]$ becomes $\delta(g)$, the Haar-measure delta-function on the de Sitter group supported at the identity.  Such limits are straightforward to construct when we take $G\rightarrow 0$ so that $\big|\psi_0\bigr]$ may contain arbitrarily large energies and momenta without inducing a large gravitational backreaction.

In contrast, if we want to approximate QFT on empty de Sitter space, at finite $G$ the backreaction effects from the state $\big|\psi_0\bigr]$ prevent us from taking a strict delta-function limit.  As a result,   the integration over $g$ in \eqref{eq:ObsdS} causes our gauge-invariant observables to be somewhat-smeared versions of local quantum $\phi$-fields, so that the desired local algebra is recovered only approximately.

\begin{figure}[H]
  \centering
  \includegraphics[width = 4in]{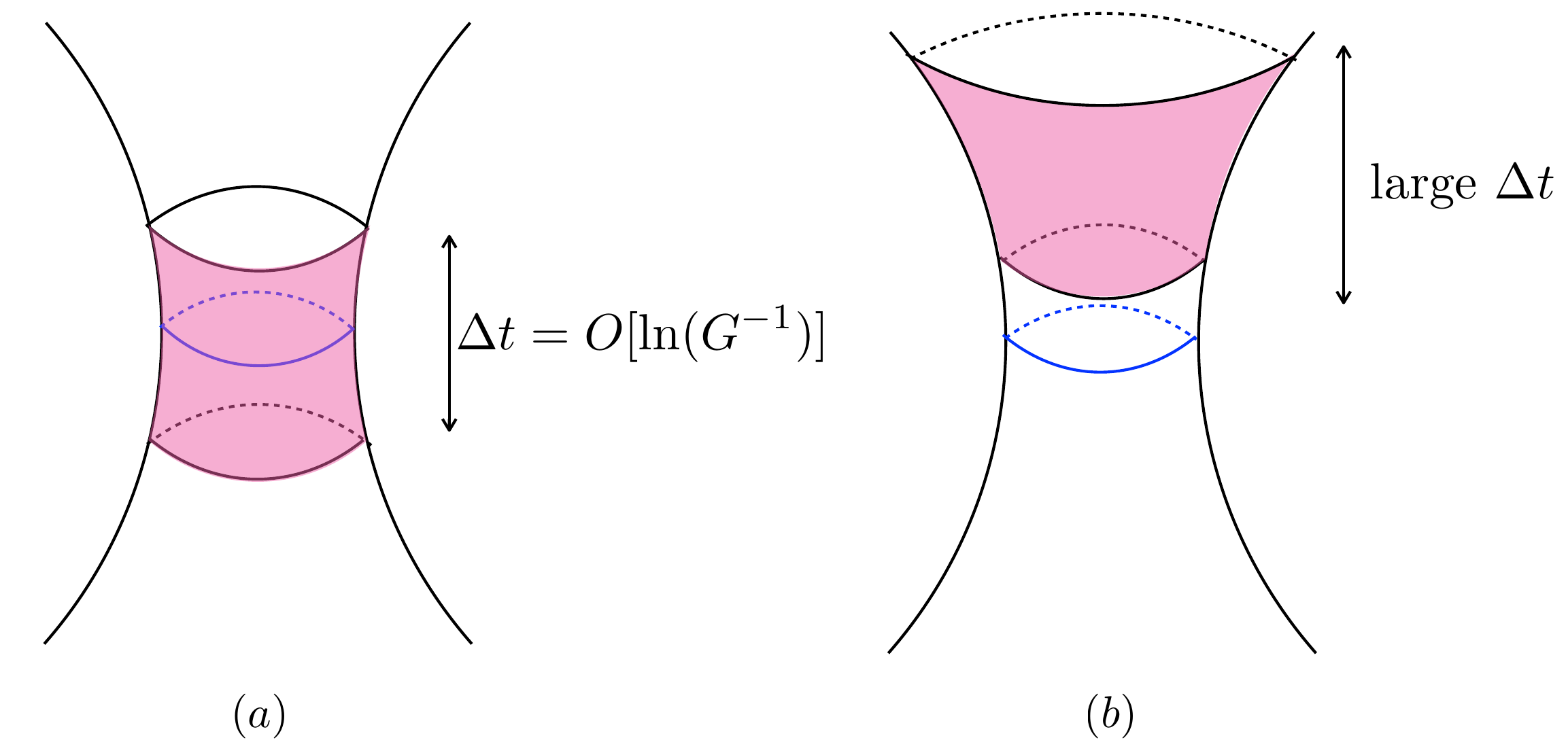}
  \caption{A sketch of global dS$_{d+1}$ indicating regions (shaded pink) where we construct good approximations to dS QFT at non-zero $G$.  {\bf (a):}  Regions that contain a minimal $S^d$ (blue) can span only global time intervals $\Delta t\lesssim O[\ln(G^{-1})]$.  {\bf (b):}
  Regions far to the future (or past) of a minimal $S^d$ (blue) can span arbitrarily large global time intervals.}
  \label{fig:regions}
\end{figure}

As we will see, the accuracy of this approximation is far from uniform across the de Sitter background.  Instead, it is typically best in a region near where the reference objects in the state $\big|\psi_0\bigr]$ are well-localized.  The approximation then degrades as one moves to more distant regions of the spacetime.  Our results also indicate that it is difficult (and likely impossible) to engineer settings where the dS QFT approximation holds to high accuracy over regions that span a global time interval of more than $O(\ell \ln G^{-1})$ that is symmetric with respect to the past and future of global de Sitter or, more generally,  which contains a minimal $S^d$ that we may call $t=0$ ; see figure \ref{fig:regions} (a).  On the other hand, because such minimal spheres describe the most fragile regions of global dS, we find that we can nevertheless obtain a good approximation to dS QFT over arbitrary spans of global time, so long as we take the associated regions to be far to the future (or far to the past) of the associated minimal $S^d$; see figure \ref{fig:regions} (b).

It will be useful to begin by describing the Hilbert space of gauge-invariant states on which the operators \eqref{eq:ObsdS} will act.  We review this construction in section \ref{sec:review}, using the group-averaging construction of \cite{Higuchi_1991, Higuchi_1991_2}.  We then apply this formalism to dS$_{1+1}$ in Section \ref{sec:1+1}.   While Einstein-Hilbert gravity is trivial in two-dimensions, it is nevertheless useful to analyze dS$_{1+1}$ as a toy model of the higher dimensional cases\footnote{While one can also study dS$_{1+1}$ in Jackiw-Teitelboim (JT) gravity,  the JT dilaton always breaks the de Sitter isometry group to a smaller (one-dimensional) group.  However, constructions analogous to those below could be studied for the case where the remaining gauge group is noncompact.}.
We first consider a reference state $\big|\psi_0\bigr]$  for which the classical limit describes having a single particle in each of two complementary static patches. We identify the regions of spacetime in which the dS QFT approximation breaks down, and we estimate the size of the region in which the dS QFT approximation holds. We then introduce additional reference particles, localizing at additional events, such that these events all lie on a single pair of antipodally-related timelike geodesics. However, we find that the size of the allowed region remains the same (or becomes slightly smaller). We then demonstrate analogous results for higher dimensions in section \ref{sec:d+1}, before finally arguing in section \ref{subsec:futureparticles} that dropping the requirement of time-symmetry {\it does} in fact allow us to approximate dS QFT well over arbitrary intervals of global time (so long as they are sufficiently far to the future or past). We  then conclude in Section \ref{sec:discussion} with comments on cosmological interpretations of our results and outlook for the future.

\section{Group averaging and perturbative dS gravity}
\label{sec:GApdS}

In a perturbative analysis of any quantum theory, one expands both the operators and the quantum states in powers of a small parameter $\epsilon$.    The expansion is typically performed about a background classical solution $s_0$, in which case the leading term in any quantum state $|\Psi\rangle$ is generally expected to be a state $|\Psi_1\rangle$ of the linearized theory around $s_0$.    However, subtleties arise when the background $s_0$ leaves some of the gauge symmetries unbroken.

The issue can be explained simply by using the Hamiltonian formalism of the classical theory.  In this formalism, the phase space is subject to constraints $C$ which generate gauge transformations by taking Poisson Brackets.  When $s_0$ leaves a gauge symmetry unbroken, there will be a corresponding constraint $C$ such that all Poisson Brackets $\{C, A\}$ vanish at $s_0$ (regardless of whether $A$ is gauge invariant).  This is of course equivalent to requiring all first order variations $\delta C$ to vanish at $s_0$; i.e., $s_0$ is a stationary point of $C$.  

As a result, the leading term in the equation of motion $C=0$ is of {\it second} (quadratic) order at $s_0$.  In particular, when passing from linear to quadratic order in perturbation theory, one encounters this new equation of motion even though it has no analogue in the linear theory.  Such new quadratic equations of motion are called {\it linearization stability constraints}.  This terminology refers to the fact that solutions to the linearized theory can be perturbatively corrected at higher orders of perturbation theory {\it only} if they satisfy such constraints.  Solutions of the linearized theory that fail to satisfy such constraints are simply spurious and do not represent linearizations of solutions to the full theory.  See e.g. \cite{Deser:1973zza,Moncrief:1975,Moncrief:1976un,Arms:1977,Arms:1979au} for discussion of such issues in classical general relativity.

\begin{figure}[H]
  \centering
  \includegraphics[width = 1.5in]{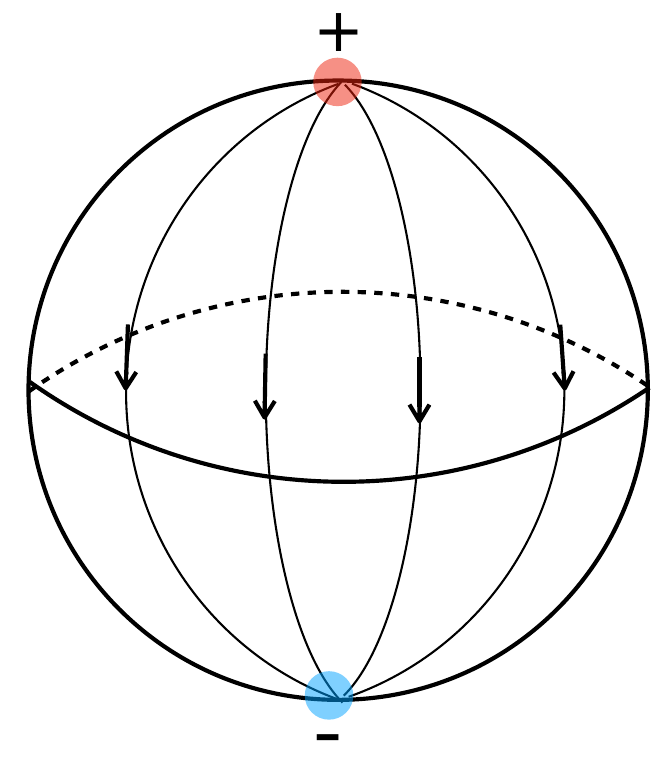}
  \caption{A positive charge (red) sources a flux of electric field (arrows) as shown.  However, if the charge lives on a sphere (say, at the north pole), the resulting field lines are forced to cross again at least at one other point (at the south pole in the example shown here).  The Gauss law then requires the resulting convergence to coincide with the location of a negative charge (blue).  As a result, only configurations of charges with zero net charge can consistently source electric fields on $S^d$.}
  \label{fig:Maxwell}
\end{figure}

A classic example of this phenomenon occurs in Maxwell theory coupled to charged fields on $S^{d} \times \mathbb{R}$ (where the $\mathbb R$ factor is the time direction).  The linearized theory will admit general linearized solutions for the charged fields.  But since the charge-density is typically quadratic in the charged fields, at quadratic order the charged fields will source the Maxwell field.  And since $S^{d}$ has no boundary, there is no way for electric flux to leave the sphere.  As a result, the  Maxwell Gauss law requires the total electric charge to vanish; see figure \ref{fig:Maxwell}.  It is thus only linearized solutions with vanishing net electric charge that can be linearizations of solutions to the full theory. 

In the Maxwell example above, it is straightforward to impose the linearization stability constraints at the quantum level as well.  After constructing the states of the linearized theory, one need only truncate that Hilbert space to the sector with vanishing total charge.  Charge conservation then prohibits such states from mixing with the states that have been discarded.  Since no new constraints arise at higher orders, we can then proceed to arbitrary orders in perturbation theory without further obstacles.

The gravitational case is qualitatively similar in many ways.  Consider in particular gravitational perturbation theory around global dS$_D$.  The SO$(D,1)$ isometries are unbroken diffeomorphisms and, since the Cauchy surfaces of global dS are compact, all diffeomorphisms are gauge symmetries.  The associated SO$(D,1)$ generators must therefore vanish and, at quadratic order, this simply sets to zero all de Sitter charges of the linearized theory.   At the classical level it is then straightforward to select linearized solutions with vanishing charges and to correct them at higher orders.

However, a subtlety arises in the quantum theory.  Since SO$(D,1)$ is non-compact, the spectra of its generators are generally continuous.  As a result, in the linearized theory, the only normalizable state with vanishing SO$(D,1)$ charges is the de Sitter-invariant vacuum $|0\rangle$.  Restricting to this state would then forbid the study of any excitations at all.

Nevertheless, a so-called group-averaging approach to constructing a larger Hilbert space for the perturbative theory was described by Higuchi in \cite{Higuchi_1991,Higuchi_1991_2}.  In essence, the idea is to first note that the linearized theory {\it does} contain states with vanishing charges, though they are non-normalizable\footnote{As described in \cite{Chen:2024rpx}, these non-normalizable states may be better thought of as well-defined weights on an appropriate algebra.}.  Since states that are annihilated by the de Sitter charges must be invariant under the de Sitter group, we will henceforth refer to these as de Sitter-invariant states.  It turns out that one may then usefully renormalize the inner product of the linearized theory to yield a well-defined Hilbert space $\mathcal H_{LPG}$ of de Sitter-invariant states satisfying the linearization stability constraints.
We will refer to this $\mathcal H_{LPG}$ as the Hilbert space of linearized perturbative gravity in the expectation that each state in $\mathcal H_{LPG}$ is indeed the linearized description of a state in the full quantum gravity theory.  

In particular, we will see that operators of the form \eqref{eq:ObsdS} are densely defined on $\mathcal H_{LPG}$.  The general theory of the Hilbert space $\mathcal H_{LPG}$ has been discussed in \cite{Marolf:1994wh,Marolf:1994ae,Ashtekar:1995zh,Giulini:1998rk,Marolf:2000iq} under a variety of names. It will be reviewed briefly in sections \ref{sec:review} and \ref{sec:GAobs} below, after which we analyze special observables of the form \eqref{eq:ObsdS} in section \ref{sec:relobsdS}.   It is useful to mention that the group averaging construction has also been called the method of coinvariants in \cite{Chandrasekaran:2022cip,Chen:2024rpx}.  See also \cite{Chakraborty:2023los} for a recent discussion of such constructions in the context of the gravitational path integral.

%\section{Group averaging in the presence of an observer}
%We saw in Section \ref{sec:intro} that perturbative gravity about a de Sitter background is subject to linearization stability constraints. To satisfy these constraints while still allowing for a nontrivial Hilbert space of states, we employ group averaging. In Section \ref{sec:review}, we review the group averaging formalism. Then, in Section \ref{sec:GAobs}, we introduce the formalism for group averaging in the presence of an observer, and discuss how it affects the standard inner products and correlators of QFT in curved (dS) spacetime.

\subsection{Review of group averaging}\label{sec:review}

It is natural to expand perturbative quantum gravity in powers of $G$.  As a result, the first-order theory will consist of linearized gravitons together with a matter quantum field theory on a fixed de Sitter background.  As mentioned above, we refer to the Hilbert space of this matter-plus-graviton theory as $\mathcal H_{QFT}$.  The matter quantum field theory can in principle be strongly coupled, though we will restrict to free theories below for simplicity.

The group averaging construction of \cite{Higuchi_1991,Higuchi_1991_2} can then be described as follows.  For a state $\big|\Psi\bigr] \in \mathcal H_{QFT}$, consider the formal integral
\begin{equation}
\label{eq:GAPsi}
    \ket{\Psi} = \int_{g \in G} dg U(g) \big|\Psi\bigr],
\end{equation}
where $G$ is the orthochronous de Sitter group SO$_0(D,1)$, $U(g)$ gives the unitary representation of $G$, and $dg$ is the Haar measure on $G$. Since $G$ is non-compact, the states $\ket{\Psi}$ are not normalizable using the standard inner product on $\mathcal{H}_{QFT}$.  Let us therfore introduce a new group-averaged inner product,
\begin{equation}\label{eq:innerproduct}
    \langle \Psi_1|\Psi_2\rangle := [\Psi_1| \cdot \ket{\Psi_2} = \int_{g \in G} dg \bigl[\Psi\big| U(g) \big|\Psi\bigr],
\end{equation}
which removes one integration over $g$.  The inner product \eqref{eq:innerproduct} thus effectively divides the old inner product by the (infinite) volume of the de Sitter group.  The Hilbert space $\mathcal H_{LPG}$ of de Sitter invariant states (which provide the linearized (L) description of valid perturbative gravity (PG) states) is then defined by choosing a useful linear space of states $V \subset \mathcal H_{QFT}$ with finite group-averaged inner products \eqref{eq:innerproduct} and completing the space spanned by their linear combinations (modulo null states).   

We note that the expression \eqref{eq:GAPsi} plays only a formal role in this construction and that one may alternately consider \eqref{eq:innerproduct} as a new inner product on the original states $\big|\Psi\bigr]$. With respect to this new inner product, states of the form  $(U(g)-1)\big|\Psi\bigr]$ are null states for all $g, \big|\Psi\bigr]$.  Using this description of the group-averaging inner product, the above construction was called the method of coinvariants in \cite{Chandrasekaran:2022cip,Chen:2024rpx}.

The group-averaging construction is useful when $V \subset \mathcal H_{QFT}$ results in a finite and positive semi-definite inner product \eqref{eq:innerproduct}.  Since \eqref{eq:innerproduct} clearly diverges for a de Sitter-invariant vacuum $\big|0\bigr]$, our $V$ can only contain states orthogonal to $\big|0\bigr]$.  This is not to say that there can be no well-defined state of quantum gravity associated with $\big|0\bigr]$, but merely that the inner product on $\big|0\bigr]$ should not be renormalized.  Furthermore, since $\big|0\bigr]$ is the unique normalizable de Sitter-invariant state in $\mathcal H_{QFT}$, any well-defined  de Sitter-invariant operator whose domain includes $\big|0\bigr]$ can only map $\big|0\bigr]$ to a multiple of itself.  Thus such observables cannot mix the state $\big|0\bigr]$ with states defined from the above domain $V$. As a result, unless one has good reason to introduce additional de Sitter invariant observables, it suffices to treat $\big|0\bigr]$ separately from all other states; see \cite{Marolf:1996gb} for general discussion of this issue.

In a theory with well-defined particle number (say, in the sense of being positive frequency with respect to global time), one might thus like to find that \eqref{eq:innerproduct} is both finite and positive definite for a natural space $V$ that is dense in the space of states with $N\ge 1$ particles.  As explained in appendix \ref{app:IP}, the full story is more complicated, and there remain holes in the existing literature associated with light scalar fields and fields with spin.  However, at least for gravitons on dS$_{3+1}$ and for scalar fields in any dimension with  mass $M > (D-1)/2\ell$, there is strong evidence that the above is essentially correct, though there are three subtleties.  

The first subtlety is that, as written, \eqref{eq:innerproduct}
is in fact ill-defined for $N=1$ particle states but, as explained in \ref{app:IP}, one should nevertheless define it to be zero for such states.  This result is the natural quantum analogue of the observation that single point-particles in dS always have at least one non-vanishing dS charge and, as a result, that a single particle never satisfies the linearization stability constraints described in the introduction.  It would be interesting to understand whether this feature might be related to complications described in \cite{Chandrasekaran:2022cip} when they attempted to introduce an observer in only one static patch.

The second subtlety is that two-particle states again have divergent group-averaging norm.  As described in appendix \ref{app:IP}, this appears to be associated with the fact that all classical 2-particle configurations with vanishing dS charges continue to leave a non-compact subgroup of SO$(D,1)$ unbroken and, as a result, well-defined de Sitter-invariant operators again cannot cause 2-particle states to mix with standard Fock states having $N\ge 3$ particles.

Finally, the third subtlety is that, while positivity for $3+1$-dimensional linearized gravitons was checked explicitly in \cite{Higuchi_1991_2}, there is not yet a complete proof that the group averaging inner product is positive semi-definite for all states of $N\ge 3$ particles of scalar fields with the masses indicated above.  See appendix \ref{app:IP} for discussion of the current status of this issue.

\subsection{Group averaging with a Reference}\label{sec:GAobs}

Let us now divide our de Sitter QFT into a target system and a reference system.  For simplicity we will assume that $\mathcal H_{QFT}$ takes the form of a tensor product,
\begin{equation}
\mathcal H_{QFT} = \mathcal H_{QFT}^\phi \otimes \mathcal H_{QFT}^\psi,
\end{equation}
where $\mathcal H_{QFT}^\psi$ describes a system to be used as a reference and $\mathcal H_{QFT}^\phi$ describes the target system whose physics we wish to more actively probe.  We will imagine that, before imposing de Sitter invariance, the system induces a definite pure state $\big|\psi_0\bigr] \in \mathcal H_{QFT}^\psi$, so that we need only consider states of the full system of the form $\big|\alpha\bigr] \otimes \big|\psi_0\bigr] \in \mathcal H_{QFT}$ for some $\big|\alpha\bigr] \in \mathcal H_{QFT}^\phi$.  Group averaging such states produces de Sitter-invariant states of the form
\begin{equation}\label{eq:alpha-psi}
    \ket{\alpha; LPG} := \int dg U(g) \big|\alpha\bigr] \otimes \big|\psi_0\bigr],
\end{equation}
which then live in the space of allowed states for perturbative gravity (PG) at order $G^0$ (at which the gravitational theory is linear (L)).
We will use $\mathcal H_{LPG,\psi_0}$ to refer to the Hilbert space defined by states of the form \eqref{eq:alpha-psi} using the group-averaging inner product.

We expect to be able to take a limit in which $\big|\psi_0\bigr]$ serves as a sharp reference within our de Sitter space, and with respect to which at least certain observables can be well-localized.  For example, for the right fields, and in the correct limit, $\big|\psi_0\bigr]$ could describe a very classical planet Earth equipped with all manner of laboratories and marked reference points with respect to which one could classically construct relational gauge-invariant observables (e.g., the average value of the Higgs field in the city of Paris during the opening ceremonies of the 2024 Olympics).  We therefore expect that, under the right conditions, we can also construct relational quantum observables which are well-described by local quantum field theory on a fixed de Sitter spacetime.

Before turning to the observables themselves, it is useful to further investigate the Hilbert space structure associated with the states \eqref{eq:alpha-psi}.
The group-averaging inner product of two such states takes the form
\begin{equation}\label{eq:fullIP}
    \bra{\beta; LPG}\ket{\alpha; LPG} = \int dg \bigl[\psi_0\big| U(g) \big|\psi_0\bigr] \bigl[\beta\big| U(g) \big|\alpha\bigr].
\end{equation}
The expression \eqref{eq:fullIP} is a convolution over the group of a state-dependent factor $ \bigl[\beta\big| U(g) \big|\alpha\bigr]$ and a factor $\bigl[\psi_0\big| U(g) \big|\psi_0\bigr]$ that will remain fixed so long as our reference system is undisturbed.  We will
refer to the fixed factor $\bigl[\psi_0\big|U(g)\big|\psi_0\bigr]$ as the {\it group averaging kernel}.

If there were a normalizable state $\big|\psi_0\bigr]$
for which this kernel was a Dirac delta-function, $\bigl[\psi_0\big|U(g)\big|\psi_0\bigr] = \delta(g)$, then the inner product \eqref{eq:fullIP} would reduce precisely to the inner product on $\mathcal{H}_{\phi}$, i.e. we would have $\bra{\beta; LPG}\ket{\alpha; LPG} = \bigl[\beta\big|\alpha\bigr]$.  While this seems unlikely to be the case for any normalizable state, it is nevertheless true that for {\it any} state $\big|\psi_0\bigr]$ with absolutely-convergent group-averaging norm
\begin{equation}
\label{eq:gaphi0}
\langle \psi_0 | \psi_0\rangle : = \int dg \bigl[\psi_0\big|U(g)\big|\psi_0\bigr],
\end{equation}
the appearance of this group-averaging kernel in \eqref{eq:fullIP} will tend to localize the integral over $g$ to a region surrounding the identity.  This follows from the fact that, since $U(g)$ is unitary, we must have $\bigl[\psi_0\big| U(g) \big|\psi_0\bigr] \le 1$ with equality only for $U(g)= \mathds{1}$.  Furthermore, if \eqref{eq:gaphi0} converges absolutely, then the kernel will suppress contributions far from the identity.

Of course,  this region may be very large for a general state $\big|\psi_0\bigr]$.  But we will study limits in which $\bigl[\psi_0\big| U(g) \big|\psi_0\bigr]$ becomes sharply peaked, so that the associated region is small.  The inner product \eqref{eq:fullIP} will then be given by the usual dS QFT inner product on $\mathcal H_{QFT}^\phi$ with small corrections associated with the finite width of the peak of
$\bigl[\psi_0\big| U(g) \big|\psi_0\bigr]$.

We will characterize these corrections more precisely in sections \ref{sec:1+1} and \ref{sec:d+1} for particular choices of reference state $\big|\psi_0\bigr]$. In particular, we will see there that the associated corrections to correlation functions are not uniformly small across the entire de Sitter space, but that their size depends on the location of the arguments of such correlators in relation to structures defined by $\big|\psi_0\bigr]$.

Let us now consider the case where $\phi$ and $\psi$ describe independent local quantum fields with no mutual interactions.  For the moment, we will still allow both $\phi$ and $\psi$ to have self-interactions.  Furthermore, we can in fact allow $\phi$ to denote a collection of mutually-interacting quantum fields, and similarly for $\psi$, so long as the fields of $\phi$ and the fields of $\psi$ do not interact with each other\footnote{We expect that the inclusion of perturbative interactions between the $\phi$ and $\psi$ will be straightforward, but we will not pursue it here.  We thus see no obstacle to including linearized gravitons in either $\phi$ or $\psi$.}.

\subsection{Relational Observables for perturbative dS gravity}
\label{sec:relobsdS}

Since operators can be built from bra- and ket-states, and since limits where $\bigl[\psi_0\big| U(g) \big|\psi_0\bigr]$  is sharply peaked make $\mathcal H_{LPG,\psi_0}$ canonically isomorphic to $\mathcal H_{QFT}^\phi$, we should also expect there to be an algebra of gauge-invariant (i.e., de Sitter-invariant) observables that reduces in this limit to the algebra of local $\phi$-fields.  The construction we will use is a direct analogue of our construction \eqref{eq:alpha-psi} of quantum states.  Given a local field $\hat \phi_{QFT}$ that acts on $\mathcal H_{QFT}^\phi$, for any point $x$ in global dS we simply define the operators
\begin{eqnarray}
    \hat{\phi}_{LPG}(x) &:=& \int \diff{g} U(g) \Bigl(\hat{\phi}_{QFT}(x) \otimes \big|\psi_0\bigr]\bigl[\psi_0\big| \Bigr) U(g^{-1}) \cr &=& \int \diff{g} \left( \hat{\phi}_{QFT}(gx) \otimes U(g)\big|\psi_0\bigr]\bigl[\psi_0\big| U(g^{-1})\right)      \label{eq:PGfield},
\end{eqnarray}
where $gx$ denotes the image of the point $x$ under the de Sitter isometry $g$.
Note that, even for a fixed value of $x$, we can use the fact that the Haar measure is invariant under $g \rightarrow g_0^{-1} g$ to write $U(g_0)\hat{\phi}_{LPG}(x) = \hat{\phi}_{LPG}(x) U(g_0)$.  The operators \eqref{eq:PGfield} are thus de Sitter-invariant and represent observables of the linearized perturbative gravity (LPG) theory for each fixed value of $x$.

As foreshadowed in the introduction, this construction makes use of non-local elements both in the integral over the group of de Sitter isometries and also through the explicit use of the global quantum state $\big|\psi_0\bigr]$.  This feature will play a critical role in ensuring that the operators \eqref{eq:PGfield} are well-defined and, in particular, that they have finite matrix elements between states of the form \eqref{eq:alpha-psi}.  Such matrix elements are computed by applying the operators \eqref{eq:PGfield} to a state $|\alpha; LPG\rangle$ and then taking the group-averaging inner product with $|\beta; LPG\rangle$.  Recalling the definition of our states \eqref{eq:alpha-psi}, and since the group-averaging inner product \eqref{eq:fullIP} removes one of the integrals over the de Sitter group,  the result then
takes the form:
\begin{align}
    &\bra{\beta;LPG}\hat{\phi}_{LPG}(x_1)\hat{\phi}_{LPG}(x_2)\dots \hat{\phi}_{LPG}(x_n)\ket{\alpha;LPG} \nonumber\\
    &= \int \diff{g_1} \diff{g_2}\dots \diff{g_{n+1}} \Bigl(\bigl[\psi_0\big|U(g_1)\big|\psi_0\bigr] \prod_{k=1}^{n} \bigl[\psi_0\big|U(g_k^{-1} g_{k+1})\big|\psi_0\bigr]\Bigr) \nonumber\\
    &\quad \times \bigl[\beta\big|\hat{\phi}_{QFT}(g_1x_1) \dots \hat{\phi}_{QFT}(g_nx_n)U(g_{n+1})\big|\alpha\bigr].
 \label{eq:PGME}
\end{align}
Convergence of the integrals in \eqref{eq:PGME} is guaranteed by the absolute convergence of \eqref{eq:gaphi0}.  Again, it is manifest that \eqref{eq:PGME} reduces to the $\phi$ correlators of dS QFT in limits where our group-averaging kernel $\bigl[\psi_0\big|U(g)\big|\psi_0\bigr]$ approaches $\delta(g)$.

In the particular case where either $\big|\alpha\bigr]$ or $\big|\beta\bigr]$ is the de Sitter-invariant $\phi$-vacuum $\big|0;\phi\bigr]$, the factor of $U(g_{n+1})$ can be dropped from the final line.  Since it will be natural to focus on this case below, we define the notation
\begin{equation}
\label{eq:LPGvac}
    \ket{0}_{LPG} := \int \diff{g} U(g) \big(\big|0;\phi\bigr] \otimes \big|\psi_0\bigr] \big).
\end{equation}
Thus we may also define
\begin{align}
    \expval{\hat{\phi}_{LPG}(x_1)\dots \hat{\phi}_{LPG}(x_n)} &:= \bra{0;LPG}\hat{\phi}_{LPG}(x_1)\dots \hat{\phi}_{LPG}(x_n) \ket{0;LPG} \nonumber\\
    &= \int \diff{g_1} \diff{g_2}\dots \diff{g_{n+1}} \left(\bigl[\psi_0\big|U(g_1)\big|\psi_0\bigr] \prod_{k=1}^{n} \bigl[\psi_0\big|U(g_k^{-1} g_{k+1})\big|\psi_0\bigr]\right) \nonumber\\
    &\quad \times \bigl[0;\phi\big|\hat{\phi}_{QFT}(g_1x_1) \dots \hat{\phi}_{QFT}(g_nx_n)\big|0;\phi\bigr]. 
 \label{eq:PGvaccorr}
\end{align}
However, we emphasize that the state \eqref{eq:LPGvac} is only vacuum with respect to $\phi$, and that the $\psi$ field is in a group-averaged version of the state $\big|\psi_0\bigr]$.  In particular, the state $\ket{0}_{LPG}$ defined above still contains our reference and is thus {\it not} the vacuum of the full perturbative gravity theory.  It will thus induce non-trivial backreaction at higher orders in $G$.

Since states $\big|\psi_0\bigr]$ with absolutely-convergent group-averaging norm will define group averaging kernels $\bigl[\psi_0\big|U(g)\big|\psi_0\bigr]$ with finite width,   choosing some of the points $x_i$ to lie on each other's light-cones will cause \eqref{eq:PGvaccorr} to differ infinitely from the corresponding correlator in dS QFT.  However, this strong difference is clearly associated with very high energies.  It is thus useful to describe the manner in which \eqref{eq:PGvaccorr} approximates correlators of dS QFT by studying smeared correlators (which are sensitive only to the physics below an energy scale set by the smearing function).  For example, in a free theory it suffices to study smeared two-point functions of the form
\begin{align}
\label{eq:smear2}
    \int \sqrt{-g(x_1)} dx_1 \,  \sqrt{-g(x_2)}  dx_2 \, F_1(x_1)F_2(x_2)\expval{\hat\phi_{LPG}(x_1)\hat\phi_{LPG}(x_2)}
\end{align}
for appropriate smearing functions $F_1,F_2$.  We will focus below on the case where $F_1,F_2$ are members of a family of functions $F_y(x)$ that are well-approximated by Gaussian functions of both the global time $t$ and of the location on the sphere $S^d$ at fixed $t$ that are peaked at the point $y$ in global dS.  We will take the width of these Gaussians in both global time $t$ and in location on the sphere to be identical as measured in terms of proper time and distance.  For example, when $y$ lies at the north pole of the sphere $S^d$ at some time $t_y$, and for an appropriate normalization factor $N$ we may take
\begin{equation}
\label{eq:GaussianF}
F_y (x) = N e^{-\frac{(t-t_y)^2}{2\sigma^2}}e^{\frac{\ell^2}{\sigma^2}[\cos(\theta)-1] \cosh^2(t/\ell)} \approx  N e^{-\frac{1}{2\sigma^2}\left[(t-t_y)^2 + \ell^2 \theta^2\cosh^2(t/\ell)\right]},
\end{equation}
where $t, \theta$ are the global coordinates of the point $x$ with $\theta$ being the polar angle on $S^d$, and where the final approximation holds for $\theta \ll 1$.  We note for future reference that the final exponent on the right-hand-side defines an effective flat Euclidean-signature metric with line element
\begin{equation}
\label{eq:Eds2}
ds^2_E :=  dt^2 + \ell^2 \cosh^2 (t/\ell) \left(d\theta^2 + \theta^2 \, d\Omega_{d-1}^2\right),
\end{equation}
so that
\begin{equation}
\label{eq:GaussianFED}
F_y (x) \approx e^{-\frac{|x-y|_E^2}{2\sigma^2}},
\end{equation}
where $|x-y|_E$ is the Euclidean distance between $x$ and $y$ defined by\footnote{The astute reader will note that, if we wish to define $F_y$ for general $y$ away from the north pole by rotating \eqref{eq:GaussianF}, then the analogue of \eqref{eq:Eds2} in fact depends on the location of $y$ on the sphere $S^d$.  While this fact is not explicitly indicated by the notation $|x-y|_E$, it will not play an important role in our analysis.} \eqref{eq:Eds2}.

 We also note that the effect of convolving smeared dS QFT correlators with our group averaging kernel will depend on the extent to which $F_y(x)$ differs from $F_y(gx)$, and thus the extent to which the exponent on the right-hand-side of \eqref{eq:GaussianF} differs between $x$ and $gx$. Since the triangle inequality bounds the change in $|x-y|_E$ in terms of $|gx-x|_E$, we see that for $|x-y|_E \lesssim \sigma$ the change in the smearing function $F_y$ is small when $|gx-x|_E \ll \sigma$.

\begin{figure}[H]
  \centering
  \includegraphics[width = 3in]{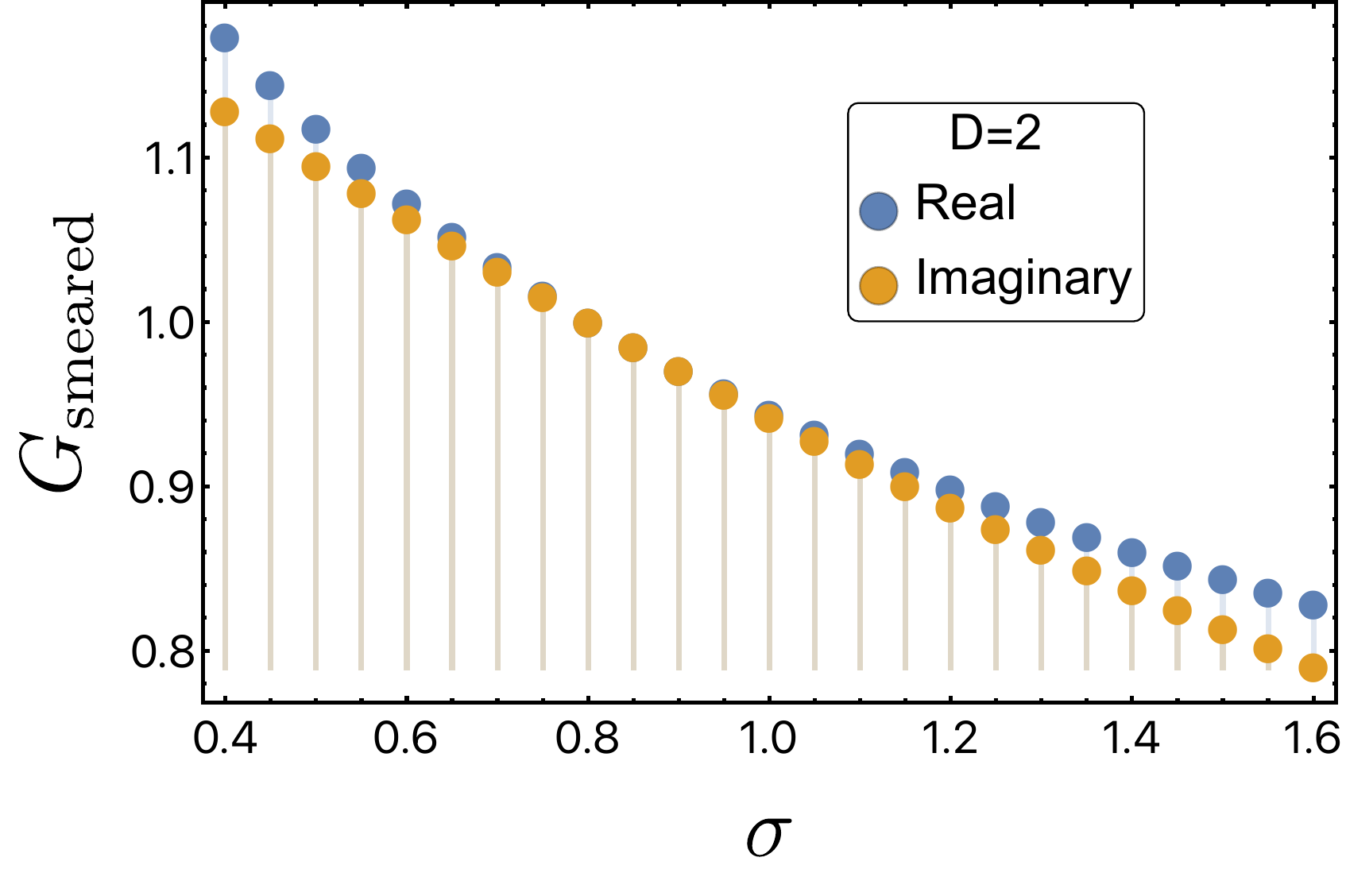}
  \includegraphics[width = 3in]{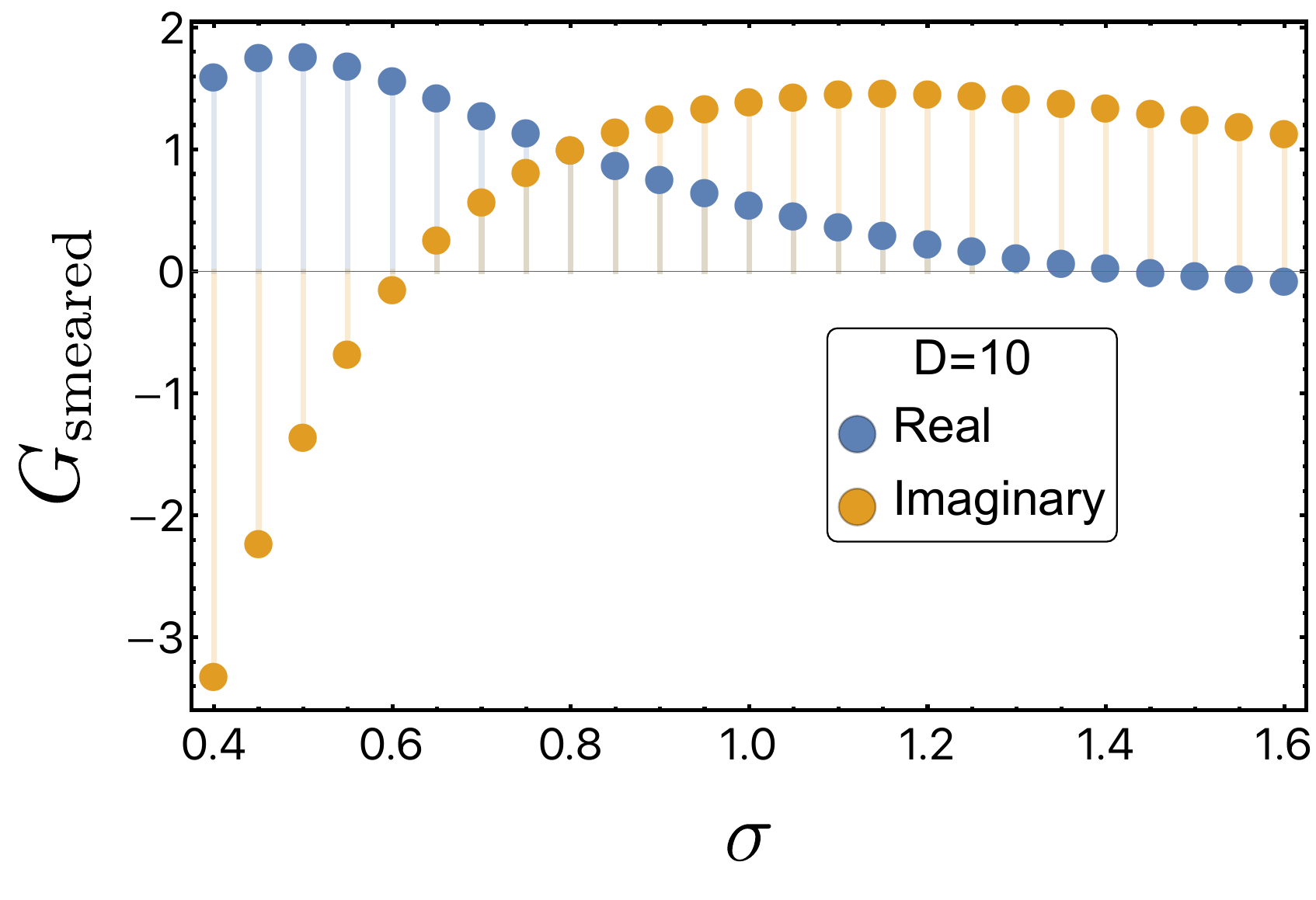}
  \caption{Numerical examples illustrating continuity of the real (blue) and imaginary (orange) parts of smeared correlators under changes of the smearing functions.   Here we consider a free scalar two-point function and take the smearing function $F_y(x)$ to be of the near-Gaussian form  \eqref{eq:GaussianF}.  Plotting $G_{smeared}(x_0, y, \sigma) : = {\cal N}^{-1} \int_{dS_D} d^Dx \sqrt{-g} F_y(x) \bigl[0;\phi\big| \hat \phi_{QFT}(x) \hat \phi_{QFT}(x_0) \big|0;\phi\bigr]$  as a function of the smearing width $\sigma$ gives results consistent with the desired continuity.  Results are shown for scalars of mass $M=d/2\ell$ (for which $\mu=0$).      The left panel shows $D=2$ while the right panel shows $D=10$.  We take the point $y$ to lie at $\theta=0$  and $t=3\ell$, with $x_0$ on the past light cone of $y$ at $t=0$. Numerical integrations were performed along a deformed contour that avoids the singularities (in a regime where the results are stable with respect to such deformations).
 Blue and orange show the real and the imaginary parts.  The normalization $\mathcal N$ was arbitrarily chosen to set $G_{smeared}=1$ at $\sigma=0.8$.   }
  \label{fig:correlator}
\end{figure}

 The question of whether this small change in the smearing function can cause a significant change in smeared correlators can be studied by using a standard partition of unity to divide the domain of integration into subregions based on whether subsets of the $x_i$ are close together or far apart.
  For example, when studying the smeared two-point function \eqref{eq:smear2}, we divide the domain of integration into a region with (Lorentz-signature de Sitter distance) $-\epsilon \lesssim |x_1-x_2|^2 \lesssim \epsilon$ and regions with
  $|x_1-x_2|^2 \gtrsim \epsilon$ and $|x_1-x_2|^2 \lesssim -\epsilon$.  In the former region, the integral is well-approximated by a smeared Minkowski-space correlator. Since the Wightman axioms  require Minkowski-space correlators to be tempered distributions \cite{Streater:1989vi}, and since tempered distributions are continuous linear functionals on the space of test functions, the integral over this region will change by only a small amount under a small change in the smearing functions $F_y$.   Here it is important to that we consider Wightman correlators rather than their time-ordered counterparts.  Similar continuity follows for the integral over the regions $|x_1-x_2|^2 \gtrsim \epsilon$ and $|x_1-x_2|^2 \lesssim - \epsilon$ since the correlator is bounded in those regions and the smearing function changes by a function of integrable norm; (i.e., by a function in $L^1(dS)$)\footnote{For free fields on dS,  one may alternatively proceed by writing the field operator as an expansion in global dS mode functions that solve the equation of motion.  Integrating any given mode against a smooth function of time yields a result that must vanish faster than any polynomial as the angular momentum of the mode becomes large. It thus follows that the relevant mode sums converge absolutely.  The desired result then follows from the fact that the above convolution makes negligible change in the high-frequency components of $F_y$.}.  {Inserted:} This continuity is illustrated by the numerical examples in figure \ref{fig:correlator}.
  As a result,  when $|gx-x|_E \ll \sigma$ for all $g$ within the peak of the group-averaging kernel, a given set of smeared dS QFT correlators will be well-approximated by the correspondingly smeared versions of the perturbative gravity correlators \eqref{eq:PGvaccorr}. 

\section{Reference states in dS$_{1+1}$}\label{sec:1+1}

The above section described our general framework for using perturbative gravity to approximate the algebra of local observables in dS QFT. There we saw that a central role is played by the group averaging kernel $\bigl[\psi_0\big|U(g)\big|\psi_0\bigr]$, and that comparison of the perturbative gravity and dS QFT correlators is controlled by i) the width of the peak of $\bigl[\psi_0\big|U(g)\big|\psi_0\bigr]$ about the identity and ii) by the effect of those isometries $g$ that lie within the above peak on the points $x$ at which we wish to evaluate such correlators.

We thus now turn to a detailed investigation of this kernel for interesting classes of states.  In this section we consider the simple-but-illustrative case of $1+1$ global de Sitter, taking the field $\psi$ to be a collection of free scalar fields with mass $M> 1/2\ell$ (so that the one-particle states lie in principal series representations of SO$(2,1)$ \cite{VK,Wong:1974cv}).  Thinking of $\psi$ as a {\it collection} of fields allows us to choose each particle to be associated with a distinct scalar field.  We may thus treat the $\psi$-particles as distinguishable, which provides a slight simplification of the calculations.
 While Einstein-Hilbert gravity is trivial for the case $D=2$, our goal is to use $D=2$ as a toy model of higher dimensional physics.  We thus simply analytically continue certain formulae from higher dimensions to $D=2$ in order to discuss versions of $D=2$ linerization stability constraints (which are again solved by group averaging), $D=2$ perturbative gravity operators, and a (dimensionless) $D=2$ Newton constant $G$.  However, we will postpone any more involved discussions of back-reaction to section \ref{sec:d+1} (where the higher-dimensional case will be addressed).

\subsection{Preliminaries}
Studying our kernel requires an understanding of how the de Sitter isometries act on our states.  It is useful to begin by recalling that, in global coordinates, the metric on dS$_{1+1}$ takes the form
\begin{equation}
    ds^2 = -dt^2 + \ell^2 \cosh^2 (t/\ell)\, d\theta^2, 
\end{equation}
where $\ell$ is the de Sitter scale.
It will sometimes also be useful to write the metric in conformal coordinates $T,\theta$, where $\cosh (t/\ell) = \sec T$ with $T \in[-\pi/2,\pi/2]$, so that the line element becomes
\begin{align}
    ds^2 = \frac{\ell^2}{\cos^2T}(-dT^2 + d\theta^2).
\end{align}
In these coordinates, the generators of the isometry group are
\begin{align}
    &B_1 \equiv \xi_1 \equiv \cos T \cos\theta \partial_T - \sin T\sin\theta \partial_\theta, \label{eq:B1action}\\
    &B_2 \equiv \xi_2 \equiv \cos T \sin\theta \partial_T + \sin T\cos\theta \partial_\theta, \label{eq:B2action} \\ 
    &R \equiv \xi_\theta \equiv \pdv{\theta}, \label{eq:Raction}
\end{align}
where the notation $B_{1}$, $B_2$, $R$ classifies the generators of SO$(2,1)$ according to their action as either boosts or rotations when one thinks of SO$(2,1)$ as Lorentz transformations on 2+1 Minkowski space.
The actions of these Killing fields on de Sitter space is shown in Figure \ref{fig:B1B2R} below.

\begin{figure}[h!]
    \centering
    \includegraphics[width=0.31\textwidth]{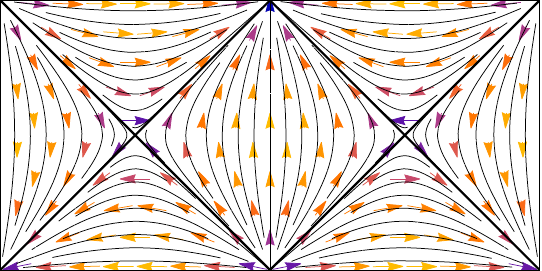}\hspace{.3cm}
    \includegraphics[width=0.31\textwidth]{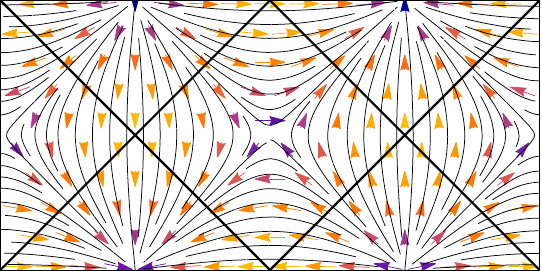}\hspace{.3cm}
    \includegraphics[width=0.31\textwidth]{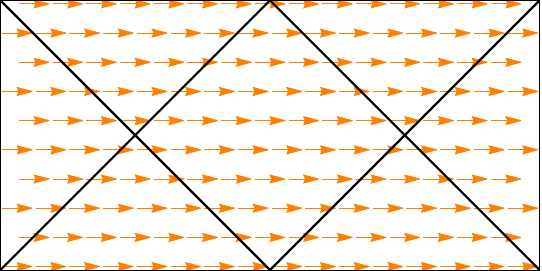}
    \caption{The Killing vector fields $B_1$ (left), $B_2$ (center), and $R$ (right),  in dS$_{1+1}$.  The figures are drawn using conformal coordinates $T, \theta$ with $-\pi<\theta<\pi$, $-\pi/2<T<\pi/2$.}
    \label{fig:B1B2R}
\end{figure}

The action of the generators on our states can be understood by defining

\begin{align}
     B_\pm \equiv B_2 \pm i B_1. \quad \text{Hence } B_1 = \frac{1}{2i}(B_+-B_-), \ \text{and}  \ B_2 = \frac12(B_++B_-).
\end{align}
We then have the commutation relations \cite{Sun:2021thf}.
\begin{align}
    [R,B_+]=B_+, \quad [R,B_-]=-B_-, \quad [B_-,B_+]=2R.
\end{align}
The Casimir operator $\mathcal{C}_2$ is given by
\begin{align}
    \mathcal{C}_2 &= B_+B_-+R(1-R) = B_1^2 + B_2^2 - R^2.
\end{align}

We now let $|m]$ be the 1-particle eigenstate of the $R$ operator with $R\big|m\bigr]=m\big|m\bigr]$, where $m$ is an integer.  In particular, we may take it to be the state created from the $\psi$-vacuum by acting with the creation operator associated with the usual mode of the scalar field $\psi$ having angular quantum number $m$.
For one-particle states, the Klein-Gordon equation gives $\mathcal{C}_2 |m] =  M^2 \ell^2|m]$, where $M$ is the mass of the scalar field and where we consider only the case $M>1/2\ell$.  It will be useful to introduce the parameter $\Delta$ through
\begin{equation}
\label{eq:DeltaCas}
\Delta(1-\Delta) = M^2 \ell^2.
\end{equation}
By convention, we take the imaginary part of $\Delta$ to be positive, so that we have $\Delta = \frac{1}{2} + i \sqrt{M^2\ell^2 -1/4}$.   It is then straightforward to check that we obtain a unitary representation of the above relations by imposing\textbf{}
\begin{align}\label{eq:B-,B+}
    &B_-\big|m\bigr] = -i|m-\Delta|\big|m-1\bigr], \quad B_+\big|m\bigr] = i|m+\Delta| \big|m+1\bigr],\\
    &B_1\big|m\bigr] = \frac1{2}\bigg[|m+\Delta| \big|m+1\bigr]+|m-\Delta|\big|m-1\bigr]\bigg], \label{eq:B1} \\
    &B_2\big|m\bigr] = \frac{i}{2}\bigg[|m+\Delta| \big|m+1\bigr]-|m-\Delta| \big|m-1\bigr]\bigg].
    \label{eq:B-,B+end}
\end{align}
Below,  we will also use the notation
\begin{equation}
\label{eq:mudef}
\mu = {\rm Im} \Delta = \sqrt{M^2\ell^2 -1/4}.
\end{equation}

Let us now conclude our discussion of preliminaries by reviewing the results of \cite{Marolf_2009} describing the asymptotics of the kernel $\bigl[\psi_0\big|U(g)\big|\psi_0\bigr]$ at large $g$.  For this purpose we will in fact consider scalar fields of any mass $M>0$,  where we take $\Delta$ to be defined by \eqref{eq:DeltaCas} with the convention that $\Delta < 1/2$ for $M< 1/2\ell$.    We will take $\big|\psi_0\bigr]$ to be an $N$-particle state and, for each particle, we take the state to be a finite linear combination of the above states $|m]$.

The results of \cite{Marolf_2009} were expressed by writing a general  $g\in \text{SO}_0(2,1)$ in the form $g=e^{i\theta_1 R}e^{i\lambda B_{1}}e^{i\theta_2 R}$. Since $\theta_1,\theta_2$ range over a compact space, and since the resulting rotations simply map any given angular momentum mode to a superposition of other such modes, the large $g$ behavior is controlled by the behavior of
$e^{i\lambda B_{1}}$ at large $\lambda$.  And since de Sitter isometries act diagonally on multi-particle states, our group averaging kernel will contain factors of $[ {m}|e^{i\lambda B_{1}} | {n}]$ for each particle, where $n$ again denotes an angular quantum number. At large $\lambda$, the asymptotic behavior of this 1-particle matrix element was shown to be
\begin{equation}
   \Bigg| [ {m}|e^{i\lambda B_{1}} | \vec{n}]  \Bigg| \sim e^{-\lambda {\rm Re}\, \Delta}
\end{equation}
so that the group averaging kernel decays exponentially with large boost parameter.   Since the relevant integration measure for dS$_{1+1}$ is  $\sinh \lambda \, d\lambda$ (see again \cite{Marolf_2009}), we see that \eqref{eq:gaphi0} converges absolutely for $N {\rm Re}\, \Delta >  1$.  In particular, for $M> 1/2\ell$ we require $N\ge 3$.  As discussed in \cite{Marolf_2009}, this is also the case for scalar fields of mass $M> d/2\ell$ in $dS_{d+1}$ for any $d\in {\mathbb Z}^+$.

\subsection{A pair of reference particles on opposite sides of dS}\label{sec:one-obs}
Having seen that our kernel strongly suppresses contributions from large $g$ for $N\ge 3$ particles with $M>1/2\ell$, we can now turn our attention to the region near the identity $(g\sim \mathds{1})$ at which the group-averaging kernel is peaked.
Though one can certainly engineer special cases where there are also important contributions from outside this peak, having a second peak of height near $1$ clearly requires fine tuning.  Furthermore, if the height of another peak is not near $1$, a different form of fine tuning is required if its contributions are to be comparable to or greater than those of the unit-height central peak. Since we do not expect this to occur for generic $\big|\psi_0\bigr]$, we will defer consideration of this possibility until introducing the particular states we wish to study.  At that point, we will display numerical results supporting the above expectations. 

For the moment, however, we will simply compute the width of the central peak for interesting classes of states by writing
$U(g) = e^{i(\lambda_1 B_1+\lambda_2B_2+\theta R)}$ and expanding to second order in $\lambda_1, \lambda_2, \theta$.
We will in fact focus on the simple case of 2-particle states (with $M> 1/2\ell$).  As noted above, to control contributions from large $g$ the full state $\big|\psi_0\bigr]$ must contain a third particle.  But we are free to choose the state of the third particle to have a much broader peak near the identity (perhaps engineered by smearing an arbitrary state over a large-but-finite range of de Sitter transformations), so that the width of the kernel is set by just the first two particles.

At the classical level, a pair of identical-mass particles can satisfy the linearization stability constraints in $dS_{1+1}$ only if their de Sitter charges cancel exactly.  This requires the geodesics followed by the two particles to be related by a rotation through an angle $\pi$ for some rotation generator.  We will take this to be the rotation $e^{i\pi R}$.  We will thus choose a quantum state $\big|\psi_+\bigr]$ for the first particle and then simply take the state of the second particle to be 
\begin{equation}
\label{eq:pmrotate}
    \big|\psi_-\bigr] = e^{i\pi R} \big|\psi_+\bigr],
\end{equation} with the full 2-particle reference state being the tensor product
\begin{equation}
\label{eq:antpart}
    \big|\psi_0\bigr]=\big|\psi_+\bigr]\otimes \big|\psi_-\bigr].
\end{equation}

\begin{figure}[t]
    \centering
    \begin{tikzpicture}

  \pgfdeclareimage[height=5.5cm]{img}{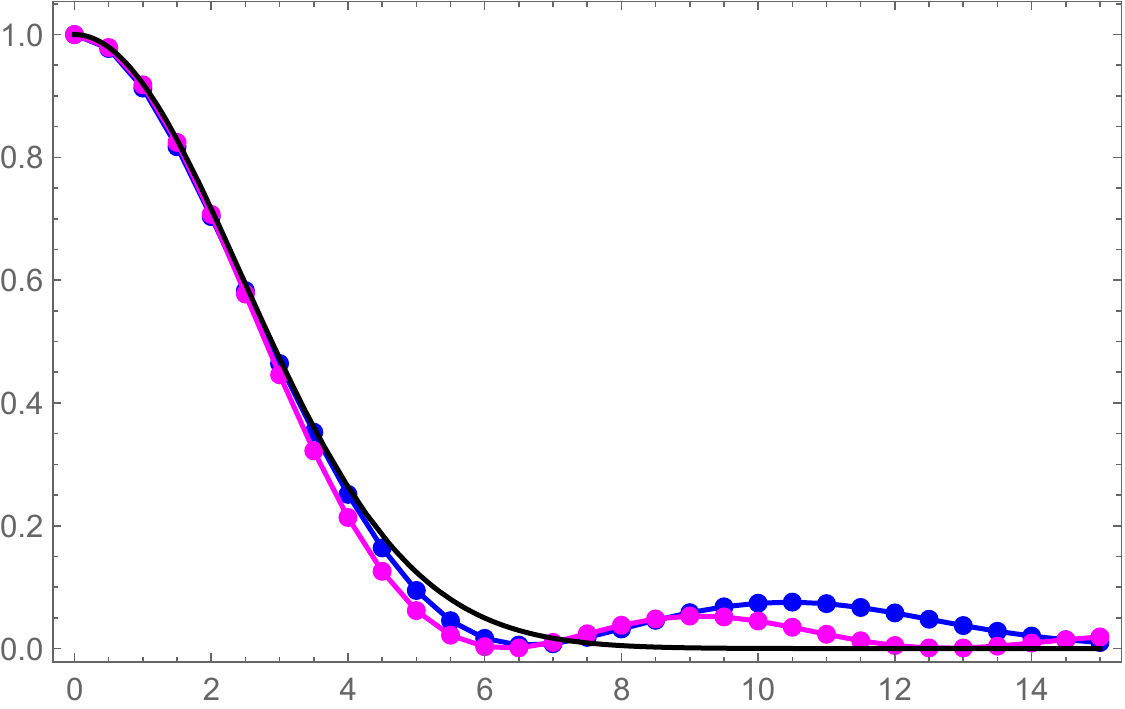}
  \node (img1) at (0,0) {\pgfuseimage{img}};

  \draw (.2,-3.1) node {\large $j_{\mathsmaller\star}\lambda_1$};
  \draw (-4.9,0) node[rotate=90] {\Large $[\psi_0|\,e^{i\lambda_1 B_1}\,|\psi_0]$};

  % \draw (2.8,1.7)  node[minimum size=1cm,draw,text width=2cm,align=left] {\tikz\draw[blue,fill=blue] (0,0) circle (.07cm); 
  % $j_{\mathsmaller \star}=10$ \\
  % \tikz\draw[pinkk,fill=pinkk] (0,0) circle (.07cm); 
  % $j_{\mathsmaller \star}=50$\\
  % \tikz\draw[black,fill=black] (0,0) rectangle ++(.17,.03); 
  % $e^{-(j_{\mathsmaller \star}\lambda)^2/12}$};

   \node [matrix,draw=black] (my matrix) at (2.7,1.4)
  {
    \fill[blue] (0,0)   circle (.7mm); & \node {$j_{\mathsmaller \star}=10$};        \\
    \fill[pinkk] (0,0)   circle (.7mm); & \node {$j_{\mathsmaller \star}=50$};        \\
    \fill[black] (-.09,-.1) rectangle ++(.2,.04); & \node {$e^{-(j_{\mathsmaller \star}\lambda)^2/12}$}; \\
  };

  \draw (.2,-4) node {(\textit{a})};

\begin{scope}[xshift=10.5cm]
    \pgfdeclareimage[height=5.5cm]{img2}{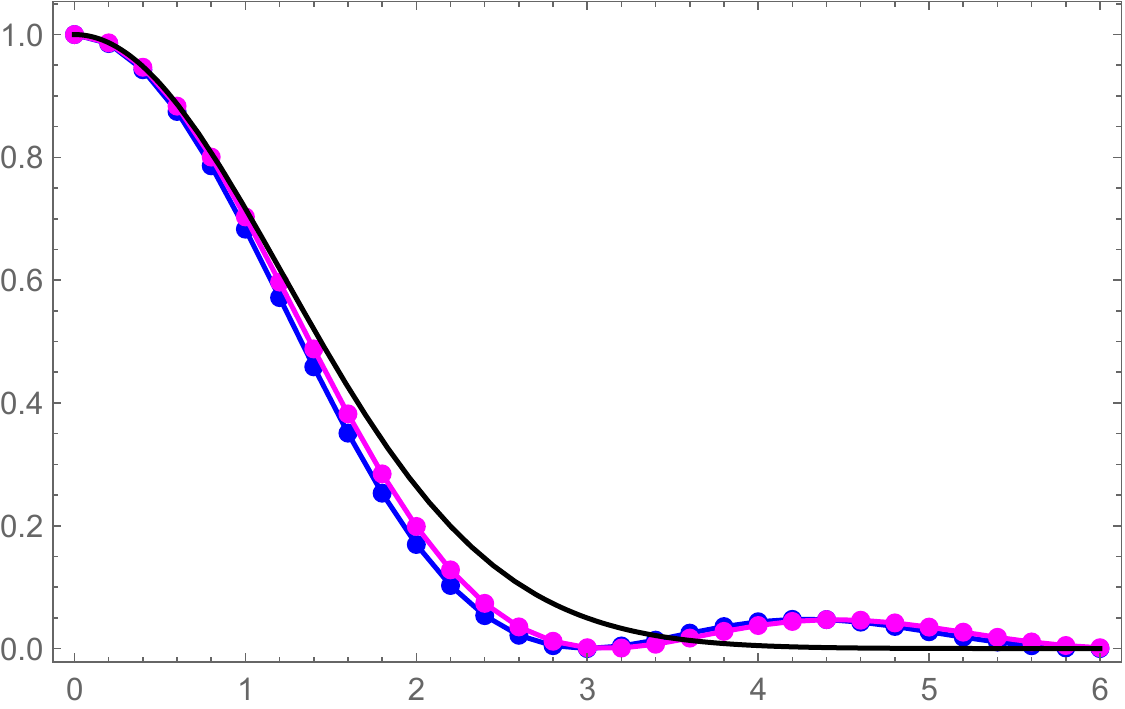}
  \node (img12) at (0,0) {\pgfuseimage{img2}};

  \draw (.2,-3.1) node {\large $j_{\mathsmaller \star} \theta$};
  \draw (-4.9,0) node[rotate=90] {\Large $[\psi_0|\,e^{i\theta R}\,|\psi_0]$};

  % \draw (3,1.5)  node[minimum size=1cm,draw,text width=1.5cm,align=center] {\tikz\draw[blue,fill=blue] (0,0) circle (.07cm); 
  % $j_{\mathsmaller \star}=10$ \\
  % \tikz\draw[pinkk,fill=pinkk] (0,0) circle (.07cm); 
  % $j_{\mathsmaller \star}=50$};

  \node [matrix,draw=black] (my matrix 2) at (2.7,1.4)
  {
    \fill[blue] (0,0)   circle (.7mm); & \node {$j_{\mathsmaller \star}=10$};        \\
    \fill[pinkk] (0,0)   circle (.7mm); & \node {$j_{\mathsmaller \star}=50$};        \\
    \fill[black] (-.09,-.1) rectangle ++(.2,.04); & \node {$e^{-(j_{\mathsmaller \star}\lambda)^2/3}$}; \\
  };

  \draw (.2,-4) node {(\textit{b})};
\end{scope}

\begin{scope}[yshift=-8cm,xshift=.5cm]
    \pgfdeclareimage[height=5.5cm]{img3}{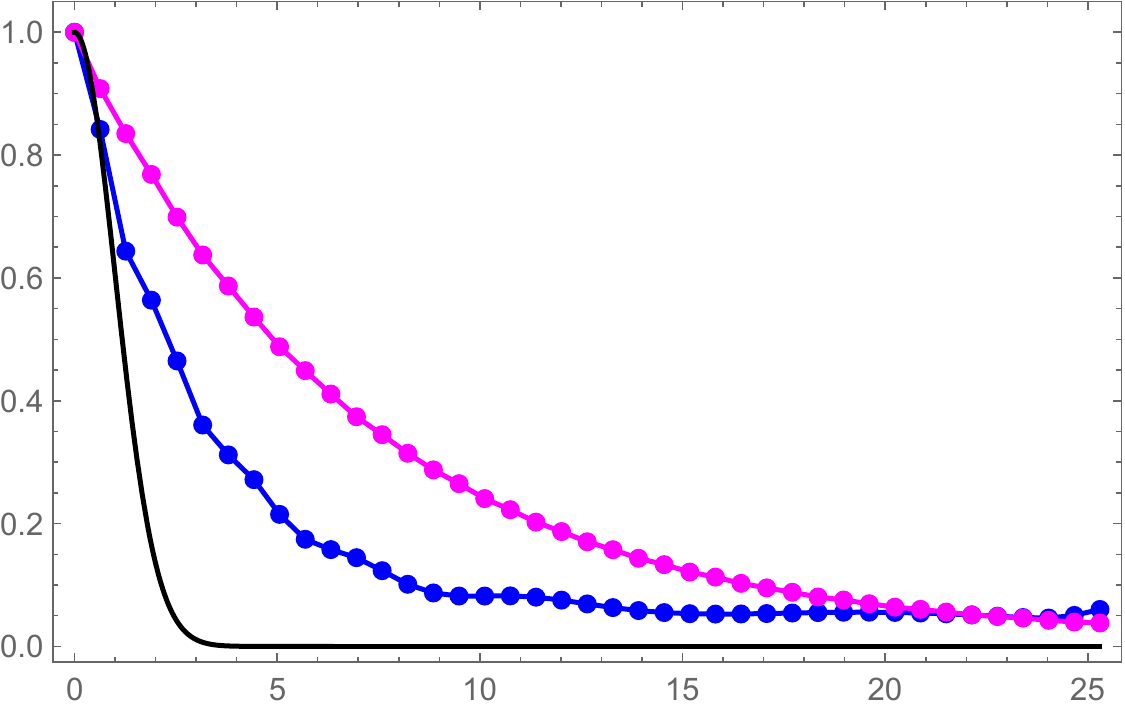}
  \node (img13) at (0,0) {\pgfuseimage{img3}};

  \draw (.2,-3.1) node {\large $\sqrt{j_{\mathsmaller \star}} \lambda_2$};
  \draw (-4.9,0) node[rotate=90] {\Large $[\psi_0|\,e^{i\lambda_2 B_2}\,|\psi_0]$};

  % \draw (3,1.5)  node[minimum size=1cm,draw,text width=1.5cm,align=center] {\tikz\draw[blue,fill=blue] (0,0) circle (.07cm); 
  % $j_{\mathsmaller \star}=10$ \\
  % \tikz\draw[pinkk,fill=pinkk] (0,0) circle (.07cm); 
  % $j_{\mathsmaller \star}=50$};

  \node [matrix,draw=black] (my matrix 3) at (2.7,1.4)
  {
    \fill[blue] (0,0)   circle (.7mm); & \node {$j_{\mathsmaller \star}=10$};        \\
    \fill[pinkk] (0,0)   circle (.7mm); & \node {$j_{\mathsmaller \star}=50$};        \\
    \fill[black] (-.09,-.1) rectangle ++(.2,.04); & \node {$e^{-(\sqrt{j_{\mathsmaller \star}}\lambda)^2/2}$}; \\
  };

  \draw[red!95!black,very thick] (-3.9,0) rectangle ++(1.6,2.7);

  \draw[dashed,red!95!black,very thick] (-2.3,2.7)-- (6.5,3.2);
  \draw[dashed,red!95!black,very thick] (-2.3,.3)-- (6.5,-2.1);

  \draw (5,-4) node {(\textit{c})};
\end{scope}

\begin{scope}[yshift=-7.5cm,xshift=10.5cm]
    \pgfdeclareimage[height=5cm]{img3}{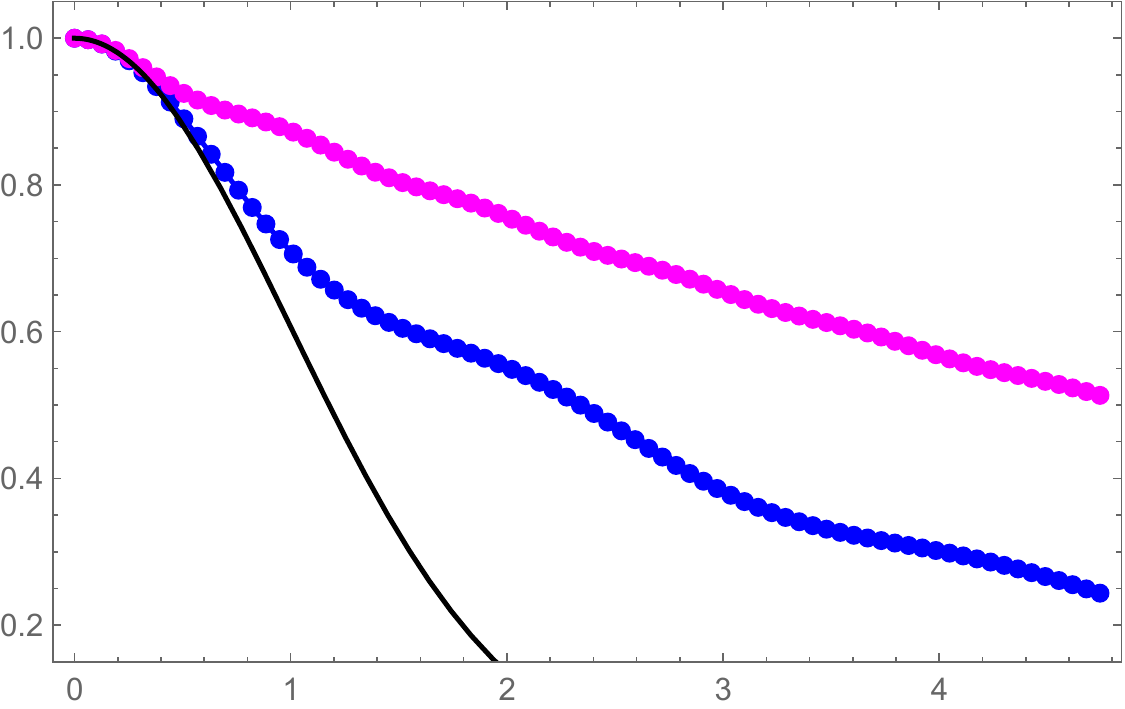}
  \node (img13) at (0,0) {\pgfuseimage{img3}};

  \draw (.2,-2.8) node { $\sqrt{j_{\mathsmaller \star}} \lambda_2$};
  \draw (-4.35,0) node[rotate=90] { $[\psi_0|\,e^{i\lambda_2 B_2}\,|\psi_0]$};

\end{scope}

\end{tikzpicture}
    \caption{The positive real group averaging kernel $\bigl[\psi_0\big| e^{i\lambda_i B_i}\big|\psi_0\bigr] = \Big|\bigl[\psi_+\big| e^{i\lambda_i B_i}\big|\psi_+\bigr]\Big|^2$ for $\mu=0$ in the 2-particle state $\big|\psi_0\bigr]$.   Panels (a,b,c) show $i=(1,R,2)$ (with $B_R:=R)$. Each panel displays results for $j_{\mathsmaller \star}=10$ (blue), $j_{\mathsmaller \star}=50$ (magenta), and for the Gaussian whose width matches the corresponding term in  \eqref{eq:chi_1}. For $B_1$ or $R$ (top row), the resulting curves largely coincide when both are plotted against $j_{\mathsmaller \star}\lambda_1$, $j_{\mathsmaller \star}\theta$.  This approximate symmetry at large $j_{\mathsmaller \star}$ reflects the scale-invariance of the corresponding Minkowski-space problem involving either time translations (analogous to $B_1$) or space translations (analgous to $R$) and massless particles localized at the origin.  Due to \eqref{eq:chi_1}, we instead plot results for $B_2$ against $\sqrt{j_{\mathsmaller \star}}\lambda_2$.  While this gives the correct scaling for the peak (see expanded view in panel (c) at right), there appears to be no scale symmetry at large $j_{\mathsmaller \star}$.}
    \label{fig:2pkernel}
\end{figure}

Rather than attempt a general classification of the possible such states $\big|\psi_+\bigr]$, we will confine our investigation to a simple choice that facilitates explicit calculations.  We take the first particle to be localized around $(T,\theta)=(0,0)$, so that the other is then localized around $(0,\pi)$.
In particular, we take $\big|\psi_\pm\bigr]$ to be of the form
\begin{align}\label{eq:psi_pm}
    \big|\psi_\pm\bigr] = \frac{1}{\sqrt{2{j_{\mathsmaller \star}}+1}} \sum_{m=-{j_{\mathsmaller \star}}}^{{j_{\mathsmaller \star}}} (\pm 1)^m \big| m\bigr]
\end{align}
for some cutoff $j_{\mathsmaller \star}$, where the coefficients $(\pm 1)^m$ are found by expanding Dirac delta-functions $\delta(\theta)$ and $\delta(\theta-\pi)$ in terms of rotational harmonics $Y_m(\theta) =\frac{1}{\sqrt{2\pi}}e^{-im\theta}$. This is a convenient choice, since Equations \eqref{eq:B-,B+}-\eqref{eq:B-,B+end} give the action of all generators on these eigenstates. As ${j_{\mathsmaller \star}}\to \infty$, the particles become perfectly localized at the points $\theta=0,\pi$ at time $t=0$. Numerical plots showing that the group averaging kernel defined by this state from a tight peak around the identity (with small secondary peaks) are shown in figure \ref{fig:2pkernel}.

Since the magnitude of our kernel is greatest at the identity, we can study the width of the peak by expanding $U(g)$ to quadratic order in $\lambda_1, \lambda_2$ and $\theta$.  To this order, the kernel $\bigl[\psi_0\big|U(g)\big|\psi_0\bigr]$ is then determined by the expectation values of $B_1, B_2, R$ and the expectation value of symmetrized products of pairs of generators.   However, many of these moments vanish even in the state $\big|\psi_+\bigr]$ since $\big|\psi_+\bigr]$ is invariant under
$\theta \rightarrow -\theta$ (which as shown in figure \ref{fig:B1B2R} maps $B_1, B_2,R$ to $B_1, -B_2, -R$).  Using $[X]$ to denote the expectation value of an observable $X$ in the state $\big|\psi_+\bigr]$ we thus find
\begin{equation}\label{eq:zero_moments}
    [B_2]=[R]= [B_1B_2+B_2B_1]= [RB_1+B_1R] =0,
\end{equation}
so that the only non-vanishing moments in $\big|\psi_+\bigr]$ at this order are $[B_1]$, $[R^2]$, $[B_2^2]$, $[RB_2+B_2R]$, and $[B_1^2]$.  Furthermore, since $\big|\psi_0\bigr]$ is invariant under   the rotation $\theta \rightarrow \theta+\pi$ (which as shown in figure \ref{fig:B1B2R} maps $B_1, B_2,R$ to $-B_1, -B_2, R$), the contributions to  $\bigl[\psi_0\big|RB_2+B_2R\big|\psi_0\bigr]$ and $\bigl[\psi_0\big|B_1\big|\psi_0\bigr]$ from $\big|\psi_\pm\bigr]$ must cancel against each other.

We first consider the case where $j_{\mathsmaller \star}\gg \mu$ (and  $j_{\mathsmaller \star}\gg 1$).  We may then use the approximation $|m\pm \Delta|\approx \text{sign}(m)(m\pm \Delta)$.
Direct computation shows that
\begin{equation}
\label{eq:chi_1}
\begin{aligned}
         \bigl[\psi_0\big|U(g)\big|\psi_0\bigr] =\ & \bigl[\psi_+\big|U(g)\big|\psi_+\bigr]\bigl[\psi_-\big|U(g)\big|\psi_-\bigr] \\
       \approx\ &1-\frac{j_{\mathsmaller \star}^2}{12}\lambda_1^2-\frac{j_{\mathsmaller \star}}{2}\lambda_2^2-\frac{j_{\mathsmaller \star}^2}{3}\theta^2.
 \end{aligned}
 \end{equation}

We may also consider the case when $1 \le j_{\mathsmaller \star}\ll\mu$. Using the approximation $|\Delta \pm m| = \sqrt{\left(\frac{1}{2}\pm m\right)^2 + \mu^2}
 = \mu + \frac{\left(\frac{1}{2}\pm m\right)^2}{2\mu} +\dots$, we find
 \begin{equation}
 \label{eq:chi_2}
\begin{aligned}
       &  \bigl[\psi_0\big|U(g)\big|\psi_0\bigr] =\  \bigl[\psi_+\big|U(g)\big|\psi_+\bigr]\bigl[\psi_-\big|U(g)\big|\psi_-\bigr] \\
       %\approx\ &1-\frac{j_{\mathsmaller \star}(j_{\mathsmaller \star}+1)}{3}\theta^2\\
      % &-\frac{2j_{\mathsmaller \star} \mu^2}{(2{\mathsmaller \star}+1)^2}\qty(1+\frac{(2{\mathsmaller \star}+5)(2{\mathsmaller \star}+1)}{12\mu^2}-\frac{(2{\mathsmaller \star}-1)(2{\mathsmaller \star}+1)^2(-45+2{\mathsmaller \star}(-39+70{\mathsmaller \star}+4{\mathsmaller \star}^2))}{11520{\mathsmaller \star}\mu^4})\lambda_1^2\\
     %  &-\frac{\mu^2}{2{\mathsmaller \star}+1}\qty(1+\frac{4{\mathsmaller \star}^2+1}{4\mu^2}-\frac{({\mathsmaller \star}^2-12{\mathsmaller \star}-5)(2{\mathsmaller \star}+1)(2{\mathsmaller \star}-1)(2{\mathsmaller \star}-3)}{1920\mu^4})\lambda_2^2\\
       \approx\ &1-\frac{j_{\mathsmaller \star}(j_{\mathsmaller \star}+1)}{3}\theta^2-\frac{2j_{\mathsmaller \star} \mu^2}{(2j_{\mathsmaller \star}+1)^2}\qty(1+\frac{(2j_{\mathsmaller \star}+5)(2j_{\mathsmaller \star}+1)}{12\mu^2})\lambda_1^2-\frac{\mu^2}{2{j_{\mathsmaller \star}}+1}\qty(1+\frac{4{j_{\mathsmaller \star}}^2+1}{4\mu^2})\lambda_2^2.
 \end{aligned}
 \end{equation}

%Direct computation of the remaining moments at this order then yields
 %\begin{equation}
  %   \begin{split}
  %       \bigl[\psi_0\big|U(g)\big|\psi_0\bigr] =& \bigl[\psi_+\big|U(g)\big|\psi_+\bigr]\bigl[\psi_-\big|U(g)\big|\psi_-\bigr] \\
  %      =&1  - \lambda_1^2 \left([B_1^2]-[B_1]^2\right) - \lambda_2^2 [B_2^2] - \theta^2 [R^2] +O([g-1]^3) \\
 %       =&1 -\frac12 \lambda_1^2 {j_{\mathsmaller \star}} \left( 1 + \frac{4\mu^2}{(2{j_{\mathsmaller \star}}+1)^2}\right) -\lambda_2^2 \frac{8{j_{\mathsmaller \star}}^3+4{j_{\mathsmaller \star}}+3+12\mu^2}{12(2{j_{\mathsmaller \star}}+1)} \\
%        &-\frac13 \theta^2 {j_{\mathsmaller \star}}({j_{\mathsmaller \star}}+1) +O([g-1]^3),
 %    \end{split}
% \end{equation}
 %where $\mu$ was defined in \eqref{eq:mudef}.
Let us therefore introduce the parameters $\chi_1$, $\chi_2$, $\chi_3$,
%\begin{equation}\label{eq:chi}
%    \chi_1 =  {j_{\mathsmaller \star}} \left( 1 + \frac{4\mu^2}{(2{j_{\mathsmaller \star}}+1)^2} \right), \,\,\,\, \chi_2= \frac{8{j_{\mathsmaller \star}}^3+4{j_{\mathsmaller \star}}+3+12\mu^2}{6(2{j_{\mathsmaller \star}}+1)}, \,\,\,\,  \chi_\theta = \frac23 {j_{\mathsmaller \star}}({j_{\mathsmaller \star}}+1),
%\end{equation}
in order to write the group averaging kernel as an approximate Gaussian
\begin{align}
\label{eq:chiGauss}
    \bigl[\psi_0\big|U(g)\big|\psi_0\bigr]
    \approx&  e^{-\chi_1\lambda_1^2/2}e^{-\chi_2\lambda_2^2/2}e^{-\chi_\theta \theta^2/2},
\end{align}
where we emphasize that  $\chi_i \geq 0$ for all $i\in\{1,2,\theta\}$. The widths of the peak in the various directions are then proportional to $\chi_i^{-1/2}$ and, in the $\chi_i \to \infty$ limit, the Gaussians become proportional\footnote{Since we have chosen $\big|\psi_0\bigr]$ to be normalized, we will always find $\bigl[\psi_0\big|U(\mathds{1})\big|\psi_0\bigr]=1$.  So, as written, the limit $\chi_i\rightarrow \infty$ gives a delta-function with a vanishing coefficient.  However, for the same reason, the norm $\langle \alpha;LPG|\alpha;LPG\rangle$ also vanishes in this limit for any normalized $\big|\alpha\bigr]\in \mathcal H_{QFT}^\phi$.   Obtaining normalized states $|\alpha;LPG\rangle$ thus requires taking $\bigl[\alpha \big|\alpha\bigr]$ to scale as $\sqrt{\chi_1\chi_2\chi_\theta}$.  Combining this factor with the Gaussian \eqref{eq:chiGauss} yields the desired delta-function.}
to a delta-function that sets $\lambda_i=0$ for all $i$. In this limit, the inner product between two physical states reduces to the standard QFT inner product without additional smearing.

However, the large $\chi_i$ limit requires ${j_{\mathsmaller \star}}$ and/or $\mu$ to become large.  In the presence of dynamical gravity, either option would induce a large gravitational backreaction. Hence the $\chi_i$ must remain bounded if we wish to keep such backreaction small.
We will characterize this backreaction more precisely in section \ref{subsubsec:back} below.

\subsubsection{Characterizing backreaction}
\label{subsubsec:back}

\begin{figure}[H]
  \centering
  \includegraphics[width = 3in]{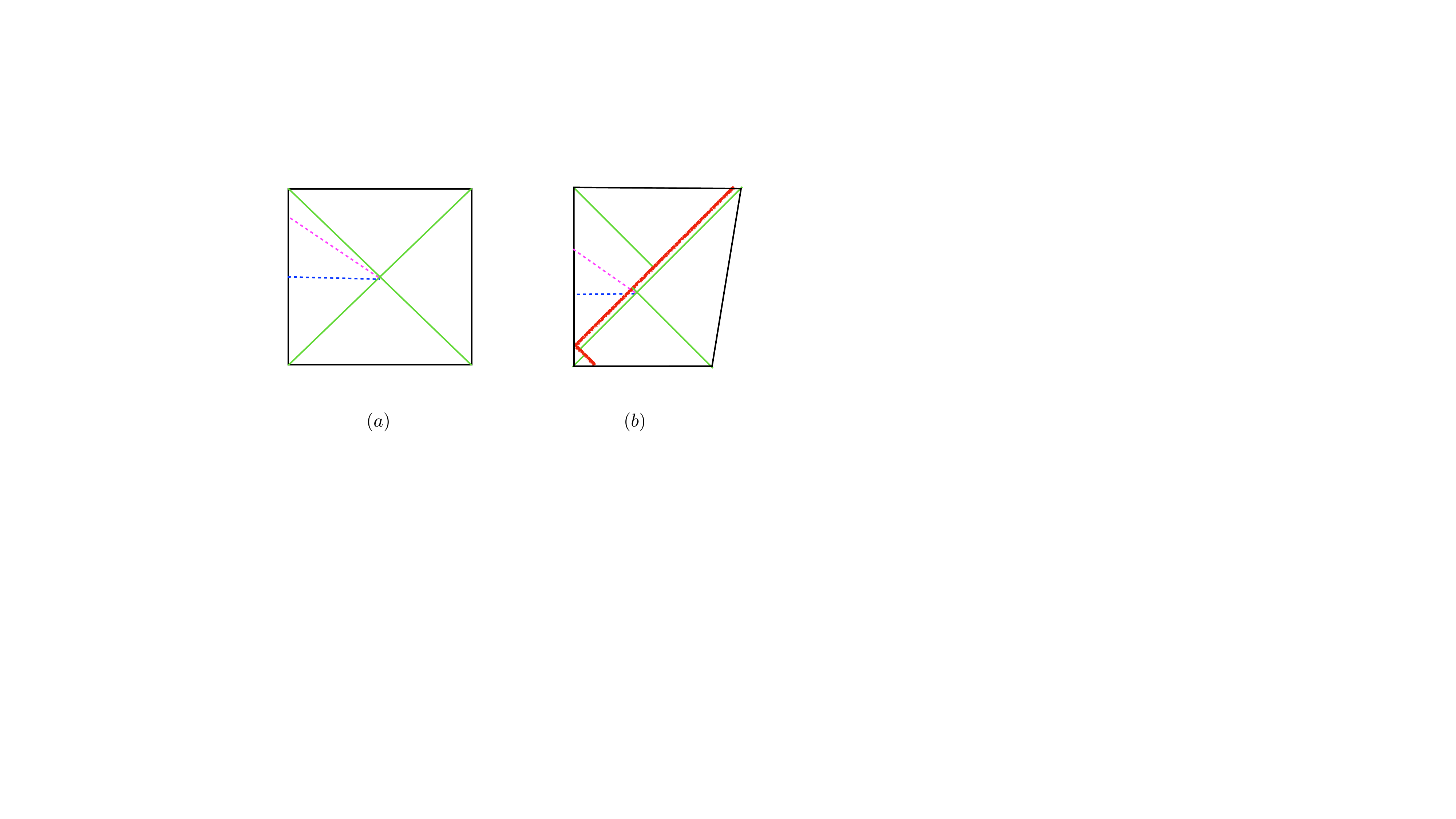}
  \caption{{\bf (a)} A conformal diagram of unperturbed dS showing light cones (green) emitted from the 4 corners of the diagram.  The dashed blue surface and its boosted image (pink) have the same proper volume.
  {\bf (b)}  
  The conformal diagram of an asymptotically de Sitter solution with a postive-energy shockwave (red). 
  Due to focussing at the shock, light cones (green) emitted from the 4 corners of the diagram no longer connect to form horizons.  As a result, even in the limit where it becomes null, the dashed pink surface hits the left edge of the diagram at finite time.  At large relative boosts, it thus has small proper volume compared with the dashed blue surface.
  }
  \label{fig:back_reaction}
\end{figure}

In order to estimate the bounds imposed on the $\chi_i$ associated with the restriction to small backreaction, we will need to say more about how this backreaction will be measured.  Gravitational backreaction on $(d+1)$-dimensional de Sitter space is generally highly non-uniform, so that for any classical perturbation of dS one can find {\it some} sense in which the backreaction is large.  This is perhaps most simply illustrated by making use of the Gao-Wald theorem \cite{Gao:2000ga}, showing that any perturbation satisfying the null energy condition forces the past (or future) of any timelike geodesic to contain a complete Cauchy surface. There is thus a sense in which applying a large boost to any spherical cross-section of the perturbed de Sitter space must give a Cauchy surface of vanishingly small total volume; see figure \ref{fig:back_reaction}.  This is in sharp contrast to the case of unperturbed dS, where applying any dS isometry to the $S^d$ at $t=0$ of course exactly preserves its finite volume.

We will choose to measure the backreaction near the round $S^d$ that passes through the spacetime points at which our $\psi$ particles are well-localized; i.e., the $S^d$ at $t=0$ with the coordinates and states defined as above.  As discussed in \cite{Giddings_2007}, a reasonable measure of the backreaction in this region is the total flux $F$ of energy through this $S^d$, where
\begin{equation}
\label{eq:FTab}
F = \int_{t=0} \sqrt{h} T_{ab}n^a n^b
\end{equation}
in terms of the QFT stress tensor $T_{ab}$, the unit normal $n^a$ to the surface $t=0$,  and the volume element $h$ of the induced metric on this surface.   If we wish to keep the level of backreaction on the geometry below some fixed cut-off, then in terms of the bulk Newton constant $G$, the maximal allowed value of $F$ will of course scale as $1/G$ in the limit $G \rightarrow 0$.  While Einstein-Hilbert gravity is not dynamical in 1+1 dimensions, we can nevertheless use our investigation of dS$_{1+1}$ as a toy model of the higher-dimensional case by introducing a (dimensionless) parameter $G$ and imposing the restriction $F\lesssim 1/G\ell.$

To understand the constraint this imposes on our $\chi_i$, we will need to estimate the contributions to $F$ arising from the mass and angular momentum of our $\psi$-particles.  This is straightforward due to the fact that the time derivative of the metric vanishes at $t=0$. As a result, the local notion of positive-frequency mode near $t=0$ associated with the standard definition of particles in global de Sitter coincides with the notion of positive frequency for the static cylinder metric
\begin{equation}\label{eq:staticcylmetric}
ds^2 =-dt^2 + \ell^2 d\theta^2.
\end{equation}
Our $F$ thus coincides with what one would call the energy $E$ on the static cylinder \eqref{eq:staticcylmetric} when computed in terms of the angular momentum $m$.  It is thus clear that, If both particles were in modes having angular momentum precisely $m$, we would find
\begin{equation}\label{eq:E_t=0}
    F = 2\sqrt{M^2 + m^2/\ell^2}.
\end{equation}
In order to limit backreaction, we must thus take the mass parameter $\mu$ and the maximum angular momentum ${j_{\mathsmaller \star}}$ to satisfy $M\ell \ll 1/G$ and ${j_{\mathsmaller \star}} \ll 1/G$ as $G\rightarrow 0$.

Let us first we examine the ultrarelativistic limit ($j_{\mathsmaller \star} \gg \mu$, though with $1/G \gg j_{\mathsmaller \star}$). The results \eqref{eq:chi_1} then simplify significantly to yield
%, the nonrelativistic limit ($\mu \gg j_{\mathsmaller \star}$ with $1/G \gg \mu$), and the case where the ratio $j_{\mathsmaller \star}/\mu$ is held fixed as $G\rightarrow 0$ (with $j_{\mathsmaller \star} \sim \mu \ll 1/G$). We will see that the primary results are similar in all three cases.
%
%\subsubsection{The ultrarelativistic case  ${j_{\mathsmaller \star}} \gg \mu$}
%Let us first consider the ultrarelativistic limit $j_{\mathsmaller \star} \gg \mu$, where we still wish to require  $1/G \gg j_{\mathsmaller \star}$.  At leading order in $1/{j_{\mathsmaller \star}}$, the coefficients $\chi$ are
\begin{equation}
    \chi_1=  \frac{1}{6}j_{\mathsmaller \star}^2 + O(j_{\mathsmaller \star}), \quad \chi_2= {j_{\mathsmaller \star}}+ O({1}), \quad \chi_\theta= \frac23 {j_{\mathsmaller \star}}^2+ O({j_{\mathsmaller \star}}).
\end{equation}
We see that the dependence on $\mu$ disappears at leading order in $j_{\mathsmaller \star}/\mu$.

As described in section \ref{sec:relobsdS} the effect of group averaging on correlation functions smeared with the global coordinate near-Gaussians $F_y$ of \eqref{eq:GaussianF}  will be small when the smearing is confined to de Sitter isometries $g$ such that
\begin{equation}
\label{eq:smallsmear}
|gx-x|_E \ll \sigma,
\end{equation}
where $|x_1-x_2|_E$ is the flat Euclidean distance defined by \eqref{eq:Eds2} and we consider all $x$ located within the peak of each near-Gaussian $F_y$.   Since the Gaussian \eqref{eq:chiGauss} gives significant weight only to group elements with $\lambda _i \lesssim \chi_i^{-1/2}$, for large-but-finite $\chi_i$ the condition \eqref{eq:smallsmear} is equivalent to
\begin{equation}
\label{eq:smallsmear2}
\chi_i^{-1/2} |\xi_i|_E \ll \sigma,
\end{equation}
where $|\xi_i|_E$ is the Euclidean norm of the appropriate vector field from \eqref{eq:B1action}-\eqref{eq:Raction}.  In particular,  we have
\begin{align}
    |\xi_1|^2_E/\ell^2 =& \cos^2\theta + \sinh^2(t/\ell) \sin^2\theta  \label{eq:xi1normE},\\
        |\xi_2|^2_E/\ell^2 =& \sin^2\theta + \sinh^2(t/\ell) \cos^2\theta \label{eq:xi2normE}, \\
    |\xi_\theta|^2_E/\ell^2 =& \cosh^2(t/\ell) .
\end{align}

\begin{figure}[h!]
    \centering
    \begin{subfigure}{\textwidth}
    \centering
    \includegraphics[width=0.31\textwidth]{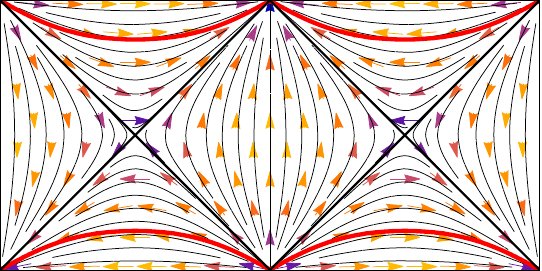}\hspace{.3cm}
    \includegraphics[width=0.31\textwidth]{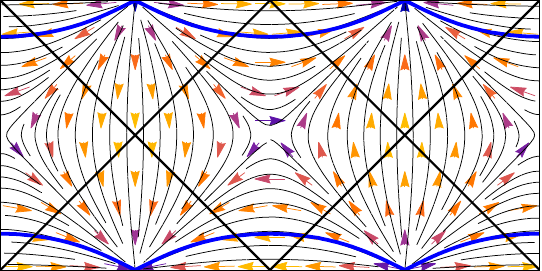}\hspace{.3cm}
    \includegraphics[width=0.31\textwidth]{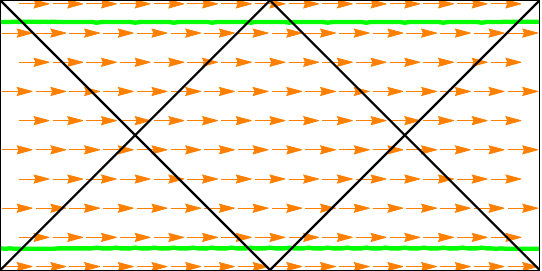}
    \caption{The KVFs and cutoff contours for $B_1$ (left), $B_2$ (center), and $R$ (right) are drawn using conformal coordinates $(T, \theta)$ for dS$_{1+1}$.}
    \label{fig:IndividualCutoffs}
    \end{subfigure}

    \vspace{.5cm}

    \begin{subfigure}{\textwidth}
    \centering
    \begin{tikzpicture}

  \pgfdeclareimage{img}{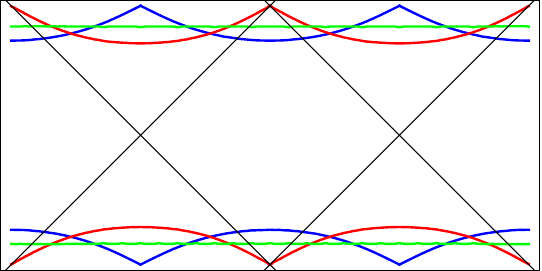}
  \node (img1) at (0,0) {\pgfuseimage{img}};

   \draw[magenta,thick] (0,1.59) -- (0,-1.59);

\end{tikzpicture}
    \caption{The above cutoff contours are displayed on together on a single  conformal diagram. We also show the part of the timelike geodesic at $\theta=0$ (pink vertical line) that is consistent with all cutoffs.}
    \label{fig:AllCutoffs}
    \end{subfigure}

    \caption{Cutoff surfaces $|\xi_i|_E^2 = \chi_i \sigma^2$  are shown for the ultrarelativistic limit of the reference state given by \eqref{eq:antpart} and \eqref{eq:psi_pm}.  For illustration purposes we have used the values $\sigma^2=1$ and ${j_{\mathsmaller \star}}=5$. At sufficient depth within these cutoffs, our perturbative gravity correlators provide good approximations to corresponding dS QFT correlators.}
    \label{fig:cutoffs}
\end{figure}

In regions of spacetime where any of the bounds \eqref{eq:smallsmear2} are exceeded, the spacetime resolution of the observables $\hat \phi_{LPG}$ is low, and the dS QFT approximation to perturbative gravity breaks down even for correlators smeared with the functions \eqref{eq:GaussianF}.
The corresponding cutoff contours (at which $|\xi_i|^2_E \sim \chi_i \sigma^2$) are shown in Figure \ref{fig:cutoffs} for specific values of $\sigma$ and $j_{\mathsmaller \star}$. At sufficient depth within the region between these contours, dS QFT correlators smeared with the function \eqref{eq:GaussianF} will be well-approximated by correspondingly-smeared perturbative gravity correlators.

Note that at large $j_{\mathsmaller \star}$ we have $\chi_2 \ll \chi_1, \chi_\theta$.  As a result, consulting \eqref{eq:xi2normE} and taking $t$ large as well, we see that over most of the spacetime the large $j_{\mathsmaller \star}$ cutoff will be set by
\begin{equation}
\label{eq:B2cutoff}
|t/\ell| \sim \frac{1}{2} \ln j_{\mathsmaller \star} + \ln  \frac{\sigma}{\ell \cos \theta}.
\end{equation}
However, this bound diverges when $\cos \theta=0$. In particular, in the static patches associated with the geodesics $\theta = \pm \frac{\pi}{2}$ 
the Euclidean norm of $\xi_2$ satisfies $|\xi_2|_E^2\le \ell^2$.  In such regions the bound is thus set by either $R$ or $B_1$.  
Since the Euclidean norms of $\xi_\theta$ and $\xi_2$ are nearly identical at $\theta = \pm \frac{\pi}{2}$, at late times \eqref{eq:chi_1} gives $\chi_1/\chi_{\theta} = \frac{1}{4} + O(1/j_{\mathsmaller \star} )$, while the cutoff is in fact instead set by  $B_1$.  In such regions we thus find
\begin{equation}
\label{eq:B2Rcutoff}
|t/\ell| =  \ln j_{\mathsmaller \star} +  \ln  \frac{\sigma}{\ell} - \frac{1}{2}\ln \frac{3}{2}+ O(1/j_{\mathsmaller \star}),
\end{equation}

So, for ${j_{\mathsmaller \star}} \sim 1/G$, the ultrarelativistic limit $j_{\mathsmaller \star}\gg \mu$ yields a good approximation to dS QFT only for global times $|t/\ell| \lesssim   \frac{1}{2}\ln G^{-1}+O(1)$, though the actual cutoff becomes $|t/\ell| \lesssim   \ln G^{-1}+O(1)$ at $\theta = \pm \frac{\pi}{2}$.  As a result, the spacetime volume of the region where our approximation is good is of order $1/G$.

We can now return to \eqref{eq:chi_2} and consider the case $ j_{\mathsmaller \star} \ll \mu \lesssim 1/G$.  If we take e.g. $ j_{\mathsmaller \star} = O(1)$ and $\mu =O(1/G)$ as $G\rightarrow 0$, we find $\chi_1, \chi_2 = O(1/G^2)$ but $\chi_\theta = O(1).$  In contrast, if we take $j_{\mathsmaller \star}\ll \mu $ but with $j_{\mathsmaller \star}, \mu = O(1/G)$, then we find $\chi_1, \chi_2 = O(1/G)$ and $\chi_\theta= O(1/G^2)$.  We can also set all 3 coefficients to be the same order in $1/G$ by choosing $\mu=O(1/G)$ and $j_{\mathsmaller \star} = O(1/G^{2/3})$, in which case we find $\chi_1, \chi_2, \chi_\theta = O(G^{-4/3})$.  Any of the above cases will again confine the region in which our dS QFT approximation is accurate to one which spans a global time interval $\Delta t$ of order $\ln(1/G)$ with a coefficient of order $1$. We will therefore confine attention to the simpler ultrarelativistic case below.

\subsection{Adding more reference particles}\label{sec:multi-obs}

For the above reference state, we found that smeared dS QFT correlators are well-approximated by our smeared perturbative gravity correlators only in a region of de Sitter space spanning global times of order $\ln G^{-1}$.  It is thus interesting to explore whether this region can be enlarged by considering a more complicated reference state.  We are particularly motivated by a desire to understand whether the region can be enlarged within a natural static patch of dS,  perhaps at the expense of shrinking the allowed region outside. In this section, we  investigate the effect of adding additional reference particles localized at points along the $\theta=0$ and $\theta=\pi$ geodesics.

\begin{figure}[H]
  \centering  \includegraphics[width = 2.8in]{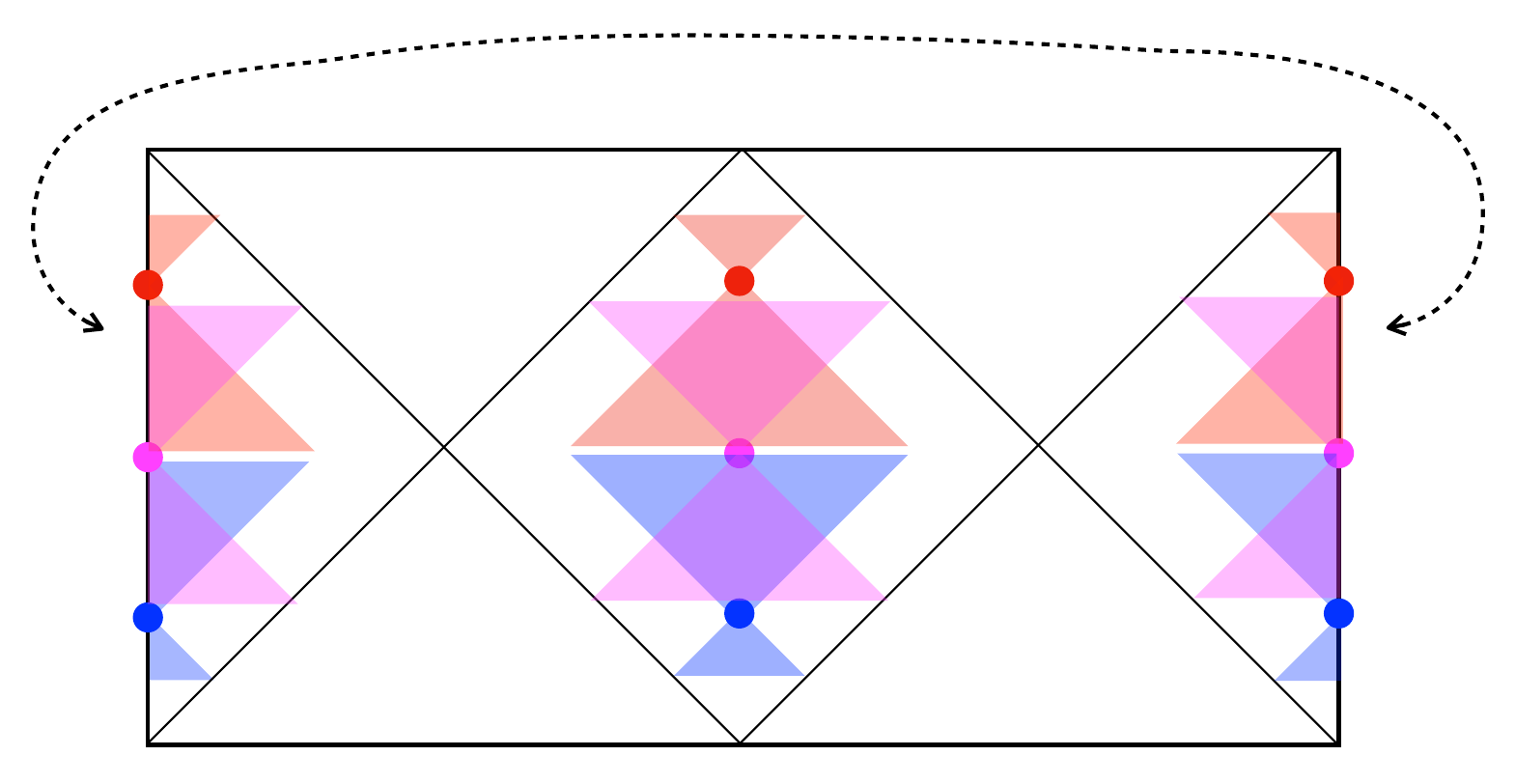}
  \caption{Events at which each particle localizes (dots) are shown together with (truncated) light cones from these events. }
  \label{fig:reference_1}
\end{figure}

As before, it will be useful to keep the full reference state $\big|\psi_0\bigr]$ properly `balanced' in the sense that it has vanishing expectation values of  $B_1,B_2,R$ to avoid giving $\big|\psi_0\bigr]$ de Sitter charges that are parametrically larger than those of the $\phi$-system (as that would then require group averaging to nearly annihilate the resulting state in order to extract a state in which the total de Sitter charges vanish).

We will consider states of the form $\big|\psi_0\bigr] = |\psi_+^3] \otimes |\psi_-^3]$ with $|\psi_-^3] = e^{i\pi R} |\psi_+^3]$, but where $|\psi_+^3]$ now contains three particles that become well localized along the geodesic $\theta=0$ at times $-t_0$, $0$, and $t_0$. In particular, we take
\begin{equation}
\label{eq:psipm3}
|\psi_\pm^3] = \left(e^{\mp i t_0B_1/\ell} \big|\psi_\pm\bigr]\right) \otimes \big|\psi_\pm\bigr] \otimes  \left(e^{\pm i t_0B_1/\ell} \big|\psi_\pm\bigr]\right),
\end{equation}
where $\big|\psi_+\bigr]$ is again given by \eqref{eq:psi_pm}.  Thinking of $B_1$ as the static patch Hamiltonian, we see that moments of $B_1, B_2, R$ in the various one-particle states will be given by moments of static-patch time-translations of $B_1, B_2, R$  in $\big|\psi_\pm\bigr]$.
It will thus be useful to compute expectation values like
\begin{equation}
\label{eq:transme}
    \bigl[\psi_\pm\big| e^{i t_0 B_1/\ell} A e^{-it_0 B_1 /\ell}\big|\psi_\pm\bigr],
\end{equation}
where $A$ is a linear or quadratic expression in $B_1, B_2, R$.
For $A=B_1$ or $A=B_1^2$, the time translation has no effect and \eqref{eq:transme} reduces to matrix elements calculated previously. For $B_2$ and $R$, it is useful to define operators $L_\pm = R \pm B_2$ corresponding to lightlike (null) rotations which have commutation relations 
\begin{align}
   [L_\pm,B_1] = \mp iL_{\pm}, \quad [L_{+},L_{-}] = -2iB_1.
\end{align}
As a result, under a time translation, these operators satisfy $e^{it_0 B_1 /\ell} L_{\pm} e^{-it_0 B_1 /\ell} = L_\pm e^{\mp t_0/\ell}$, from which we obtain the time translations of $R$ and $B_2$:
\begin{align}\label{eq:Rintime}
    e^{it_0 B_1 /\ell}R e^{-it_0 B_1 /\ell} =& R\cosh (t_0/\ell) - B_2 \sinh (t_0/\ell), \\
    e^{it_0 B_1 /\ell}B_2 e^{-it_0 B_1 /\ell} =& B_2 \cosh (t_0/\ell) - R \sinh (t_0/\ell). \label{eq:B1intime}
\end{align}

We can now use the above results to compute the group averaging kernel. Since our new $\big|\psi_0\bigr]$ still enjoys the symmetries discussed near \eqref{eq:zero_moments}, the moments listed in \eqref{eq:zero_moments} once again vanish.  It thus remains only to compute expectation values of $B_1^2$, $B_2^2$, and $R^2$.  These moments receive contributions from the corresponding moments of each 1-particle state.  The expectation value of $B_1^2$ also receives contributions from cross terms between various pairs of particles, associated with the fact that $\bigl[\psi_+\big|B_1\big|\psi_+\bigr]$ is non-zero in all one-particle states.  Other cross terms vanish since $\bigl[\psi_+\big|B_2\big|\psi_+\bigr] = \bigl[\psi_+\big|R\big|\psi_+\bigr]=0.$
The final result for our kernel is thus 
\begin{equation}
\begin{split}
    [\psi_0^{3}|U(g)|\psi_0^{3}] =&1 + 3\lambda_1^2([B_1]^2 - [B_1^2]) - \lambda_2^2 \bigg([B_2^2](1+2\cosh^2(t_0/\ell))+2[R^2]\sinh^2(t_0/\ell)\bigg)  \\
    &- \theta^2 \bigg([R^2](1+2\cosh^2(t_0/\ell))+2[B_2^2]\sinh^2(t_0/\ell)\bigg) +O([g-1]^3) \\
    =&1 - \lambda_1^2 \tilde{\chi}_1/2 - \lambda_2^2 \tilde{\chi}_2/2 - \theta^2 \tilde{\chi}_\theta/2 +O([g-1]^3), 
\end{split}
\end{equation}
where $[{\cal O}]$ again refers to the expectation value of ${\cal O}$ in $\big|\psi_+\bigr]$. 

For $j_{\mathsmaller \star}\gg\mu$, these coefficients are given by
\begin{align}
    &\tilde \chi_1 \approx\frac{j_{\mathsmaller \star}^2}{2}, \\
    &\tilde \chi_2 \approx j_{\mathsmaller \star}\qty(1+2\cosh^2(t_0/\ell))+\frac{4}{3}j_{\mathsmaller \star}^2\sinh^2(t_0/\ell),\\
    &\tilde \chi_\theta \approx \frac{2}{3}j_{\mathsmaller \star}^2\qty(1+2\cosh^2(t_0/\ell))+2j_{\mathsmaller \star}\sinh^2(t_0/\ell).
\end{align}
%\begin{align}
  %  \tilde{\chi}_1 =& 3 {j_{\mathsmaller \star}} \left( 1 + \frac{4\mu^2}{(2{j_{\mathsmaller \star}}+1)^2}\right), \\
 %   \tilde{\chi}_2 =& \frac{8{j_{\mathsmaller \star}}^3+4{j_{\mathsmaller \star}}+3+12\mu^2}{6(2{j_{\mathsmaller \star}}+1)}(1+2\cosh^2(t_0/\ell)) + \frac43{j_{\mathsmaller \star}}({j_{\mathsmaller \star}}+1)\sinh^2(t_0/\ell), \\
 %   \tilde{\chi}_\theta =& \frac23{j_{\mathsmaller \star}}({j_{\mathsmaller \star}}+1)(1+2\cosh^2(t_0/\ell)) + \frac{8{j_{\mathsmaller \star}}^3+4{j_{\mathsmaller \star}}+3+12\mu^2}{3(2{j_{\mathsmaller \star}}+1)} \sinh^2(t_0/\ell).
%\end{align}
As in the two particle case, the QFT approximation holds exactly in the limit $\tilde{\chi}\to \infty$. The coefficients $\tilde{\chi}$ still grow with increasing ${j_{\mathsmaller \star}}$, but now they also grow with increasing $t_0$.

Let us thus investigate how large we can take $t_0$ while keeping the backreaction small as measured by \eqref{eq:FTab}; i.e., for  $F \lesssim 1/G$. Recall that $F$ gives the total energy $E$ that the state would have if it were placed in a static cylinder spacetime of radius $\ell$ (keeping the state unchanged in the Fock basis defined by angular momentum modes). The particles at $t=0$ will contribute to $F$
according to \eqref{eq:FTab}
 as they did before. However, the contributions of the time translated particles are easiest to study by evolving the particles from $t=\pm t_0$ to $t=0$.

For the particles that localize at $\theta=0$, it is convenient to perform this evolution using the description provided by the static patch centered at $\theta=0$.  The time translation from $t=0$ to $t=\pm t_0$ is then trivial but, as we evolve them back to $t=0$, the particles move to higher energies as they fall away from $\theta=0$ toward the static patch horizon.

Let us first consider the limit ${j_{\mathsmaller \star}} \gg \mu$ so that the particles are relativistic, and so travel along null rays.  Such particles rapidly approach the de Sitter horizon and then blueshift exponentially with respect to the vector field $n^a$ in \eqref{eq:FTab}.    We thus find that the total flux of energy through $t=0$ is
\begin{equation}
    F \sim (2+4\cosh(t_0/\ell)){j_{\mathsmaller \star}} \sim 2{j_{\mathsmaller \star}}e^{t_0/\ell},
\end{equation}
where the final right-hand-side gives the leading behavior at large ${j_{\mathsmaller \star}}$ and $t_0$.  

%We take the large $t_0$ limit as well because we know $\tilde{\chi}$ grow with $t_0$, and we are allowed to take $t_0$ large as long as we still satisfy the bound $F \approx {j_{\mathsmaller \star}}e^{t_0} \sim 1/G$. Now, let us see what this bound implies for the coefficients $\tilde{\chi}$, and then, through Eq. \eqref{eq:bound}, the size of the allowed region.

At leading order in $1/{j_{\mathsmaller \star}}$ and $e^{-t_0/\ell}$ we also find
\begin{equation}
    \tilde{\chi}_1 \sim \frac{1}{2} {j_{\mathsmaller \star}}^2 , \quad \tilde{\chi}_2 \sim \frac13 {j_{\mathsmaller \star}}^2 e^{2t_0/\ell}, \quad \tilde{\chi}_\theta \sim \frac13 {j_{\mathsmaller \star}}^2 e^{2t_0/\ell}.
\end{equation}
The cutoff contours will look very similar to the ones we found before, except that there is now an extra parameter to vary. Outside the static patch, we expect the cutoffs to again be set by $\chi_1$ since it remains of order $j_{\mathsmaller \star}^2$ (since particles related by static-patch time-translations must contribute equally to $\chi_1$). 
Inside the static patch, estimating the cutoff time $t_c$ using either $\chi_2$ or $\chi_\theta$ and setting $F\sim 1/G$ leads to
\begin{equation}
    \cosh^2 (t_c/\ell) = \frac13 {j_{\mathsmaller \star}}^2 e^{2t_0/\ell} \sim \frac{1}{G^2},
\end{equation}
so that we again find $t_c \lesssim \ell \ln(1/G)$ for any allowed choice of $j_{\mathsmaller \star},t_0$.  We thus see that, with a given finite energy budget measured by $F$, adding localized particles along the geodesics $\theta=0, \pi$ fails to increase the size of the allowed region.

While we have not investigated other choices of reference states in detail, the exponential increase of kinetic energies with static patch time is typical of any particles falling toward a de Sitter horizon. This suggests that the above behavior is generic when we require backreaction to be small at $t=0$ for our global time $t$; i.e., at a minimal $S^d$.  For example, the same exponential factors arise in the nonrelativistic limit $\mu \gg {j_{\mathsmaller \star}}$.   However, in section \ref{subsec:futureparticles} we will explore the possibility of allowing backreaction at such minimal spheres to be large, and thus allowing $F$ to be large, while requiring backreaction to be small in other regions of de Sitter space.   Due to that fact that it will require a slightly more involved discussion of backreaction in Einstein-Hilbert gravity, and  since our discussion of `backreaction' for $D=2$ was simply a convenient fiction designed to provide a toy model of well-known results for Einstein-Hilbert gravity in the higher dimensional case, we postpone that discussion to section \ref{subsec:futureparticles} and, in particular, until after treating group averaging in higher dimensions in section \ref{sec:d+1}.

%In dS$_{1+1}$, there are many other candidate $\big|\psi_0\bigr]$ that could be tested. For instance, we could test the  limit or the double scaling limit for these multiple observer states, or we could test instead coherent states. However, we take the calculation above as evidence that the details of the observer state do not drastically affect the size of the allowed region. All states will have cutoffs determined the same bound on the energy flux through the dS neck. While it is possible there are states which slightly enlarge the region, say from $2\ln(1/G)$ to $3\ln(1/G)$, we find it unlikely that any state could give an exponential enhancement.

\section{Reference states in dS$_{d+1}$}\label{sec:d+1}
We will now see that essentially the same results found above for $dS_{1+1}$ also hold for dS$_D$ with $D=d+1 > 2$.  We begin with a discussion of particle states and the associated action of SO$(D,1)$ generators following \cite{Marolf_2009}.  To this end, we consider a sphere $S^d$, with metric
\begin{equation}
    d\Omega_d^2=d\theta_{1}^2+\sin^{2}\theta_{1}^2d\Omega_{d-1}^2,
\end{equation}
where $\theta_{1}$ is the polar angle and we use coordinates $\Omega_{d-1}=(\theta_{2},...,\theta_{d})$. In global dS we use the corresponding global coordinates (with global time $t$) in which the metric takes the form
\begin{equation}
    ds^2 = -dt^2 + \ell^2 \cosh^2 (t/\ell) (d\theta_{1}^2 + \sin^2\theta_{1}d\Omega_{d-1}^2).
\end{equation}

Bosonic one-particle wavefunctions on $S^d$ can be written in terms of spherical harmonics labelled by angular momentum vectors $\vec j = (j_d,\dots, j_1)$  with  $j_k \in {\mathbb Z}^+$ for $k\ge 2$. For $k \ge 1$, we take $j_k$ to be the total angular momentum quantum number for the SO$(k+1)$ subgroup of SO$(D,1)$ associated with the $S^k$ sphere at constant $\theta_{n}$ for $n\le d-k$. The above quantum numbers thus satisfy
\begin{equation}
    j_d \geq j_{d-1} \geq ... \geq j_2 \geq |j_1|.
\end{equation}

In analogy with the construction in section \ref{sec:one-obs}, we begin by considering a reference state $\big|\psi_1\bigr] = \big|\psi_+\bigr]\otimes \big|\psi_-\bigr]$ where each state describes a particle that is well-localized at $t=0$ at one of the poles of the $S^d$.  For simplicity, we take each particle to be invariant under the SO$(d)$ rotations that preserve the poles.  As a result, two particles will not suffice to break all of the dS isometries, so we will need to add more particles later.   Indeed, we will soon define $\big|\psi_0\bigr] = \big|\psi_1\bigr] \otimes \big|\psi_2\bigr] \otimes \dots \big|\psi_D\bigr]$, where the $\big|\psi_i\bigr]$ for $i\ge 2$ are constructed from $\big|\psi_1\bigr]$ by applying rotations by $\pi/2$ in $d=D-1$ orthogonal directions; see the discussion below.

In the state $\big|\psi_1\bigr] = \big|\psi_+\bigr]\otimes \big|\psi_-\bigr]$, the only non-zero angular momentum will be $j_d$, for which we henceforth use the simplified notation $j=j_d$. We will again consider a free scalar field with 1-particle states in the principle series, with $\Delta = d/2 + i\mu$. We take the state of each particle to be of the form
\begin{equation}
\label{eq:psi_pmD}
    \big|\psi_\pm\bigr]=N \sum_{j=0}^{j_{\mathsmaller \star}} c_j^\pm |\Delta, j, \vec{0}],
\end{equation}
where $+$ denotes a particle at the north pole and $-$ denotes a particle at the south pole, and where $N$ is a normalization constant.

The coefficients $c_j^\pm$ will be given by the spherical harmonic expansion of an $S^d$ Dirac $\delta$-function localized at the relevant pole. It is natural to write this delta function $\tilde \delta(\theta_1)$ in the form
\begin{equation}
    \tilde{\delta}(\theta_{1})=\frac{\delta(\theta_{1})}{V_{d-1}\sin^{d-1}\theta_{1}},
\end{equation}
where $\delta(\theta_1)$ is the standard Dirac delta-function associated with the measure $d\theta_1$ and
$V_{d-1}=\int_{\Omega_{d-1}} d\Omega_{d-1}=\frac{2\pi^{d/2}}{\Gamma(d/2)}$ is the volume of the unit $(d-1)$-sphere. There is an analogous result for the $\delta$-function at the south pole. The $d$-dimensional spherical harmonics for $j_{d-1}=j_{d-2}=...=j_1=0$ are given (see e.g. \cite{Marolf_2009}) by
\begin{equation}
    \leftindex^{(d)\, } {Y}_{\vec{j}}(\Omega_d) = \frac{1}{\sqrt{2\pi}} \leftindex^{(d)} {\mathcal{Y}}_{j0}(\theta_{1})\prod_{n=2}^{d-1} \leftindex^{(n)} {\mathcal{Y}}_{00}(\theta_{d+1-n}),
\end{equation}
where there is no $\theta_d$ dependence, since the associated harmonic simplifies to $1/\sqrt{2\pi}$.  The other harmonics are given by
\begin{equation}
\begin{split}
    \leftindex^{(d)} {\mathcal{Y}}_{j0}(\theta_{1}) =& \frac{1}{{2^{(d-2)/2}\Gamma(\frac{d}{2})}} \left[ \left(j+\frac{d-1}{2}\right) \frac{\Gamma(j+d-1)}{\Gamma(j+1)}\right]^{1/2} \cos^j\theta_{1} \\
    &\times\leftindex_{2} {F}_1\qty(-\frac{j}{2},\frac{1-j}{2};\frac{d}{2};-\tan^2\theta_{1}),
\end{split}
\end{equation}
and by
\begin{equation}
    \leftindex^{(n)} {\mathcal{Y}}_{00}(\theta_{d+1-n}) = \frac{\left[(n-1)\Gamma(n-1)\right]^{1/2}}{2^{(n-1)/2}\Gamma(\frac{n}{2})},
\end{equation}
where we see the $\leftindex^{(n)} {\mathcal{Y}}_{00}(\theta_{d+1-n})$ above are independent of $\theta_{d+1-n}$. Thus the relevant spherical harmonics depend only on $\theta_{1}$. We can now determine the coefficients $c_j^\pm$ in the expansion of $\tilde{\delta}(\theta_{1})$ and $\tilde{\delta}(\theta_{1}-\pi)$ in terms of the spherical harmonics $\leftindex^{(d)\, } {Y}_{\vec{j}}(\theta_{1})$ noting that, since we will normalize the answer afterwards,  we care only about the $j$-dependent factors. We find
\begin{equation}
    c_j^\pm = (\pm 1)^j \left[ \left(j+\frac{d-1}{2}\right) \frac{\Gamma(j+d-1)}{\Gamma(j+1)}\right]^{1/2}.
\end{equation}
With these coefficients, the normalizations $N$ for the $\big|\psi_\pm\bigr]$ states are
\begin{equation}
    N= \sqrt{\frac{2d\Gamma({j_{\mathsmaller \star}}+1)}{(2{j_{\mathsmaller \star}}+d)\Gamma({j_{\mathsmaller \star}}+d)}}.
\end{equation}

The generators of the  de Sitter group $U(g)$ consist of  the $\frac{D(D-1)}{2}$ rotations $J_{ik}$ about each spatial direction (with $i<k$ and $i,k=1,...,D$), and $D$ boosts $B_k$. It is convenient to use the embedding space formalism to find the expressions for the corresponding Killing vector fields in terms of global coordinates. In this formalism, we represent our de Sitter space as the hypersurface $X_\mu X^\mu = \ell^2$ in a $(D+1)$-dimensional Minkowski space with metric $ds^2 = -dX_0^2 + dX_1^2 + ... + dX_{D}^2$. On the hyperboloid, the Minkowski coordinates are then related to  global coordinates through $X_0=\ell \sinh (t/\ell)$, $X_i = \ell z_i \cosh (t/\ell)$, where  the $z_i$ are functions of the angles on $S^d$ that define the standard embedding of $S^{D-1}$ in $\mathbb{R}^D$; e.g. $z_1=\cos\theta_1$, $z_2=\sin \theta_1 \cos \theta_2$, etc. The Killing fields are thus 
\begin{align}
    B_k =& z_k \ell \partial_t + \sum_{l=1}^k \frac{z_k}{z_l^2} \tanh (t/\ell) \cot\theta_l(\cos^2\theta_l - \delta_{lk})\partial_{\theta_l}, \ \ \ {\rm and} \label{eq:B_KVF} \\
    J_{ik} =& \sum_{l=i}^k \frac{z_i z_k }{z_l^2}\cot\theta_l (\cos^2 \theta_l + \sin^2\theta_i \delta_{li} - \delta_{lj})\partial_{\theta_l}. \label{eq:J_KVF}
\end{align}
Note that the action of $B_1$ is the same as in dS$_{1+1}$ given by Eq. \eqref{eq:B1action}, but with $\theta$ replaced by $\theta_{1}$ and with \eqref{eq:B2action} rewritten in terms of global coordinates.

As in Section \ref{sec:1+1}, we use the above description of $U(g)$ to compute the group averaging kernel to order $(g-1)^2$ in order to determine its width around the identity. The calculation of the kernel is greatly simplified by the symmetries of our reference state. First, $J_{ik}\big|\psi_\pm\bigr]$ vanishes for $i,k \neq d-1$, since these rotations have no effect on scalar particles at the poles. Second, the expectation values of all rotation generators $J_{ik}$ also vanish, and so too will the expectation values of all $B_l$ for $l=2,...,d$ due to the invariance of our states under reflections.  In particular, the states  $\big|\psi_\pm\bigr]$ are each individually invariant under reflections defined by choosing some $i\ge 2$ and mapping $X_i \to -X_i$ while holding fixed all $X_k$ with $k\neq i$. Additionally, under the reflection $X_1 \rightarrow -X_1$, the $B_1$ generator transforms as $B_1 \to - B_1$. This leads to a cancellation between the remaining terms of  order $(g-1)$  (since the only such terms were those associated with the expectation value of $B_1$).

Finally, we consider the cross terms $[J_{1i}J_{1k}]$, $[B_1B_i]$, $[B_i B_k]$, and $[B_k J_{1i}]$, for $i,k \neq 1$, and where the expectation values $[...]$ are taken in either the $+$ or $-$ state. That these all vanish can be seen by applying the reflection symmetries $X_i \to -X_i$, under which each of the above combinations of generators picks up a sign, but under which the states $\big|\psi_\pm\bigr]$ are individually invariant. Due to the above vanishing moments and cancellations, the group averaging kernel defined by $\big|\psi_1\bigr]$ becomes just
\begin{equation}
\label{eq:coeffeq}
    \begin{split}
        \bigl[\psi_1\big|U(g)\big|\psi_1\bigr] =& \bigl[\psi_+\big|U(g)\big|\psi_+\bigr] \bigl[\psi_-\big|U(g)\big|\psi_-\bigr] \\
        =& 1 - \lambda_{1}^2 \left([B_1^2]- [B_1]^2\right)  - \sum_{l= 2}^D (\lambda_{\perp}^l)^2 [B_l^2] - \sum_{1\le i<k\le D} (\theta^{ik})^2 [J_{ik}^2] + O([g-1]^3).
    \end{split}
\end{equation}

The analysis of the above coefficients is somewhat tedious.  We therefore relegate the details to appendix \ref{app:coeffs} and merely quote the leading results at large $j_{\mathsmaller \star}$ from \eqref{eq:leadingjcoeffs}:

\begin{align}
\label{eq:leadingjcoeffs2}
\left([B_1^2]-[B_1]^2\right) &=\frac{d}{(d+2)(d+1)^2}{j_{\mathsmaller \star}}^2 + O(j_{\mathsmaller \star}), \cr  [B_l^2] &= O(j_{\mathsmaller \star}), \cr 
 [J_{ik}^2] &= \left(\frac{1}{d+2} {j_{\mathsmaller \star}}^2 + \frac{d}{d+2} {j_{\mathsmaller \star}}\right) \delta_{i,1} \ \ \ {\rm for \ } k>i.
\end{align}
It is reassuring to note that for $d=1$ the results \eqref{eq:leadingjcoeffs2} match exactly with \eqref{eq:chi_1}.

The exact zeros of $[J_{ik}^2]$ for $i, k\neq 1$ are due to the fact that -- for simplicity -- we chose our state to preserve certain rotational symmetries.  Similarly, the small values of $[B_l^2]$ are due to the fact that our particles are localized near the corresponding horizons, so that they also preserve those symmetries at leading order.

\begin{figure}[H]
  \centering
  \includegraphics[width = 1.8in]{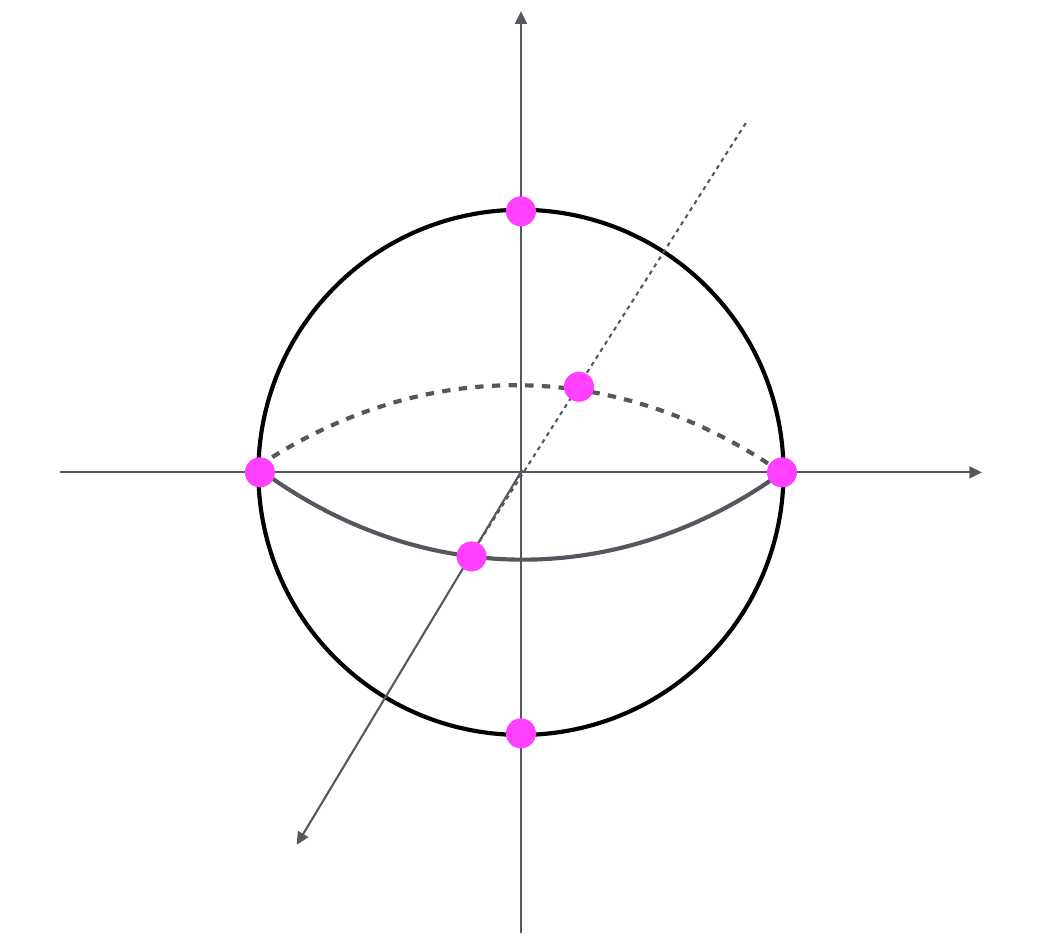}
  \caption{The state 
  \eqref{eq:2Dparticles} with particle number $2D$.  Each particle localizes at a
 point (pink dot) on the $S^d$ at global time $t=0$ along one of the coordinate axes of ${\mathbb R}^D$.}
  \label{fig:particles_higherD}
\end{figure}

However, as noted above, we wish to add additional particles to break these symmetries.   It is convenient to take the additional particles to be obtained by taking the above pair of particles (which localize along the $i=1$ axis) and rotating them so as to instead localize along the $k$th axis.  In particular, 
for $k =2, \dots, D$, we define the 2-particle states $\big|\psi_k\bigr] = e^{i\frac{\pi}{2}J_{k1}}\big|\psi_1\bigr]$.  We then combine the above states $\big|\psi_k\bigr]$  to construct a state
\begin{equation}
\label{eq:2Dparticles}
\big|\psi_0\bigr] := \bigotimes_{k=1}^D \big|\psi_k\bigr], 
\end{equation}
now with particle number $2D$ such that each particle is localized at $t=0$ along a different positive or negative axis in the embedding space ${\mathbb R}^D\supset S^{D-1}$; see figure \ref{fig:particles_higherD}.  

For the state \eqref{eq:2Dparticles} we find
\begin{equation}
\bigl[\psi_0\big|U(g)\big|\psi_0\bigr] = 1 - \frac{1}{2}\sum_{i=1}^D \lambda_i^2 \chi_i - \frac{1}{2}\sum_{1 \le i < k \le D} \lambda_{ik}^2 \chi_{ik} + O([g-1]^3),
\end{equation}
with
\begin{eqnarray}
    \chi_{i} &=& 2\left([B_1^2]-[B_1]^2\right) + 2\sum_{k=2}^D [B_k^2]
   = 
   \frac{2d}{(d+2)(d+1)^2}{j_{\mathsmaller \star}}^2 + O(j_{\mathsmaller \star}), \ \ \ {\rm and} \cr \chi_{ik} &=& 2[J_{1k}^2]  = \frac{2}{d+2}{j_{\mathsmaller \star}}^2  + O(j_{\mathsmaller \star})\ \ \ {\rm for \ } k>1.
\end{eqnarray}
Here $[X]$ still denotes the expectation value of $X$ in the original state $\big|\psi_+\bigr]$ and the approximation is valid at leading order in $j_{\mathsmaller \star}$. We see that all $\chi_i, \chi_{ik}$ are manifestly positive.

 By the same argument as in dS$_{1+1}$, our dS QFT correlators will be well-approximated by our perturbative gravity correlators in regions where the Euclidean norms (see \eqref{eq:Eds2}) of the killing vectors are much smaller than the $\chi_i^{1/2}$, $\chi_{ik}^{1/2}$.   To find the relevant $|B_k|_E^2$, we use \eqref{eq:B_KVF}. To find the magnitude of the $|J_{ik}|^2$ in Euclidean signature, we can use \eqref{eq:J_KVF}, or we can make direct use of the embedding space coordinates.  The results are
\begin{align}
\label{eq:Enorms}
    |B_k|_E^2/\ell^2 =& z_k^2  + (1-z_k^2)\sinh^2\, (t/\ell), \ \ \ {\rm and} \\
    |J_{ik}|_E^2/\ell^2 =& (z_i^2  + z_k^2 )\cosh^2\, (t/\ell).
\end{align}

In order to describe the region in which our dS QFT correlators are well-approximated by our perturbative gravity correlators, let us note that since $\chi_i$ is independent of $i$, and since $\chi_{ik}$ is independent of $i,k$, the region in which our dS QFT approximation holds to any fixed accuracy $\epsilon$ will be invariant under the full SO$(D)$ group of rotations that preserve the global time $t$.    It therefore suffices to test the conditions $|gx-x|_E \ll \sigma$ only at the pole where $\theta_1=0$. Furthermore, we may focus on the generators $B_2$ and $J_{12}$ since, for large $t_0/\ell$, we see from \eqref{eq:Enorms} that all other boosts and rotations have equal or smaller Euclidean norm at $\theta=0$.  Using either $B_2$ or $J_{12}$ leads to the condition
\begin{equation}
e^{t/\ell} \ll \chi^{1/2} \sigma \sim j_{\mathsmaller \star} \sigma,
\end{equation}
where the last step extracts the leading behavior and drops coefficients of order $1$.
Thus, as in our 1+1 toy model, for $j_{\mathsmaller \star} \lesssim 1/G$, we find that our dS QFT approximation can hold only in a region spanning a global time interval of size $\Delta t \sim \ell \ln \frac{\ell^{d-1}}{G}$.

\section{Reference particles in future dS}
\label{subsec:futureparticles}

In section \ref{sec:multi-obs} we found that, with a fixed energy budget measured by the flux $F$ through a minimal $S^1$ in dS$_{1+1}$, adding boosted particles in states $e^{\pm i t_0 B_2/\ell}\big|\psi_\pm\bigr]$ did not improve our approximation of the local algebra dS QFT in any region of dS$_{1+1}$. Since the results in section \ref{sec:d+1} for particles localized at $t=0$ in $dS_{d+1}$ are quite similar to those in section \ref{sec:one-obs}, it is again clear that with fixed total energy-flux $F$ through a minimal $S^d$, adding more particles localized at other times will again fail to improve our approximation of local algebras in $(d+1)$-dimensional dS QFT.

However, the key limitation in section \ref{sec:multi-obs} arose from  fixing $F$.  Furthermore, as discussed in section \ref{subsubsec:back}, there is generally no sense in which backreaction can remain small across {\it all} of dS, so one must make a choice of both where,  and in what sense, one wishes perturbation theory to hold.  Finally, since global de Sitter space is exponentially large in both the far future and the far past, the energy carried by perturbations tends to become extremely diluted in such regions and backreaction tends to be much smaller than at a minimal $S^d$.

Let us therefore investigate what we can do if we decide to allow large backreaction near the minimal $S^d$ at $t=0$ (thus dropping the constraint on $F$), though we will still require backreaction to be small to the future of some global time slice $t=t_0>0$ where the particles all localize.    We will do so using the reference state
\begin{eqnarray}
\label{eq:psifuture}
\big|\psi_0\bigr] &=& \bigotimes_{k=1}^D \big|\psi_k\bigr], \ \ \ \text{with} \\
\big|\psi_1\bigr] &:=& e^{i\frac{t_0}{\ell}B_1}\big|\psi_+\bigr] \otimes e^{-i\frac{t_0}{\ell}B_1}\big|\psi_-\bigr], \\
\big|\psi_k\bigr] &:=& e^{i\frac{\pi}{2}J_{k1}}\big|\psi_1\bigr], \ \ \ {\rm for} \ 2\le k \le D, 
\end{eqnarray} where $\big|\psi_\pm\bigr]$ are again given by \eqref{eq:psi_pmD}; i.e., we use {\it only} particles that become localized at global time $t=t_0 >0$ and we again impose $j_{\mathsmaller \star} \ll 1/G$ in \eqref{eq:psi_pmD}.  At time $t=t_0$, each particle thus gives only a small perturbation, and the perturbation toward the future should be even smaller.  We will investigate later the extent to which the resulting perturbations can remain small at times $t< t_0$.

\begin{figure}[H]
  \centering
  \includegraphics[width = 1.3in]{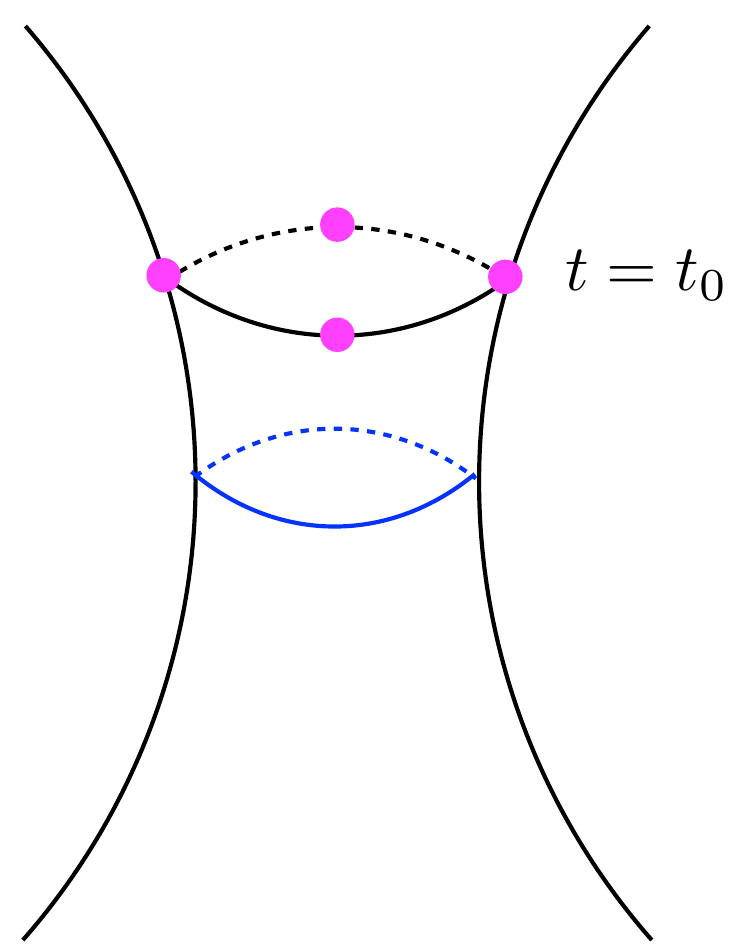}
  \caption{The state \eqref{eq:psifuture} has particle number $2D$.  Each particle localizes at a
 point (pink dot) on the $S^d$ at global time $t_0 >0 $ along one of the coordinate axes of ${\mathbb R}^D$. }
  \label{fig:reference_future}
\end{figure}

Since each state $\big|\psi_k\bigr]$ is invariant under the same set of (spatial) reflection and rotation symmetries as the similarly-labeled state in section \ref{sec:multi-obs},  the non-vanishing moments that contribute to the group-averaging kernel are again just the expectation values of $B_k^2$ and $J_{ik}^2$, for which will again define squared-widths $\chi_k$ and  $\chi_{ik}$.  Furthermore,
while it is straightforward to compute these expectation values from the results in section \ref{sec:multi-obs}, the analyses of the previous sections show that we do not in fact need the detailed forms of the results.  Instead, the important point is that, due to the boost transformations of $B_k, J_{in}$ (analogous to those in \eqref{eq:Rintime}), for the particles in state $\big|\psi_k\bigr]$ and for $i\neq k$ we will find $\chi_i$ and $\chi_{kn}$ to be proportional to $e^{2t_0/\ell}$ for large $t_0$, though $\chi_k$ and $\chi_{in}$ for $i,n\neq k$ will be unchanged by the boost.

As a result, in the full state $\big|\psi_0\bigr]$ we will find all $\chi_i, \chi_{ik}$ to be proportional to $e^{2t_0/\ell}$ for large $t_0$.
This will then compensate for the fact that the Euclidean norms $|B_i|^2_E$ and $|J_{ik}|_E^2$ are exponentially large at $t=t_0$, and it will similarly allow these vector fields to satisfy our criterion for a good approximation to dS QFT for a global time interval of order $\ln G^{-1}$ to the future of $t=t_0$.  This condition is also satisfied whenever $|t|< t_0$, so long as we consider a region of dS in which backreaction is small.

It thus remains only to estimate the backreaction from our state.  There are several pieces to this discussion.  First, we may note that at time $t=t_0$ we have only $2D$ particles on a sphere $S^d$ of volume $\cosh^d(t_0/\ell)\,  V_d \ell^d$, where $V_d$ is the volume of the unit sphere.  Furthermore, by construction each particle alone has small backreaction, meaning that it can be modeled as a Schwarzschild black hole of radius much less than $\ell$.  As a result, for all $t \gg \ell$ the particles are exponentially far apart, and -- in some reference frame\footnote{For $t_0-t \gg \ell$, the particle will be highly relativistic in the reference frame associated with slices of constant global time $t$.  It should then be described as a de Sitter version of the Aichelberg-Sexl solution \cite{Aichelburg:1970dh}.   In 2+1 dimensions there is no curvature away from the particle, and for $d\ge4$ the Aichelberg-Sexl solution decays at long distance.  But in 3+1 dimensions the Aichelburg-Sexl metric grows logarithmically.  Of course, curvatures still decay and, more importantly, the gravitational field is entirely confined to a shock wave in a null plane.  As a result, for $t\gg \ell$ the vast majority of de Sitter space remains exponentially far from such shock waves and, indeed, the total probability for an object in dS to encounter such a shock wave between global time $t$ and global time $t=\infty$ is exponentially small.} -- the region near each particle can be modeled as a Schwarzschild de Sitter solution (again with Schwarzschild radius much less than $\ell$).  In this sense the backreaction is small at $t=t_0$ unless one probes the small region very close to one of the particles.  Recalling that metric perturbations can in principle be included in the field we call $\phi$, such comments can be promoted to statements about gauge-invariant operators of the form \eqref{eq:PGfield} if desired.

It is important to realize, however, that there are small perturbations to de Sitter at $t=t_0$ even far away from the expected locations of the particles.  Some of this effect is due to the fact that, since we cut off the mode sum defining each particle's state at some $j_{\mathsmaller \star}$, the particles are not perfectly localized and their wavefunctions have long tails that extend across all of de Sitter space.  That, however, is a minor issue as, at large $j_{\mathsmaller \star}$, those long tails correspond to only a tiny net probability for the particles to be far from their expected location.  Furthermore, one could remove the long tails by replacing each particle by the coherent state obtained by making a unitary transformation of the $\psi$-vacuum using $e^{i\int \psi(x) f(x)}$ for some compactly-supported $f(x)$.

While $e^{i\int \psi(x) f(x)}$ is invariant under linearized gauge-transformations, it will not be gauge-invariant at the higher orders in perturbation theory used to compute backreaction.  Instead, as usual, it must be `gravitationally dressed'.  In the present context, the only structure to which such operators can be dressed is the background itself.  This is simply a set of words which means that the backreaction need not vanish in regions that are causally separated from the support of $f$, and that one must instead solve the gravitational constraints to analyze what happens in such regions.

It is reasonable to expect that the overall effect on the expansion/contraction of the spheres $S^d$ will be well-approximated by smearing the particles over the sphere.  By this we mean that we will simply solve the Friedman equations for homogeneous isotropic cosmologies using the homogeneous energy density $\rho$ that gives a flux $F_t= \int_t \sqrt{h} T_{ab} n^a n^b =  \rho \cosh^d(t_0/\ell)\, V_d \ell^d $ of energy
through the given sphere $S^d$ that agrees with the flux $F_t$ computed perturbatively for our particles. Since angular momentum is conserved, assuming that $\mu$ and $j_{\mathsmaller \star}$ are of the same order in $1/G$, once $t$ is significantly less than $t_0$ the particles will be relativistic and we will have
\begin{equation}
F_t \sim \frac{2D j_{\mathsmaller \star}}{\ell} e^{\frac{t_0-t}{\ell}}.
\end{equation}
Replacing our particles with a uniform energy density
\begin{equation}
\rho(t) \sim F_t e^{-dt/\ell} \ell^{-(d+1)} \sim j_{\mathsmaller \star} e^{\frac{t_0-(d+1)t}{\ell}}\ell^{-(d+2)},
\end{equation}
and comparing this with the energy density $\rho_\Lambda = \frac{d(d-1)}{16\pi G\ell^2}$ associated with the de Sitter cosmological constant, we find that $\rho(t) \ll \rho_\Lambda$  for
\begin{equation}
e^{(d+1)\frac{t}{\ell}} \gg
 \frac{ G j_{\mathsmaller \star}}{\ell^{d-1} }  e^{t_0/\ell};
\end{equation}
i.e., the backreaction from our homogeneous $\rho$ will be small whenever $t$ exceeds $\frac{t_0}{d+1} - \frac{\ell}{d+1}\ln \left(\frac{\ell^{d-1}}{Gj_{\mathsmaller \star}}\right)$ by at least a few e-folding times.

Now, since the state $\big|\psi_0\bigr]$ contains only a small number ($2D$) of particles, it is clear that the actual energy density is far from homogeneous.  Some of the issues involving the inhomogeneous part of $\rho$ were discussed above and relate to probing the local spacetime near each particle.  However, additional effects arise when, e.g. by random chance, some subset of the particles finds themselves closer together than other subsets.   It is then natural to model such circumstances by a $\rho$ that is again smooth, but where the local energy density in that region is larger than in other regions.  Comparing with our analysis above, we see that this will then increase the backreaction in such regions, though this can only be the case in small regions of spacetime.  Qualitatively, then, this is similar to the comments above about probing small regions near each particle.  In this sense, then, we expect backreaction to be small over the vast majority of the region of our de Sitter space at global times
\begin{equation}
t> \frac{t_0}{d+1} -\frac{\ell}{d+1}\ln \left(\frac{\ell^{d-1}}{Gj_{\mathsmaller \star}}\right).
\end{equation}
As a result, our perturbative gravity correlators will be a good approximation to our dS QFT correlators over the vast majority of the region satisfying
\begin{equation}
t_0 + \ell \ln \left(\frac{j_{\mathsmaller \star}\sigma}{\ell}\right) > t> \frac{t_0}{d+1} -\frac{\ell}{d+1}\ln \left(\frac{\ell^{d-1}}{Gj_{\mathsmaller \star}}\right).
\end{equation}
Taking $\sigma = \epsilon_1 \ell$, $j_{\mathsmaller \star} = \epsilon_2 \ell^{d-1}/G$ then yields
\begin{equation}
t_0 + \ell \ln \left(\epsilon_1 \epsilon_2   \frac{\ell^{d-1}}{G}\right) > t> \frac{t_0}{d+1} +\frac{\ell}{d+1}\ln \epsilon_2.
\end{equation}
Taking $t_0$ large (say, with $\epsilon_1$ and $\epsilon_2$ small but with $\epsilon_1 \epsilon_2 \frac{\ell^{d-1}}{G}$ large)
then allows us to make our approximation highly accurate over a region with arbitrarily large spacetime volume and which spans an arbitrarily large interval of global time.

\section{Discussion}\label{sec:discussion}

Our work above studied the use of the perturbative gravity observables \eqref{eq:PGfield} in approximating algebras of local quantum fields on a fixed de Sitter spacetime $dS_{d+1}$.  In the limit $G\rightarrow 0$, one can approximate such local fields well over arbitrarily large regions of dS.  However, if the region of interest includes a minimal $S^d$, we found this approximation to fail at small $G$ when the region spanned a global time interval significantly larger than $\ln(\ell^{d-1}/G)$ (plus subleading corrections).
On the other hand, we argued that the approximation could hold to high precision in regions spanning arbitrarily large global time intervals so long as they are located far to the future (or far to the past) of the associated minimal $S^d$.  This in particular includes arbitrarily large regions of any static patch of the de Sitter space.

Although our analysis of the possible constructions was far from exhaustive, and although our detailed computations were performed only for free scalar fields with masses $M> 2/d\ell$, we saw that the main results depended only on the presence of certain exponential behaviors that follow from basic de Sitter kinematics.  We therefore expect our conclusions to be quite robust.
It would nevertheless be useful to make the analysis more complete with respect to possible choices of reference $\big|\psi_0\bigr]$, and to incorporate perturbative interactions, gravitational or otherwise.   Similarly, for simplicity we treated our $\psi$-particles as distinguishable but, since no two $\psi$-particles occupy the same mode, it is clear that symmetrizing/antisymmetrizing over particles in $\big|\psi_0\bigr]$ will not affect our results.

We emphasize that our interest here concerned {\it algebras} of local fields.  In particular, we may consider arbitrary products of the perturbative gravity observables $\hat \phi_{LPG}(x)$ defined in  \eqref{eq:PGfield}. While there are subtleties related to the fact that our $\hat \phi_{LPG}(x)$ are unbounded, this is easily remedied by replacing the operators $\hat \phi_{QFT}(x)$ in the integrand of \eqref{eq:PGfield} with bounded functionals of $\hat \phi_{QFT}(x)$.

Of course, our dS QFT approximation does not hold uniformly for all elements of the resulting algebra, nor does it hold uniformly in all states.  In particular, at any fixed value of $G$, operators formed by taking sufficiently large products will create states with large back-reaction. Nevertheless, it is clear that as $G \rightarrow 0$ one can choose parameters such that the approximation holds for a larger and larger subset of elements of the algebra, and such that it holds well over a larger and larger set of states.  Thus as $G\rightarrow 0$ one should recover the entire algebra of local quantum $\phi$-fields, though filling in the technical details and characterizing the various rates of convergence remains a project for future investigation.

In contrast, had we been interested only in computing vacuum correlation functions (without first constructing an algebra), we could have approximated the results of dS QFT to much higher precision.  At the physical level, this relates to the point often made by cosmologists that, since the vacuum is de Sitter invariant, if one wishes to compute the vacuum two-point function $\bigl[0\big|\phi(x) \phi(y)\big|0\bigr]$, then there is no need to sharply define the location of both points $x$ and $y$ so long as the geodesic distance between the two is sharply defined.  Mathematically, we may note that we can construct a de Sitter-invariant perturbative-gravity observable
\begin{equation}
\label{eq:2ptO}
{\cal O}(x,y) := \int dg \, \phi(gx) \phi(gy) \otimes U(g)\big|\psi_0\bigr]\bigl[\psi_0\big|U(g^{-1})
\end{equation}
whose expectation value in our $\phi$-vacuum state $|0;LPG\rangle$ is
\begin{equation}
\label{eq:2ptOcorr}
\langle 0; LPG| {\cal O}(x,y)|0;LPG\rangle = \bigl[0\big|\phi(x) \phi(y)\big|0\bigr]\, \left(\int dg   \Big|\bigl[\psi_0\big|U(g)\big|\psi_0\bigr]\Big|^2\right) ,
\end{equation}
so that it {\it exactly} reproduces the two-point function of dS QFT at all $x,y$ for any value of $G$.  However, since \eqref{eq:2ptO} is not the product of two perturbative gravity observables \eqref{eq:PGfield}, the result \eqref{eq:2ptOcorr} says nothing about the accuracy of approximating the {\it algebra} of local fields.

Our interest in local algebras was in part motivated by recent works constructing type II von Neumann algebras of local fields \cite{Chandrasekaran:2022cip,Chandrasekaran:2022eqq,Penington:2023dql,Jensen:2023yxy,Chen:2024rpx,Kudler-Flam:2024psh}.  Specifically, we wished to investigate the way in which the static patch algebra of \cite{Chandrasekaran:2022cip} could emerge from perturbative gravity observables.    In this regard, there are several aspects of our approach on which we wish to remark.

The first of these is that the algebra of perturbative-gravity observables generated by $\hat \phi_{LPG}(x)$ does not directly reproduce the {\it full} algebra of dS QFT, but only the algebra of $\phi$-fields.   In particular, it does not include the algebra of $\psi$-fields which can change our reference state $\big|\psi_0\bigr]$.  Nevertheless, this precisely matches the structure of the Hilbert space used in \cite{Chandrasekaran:2022cip} in the sense that that work assumed the existence of a so-called observer, and that the operator algebra was not allowed to either create or destroy such observers.  Furthermore, while the observer's clock operator was used at an intermediate point of the construction, the observables constructed in \cite{Chandrasekaran:2022cip} can be described as what we would call $\phi$-observables defined at times relative to the observer's clock.  This is clearly in direct parallel with our   perturbative-gravity observables, which describe $\phi$-fields relative to our reference state $\big|\psi_0\bigr]$.

However, there should be no problem with including additional operators that allow the creation/annihilation of $\psi$-quanta which are {\it not} to be considered part of our reference state; e.g. which act on modes with much smaller angular momentum on each $S^d$.  Indeed, to the extent that we can treat $\psi$-particles as distinguishable, one may simply consider a space of states that is the tensor product of a high-angular-momentum $\big|\psi_0\bigr]$ state with arbitrary low-angular-momentum states of the $\psi$-field, and one may then define perturbative-gravity observables that act on this space in direct analogy with \eqref{eq:PGfield}.    Taking into account that $\psi$-particles are identical then involves a formal symmetrization/antisymmetrization depending on the bosonic/fermionic nature of the $\psi$-particles but, as usual, this has little effect when the relevant two sets of particles occupy very different modes.

A more interesting point is that, while they clearly reproduce a local algebra in the limit $G\rightarrow 0$, from the perspective of the dS quantum field theory that acts on the Hilbert space $\mathcal H_{QFT}=\mathcal H_{QFT}^\phi \otimes \mathcal H_{QFT}^\psi$ the operators $\hat \phi_{LPG}$ are highly non-local even before they are group-averaged.  This is due to the fact that, in contrast to the constructions used in \cite{DeWitt:1962,Marolf:1994nz,Giddings:2005id,Marolf:2015jha,Chen:2024rpx},  the integrand in \eqref{eq:PGfield} contains a factor of $\big|\psi_0\bigr]\bigl[\psi_0\big|$, which is an operator not contained in the local algebra of quantum $\psi$-fields for any  subregion of dS which cannot describe a complete Cauchy surface.  As discussed in the introduction (following \cite{Giddings_2007}), this property is absolutely essential if the group averaging integral in \eqref{eq:PGfield} is to converge in the presence of a long-lived de Sitter vacuum state.  While it is naturally viewed as a surprise, and perhaps in fact a distasteful one from the perspective of local quantum field theory, we emphasize that perturbative gravity about a background that fails to break all diffeomorphism gauge symmetries is {\it not} a local quantum field theory.  In particular, since observables in perturbative gravity about dS must be invariant under the entire de Sitter group (see again the discussion at the beginning of section \ref{sec:GApdS}),  they are in some sense necessarily as far from local operators as one can get.  In any case, whatever the philosophical issues may be regarding our construction, we see that it does in fact reproduce local quantum field theory in the limit $G\rightarrow 0$.

The idea of using highly non-local ingredients like quantum states to define observables which, in some limit, nevertheless reproduce local physics seems likely to be extremely useful in quantum gravity more generally, and especially in attempts to go beyond perturbation theory.  The point here is that the path integral is naturally taken to define an inner product on quantum states that projects onto gauge-invariant states, and which can then be used to build a gauge-invariant Hilbert space; see e.g. the discussions in \cite{Halliwell:1990qr,Marolf:1996gb,Reisenberger:1996pu}.   As a result, this path integral inner product automatically implements group averaging when expanded perturbatively (see e.g. the recent discussion in \cite{Chakraborty:2023los}).  Thus, to the extent that we understand how to compute gravitational path integrals, we already have the desired Hilbert space of states at hand.  We may thus use such states $|\Psi_1\rangle, |\Psi_2\rangle$ to directly construct gauge-invariant operators $|\Psi_1\rangle \langle \Psi_2|$ rather than take on the technical challenge of attempting to perform an integral over the diffeomorphism group of some more local operator expression (and then needing to worry further if e.g. topology-changing processes further enlarge the gauge group in the non-perturbative quantum theory; see e.g. the discussion in \cite{Marolf:2020xie}).

Our work also reported some technical progress regarding the de Sitter group-averaging inner product that underlies our analysis.   Appendix \ref{app:IP} proposed a potential alternate formulation of this inner product that, if it is finite and non-zero, must be equivalent to group averaging by the uniqueness theorem of  \cite{Giulini_1999} and the freedom to tune a parameter $(\alpha)$ to make the result finite and non-zero. However, the alternate inner product is manifestly non-negative.  We also argued that any inner product for perturbative gravity on dS must map one-particle states of our $\mathcal H_{QFT}$ to null states.  The argument was a direct quantum analogue of the fact that classical one-particle states cannot satisfy the linearization stability constraints.  Finally, we argued that the divergence of group averaging for Fock-basis 2-particle states of massive scalar fields is related to the fact that all classical 2-particle solutions preserve a notion of static-patch time-translation symmetry and thus, like the de Sitter-invariant vacuum $\big|0\bigr]$, leave a non-compact gauge group unbroken.  The fact that the unbroken gauge group is now only $\mathbb{R}$ is then naturally associated with the fact that group-averaging diverges only linearly for such 2-particle states, while it diverges exponentially for $\big|0\bigr]$.  While it would be useful to sharpen this last argument, and also to rigorously prove equivalence of group-averaging with our alternate inner product, for heavy fields in dS (e.g., for scalars with $M>d/2\ell$), this gives a rather complete understanding of the group averaging inner product.

This technical progress again has implications for the construction of local algebras and the connection between our work and that of \cite{Chandrasekaran:2022cip,Jensen:2023yxy,Chen:2024rpx,Kudler-Flam:2024psh}.  In particular, the two-particle states with linearly-divergent group-averaging norm naturally play the role of a clock-less version the observer assumed in \cite{Chandrasekaran:2022cip,Jensen:2023yxy,Chen:2024rpx,Kudler-Flam:2024psh}.  Indeed, without a clock, group-averaging would again diverge linearly in the context studied in those works.   Adding an appropriate clock degree of freedom, whether realized as the relative motion of a 3rd particle or as the addition of infinitely many internal states that mix under time evolution, will thus cause group averaging to converge.  The resulting states can then be used as a reference $\big|\psi_0\bigr]$ in precisely the manner described here.

\begin{figure}[H]
  \centering
  \includegraphics[width = 1.3in]{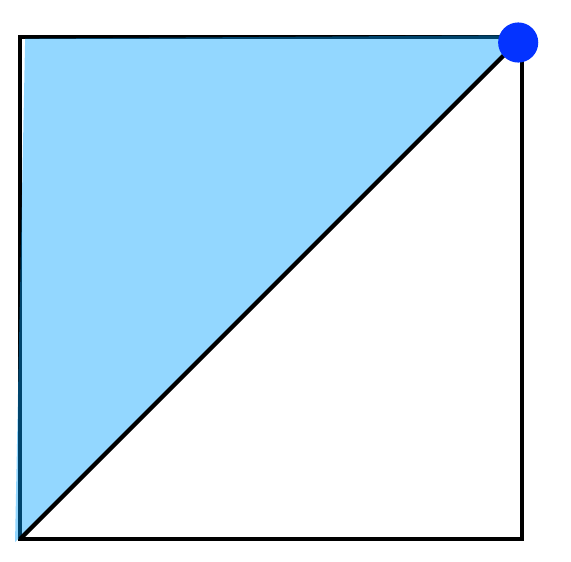}
  \caption{ The inflating patch (blue) is shown on a conformal digram of unperturbed global dS.  Discussions of physics in this patch require boundary conditions at the indicated point (dot in upper right corner of the diagram).}
  \label{fig:inflating_patch}
\end{figure}

Finally, some readers may be surprised that we have focused so heavily on the study of {\it global} de Sitter space.  In contrast, many treatments of de Sitter or inflation discuss only the inflating patch of de Sitter.  Since the inflating patch has {\it noncompact} Cauchy surfaces, its Killing fields are not normally treated as generating gauge symmetries.  In effect, one typically assumes (perhaps implicitly) that boundary conditions are imposed at the upper corner of the inflating patch (see figure \ref{fig:inflating_patch}.) such that the associated diffeomorphisms are non-trivial asymptotic symmetries rather than gauge.  However, we are not aware of a complete technical specification and treatment of such boundary conditions and, perhaps as a result, many longstanding questions and confusions remain regarding the detailed relationship between analyses of global dS and analyses in the inflating patch.  Nevertheless, it seems likely that, as discussed briefly in \cite{Higuchi:2011vw}, 
the physics of global dS linearization stability constraints is directly related to large logarithms that arise when studying gravitational perturbation theory in the inflating patch; see, e.g., \cite{Seery_2010} and reference therein, \cite{Linde_2005,Giddings_2011}.   It would thus be very interesting to better understand the implications of constructions described here for physics in the inflating patch and, in particular, to understand if (despite the seemingly different time-dependence involved) the above-mentioned large logarithms might be related to the fact that our perturbative gravity correlators become highly smeared versions of dS QFT correlators at late times.

\acknowledgments

We thank Xi Dong, Gary Horowitz, and Geoff Penington for useful discussions.
This research was supported by NSF grants PHY-2107939 and PHY-2408110, and by funds from the University of California. X.Y. was supported by the National Science Foundation Graduate Research Fellowship under Grant No. 2139319. Y.Z. was supported in part by the National Science Foundation under Grant
No. NSF PHY-1748958, by a grant from the Simons Foundation (815727, LB), by Heising-Simons Foundation award-6950153, and by David and Lucile Packard Foundation Fellowship-2749255. Any opinion, findings, and conclusions or recommendations expressed in this material are those of the authors(s) and do not necessarily reflect the views of the National Science Foundation.

\appendix
\section{Group averaging and its extensions}
\label{app:IP}

This appendix addresses the finiteness and positivity properties of the group averaging inner product.    We begin in section \ref{subsec:summary} with a summary of results in the existing literature.  We then propose a family (labeled by a single parameter $\alpha$) of potential alternative  inner products in section \ref{subsec:positivity}. We use the uniqueness theorem of \cite{Giulini_1999} to  argue that, if there is a value of $\alpha$ such that this alternate inner product is well-defined on the space $V\subset \mathcal H_{QFT}$ of states on which group-averaging converges absolutely, then the two inner products must coincide on $V$, so that the alternate inner product is in fact an extension of the group-averaging inner product.    The extended definition is manifestly positive semi-definite
and, as discussed in section \ref{subsec:ext}, it assigns vanishing norm to all 1-particle states.  One may also independently show that any allowed extension must assign vanishing norms to these states.   For scalar fields with masses $M>d/2\ell$, this would thus suffice to show that the extended inner product is finite and positive semi-definite for seed states in a dense subspace of the space that is both orthogonal to the dS-invariant vacuum $\big|0\bigr]$ and orthogonal to all 2-particle states.  Finally, the case of 2-particle states is discussed in section \ref{subsec:2ps}, where we suggest that their divergent group-averaging norm is related to the fact that the corresponding classical solutions leave unbroken a non-compact subgroup of SO$(D,1)$.

\subsection{Summary of previous results}
\label{subsec:summary}

The case of linearized gravitons on dS$_4$ turns out to be exactly solvable, and \eqref{eq:innerproduct} was shown in \cite{Higuchi_1991_2} to be finite and positive semi-definite when $V$ is the space spanned by Fock basis-states with $N\ge 2$ particles associated with the standard linearized graviton modes on global dS$_4$.  While it was not obviously finite for $N=1$ particle states, we will return to that case in section \ref{subsec:ext} below.

For general massive free minimally-coupled scalar fields, the group averaging integral \eqref{eq:innerproduct} was shown to be finite in \cite{Marolf:2010zp} for standard Fock states that contain a sufficient number of particles.  For scalar fields in the so-called principal series of SO$(D,1)$ representations (having $M^2> (D-1)^2/4\ell^2$ where $\ell$ is the de Sitter length scale), this result holds for $N > 2$ particles.  The threshold particle number  is higher for scalar fields than for the free gravitons studied in \cite{Higuchi_1991_2} because convergence is aided by adding angular momentum and because there are no gravitons with angular momentum quantum number $j_{d-1}$ (as defined in section \ref{sec:d+1}) satisfying $j_{d-1}< 2$.  The reader may think of this result as a generalization of Birkhoff's theorem (which immediately excludes gravitons with $j_{d-1}=0$).   

Larger numbers of particles are required for fields with smaller masses. The required number $N$ diverges as $M\rightarrow 0$ due to the fact that massless fields on dS have a zero mode, though sufficient excitations of this zero-mode also provide convergence; see \cite{Marolf:2008it} for discussion of the dS$_{1+1}$ case.  For smaller numbers of particles the integral \eqref{eq:innerproduct} fails to converge absolutely, though for the $N=1$ particle case this can be dealt with as described in section \ref{subsec:ext} below.

\subsection{An alternate definition?}
\label{subsec:positivity}

An important part of justifying the group-averaging inner product \eqref{eq:innerproduct} is that it satisfies a certain uniqueness result \cite{Giulini_1999}.  We will now use this uniqueness to outline an argument that \eqref{eq:innerproduct} is positive semi-definite when it converges absolutely.  The argument involves the introduction of an alternate inner product which, if it is finite and non-zero on {\it any} state in the space $V$ on which group-averaging converges absolutely (and for which the group-averaging norm is non-zero), must agree with group-averaging on all of $V$ by the argument of \cite{Giulini_1999}.  We in fact introduce a one-parameter family of potential alternate inner products in the expectation that there will be a value of this parameter which gives a finite no-zero result on $V$.  Unfortunately, however, the alternate inner products are difficult to compute.  We thus leave detailed investigation of this expectation for future work.

Translating to the language of the current paper, the uniqueness theorem of \cite{Giulini_1999} shows that if $V \subset \mathcal H_{QFT}$ is a space of states on which \eqref{eq:innerproduct} converges absolutely, then there is an associated algebra $\mathcal A_V$ of de Sitter-invariant linear operators defined by
\begin{equation}
a_{\psi_1, \psi_2} : = \int dg U(g) \big|\psi_1\bigr] \bigl[\psi_2\big| U(g^{-1})
\end{equation}
for $\psi_1, \psi_2 \in V$.  Since the Haar measure $dg$ is invariant under $g \rightarrow g^{-1}$, these operators map $V$ to itself and also satisfy
\begin{equation}
\label{eq:Herm}
\bigl[\phi_1\big|a_{\psi_1, \psi_2}\phi_2] = [a_{\psi_2, \psi_1}\phi_1\big|\phi_2\bigr]
\end{equation}
for all $\phi_1, \phi_2 \in V$.
Here as usual we have defined $|a\phi_1] := a\big|\phi_1\bigr]$.
The theorem then shows that, up to an overall normalization, \eqref{eq:innerproduct} is the unique product on $V$ that satisfies linearity with respect to the 2nd argument,
the complex conjugation condition
\begin{equation}
\label{eq:IPconj}
\langle \phi_1|\phi_2\rangle^* = \langle \phi_2 | \phi_1 \rangle,
\end{equation}
the $*$-representation condition\footnote{The theorem of \cite{Giulini_1999} is stated in terms of so-called rigging maps which are analogues of \eqref{eq:GAPsi}.  In that language, the main result is that the rigging map commutes with all $a\in\mathcal A_V$.  But the reality property of rigging maps together with \eqref{eq:Herm} makes this equivalent to \eqref{eq:hermAPhi}.}
\begin{equation}
\label{eq:hermAPhi}
\langle \phi_1|a_{\psi_1, \psi_2}\phi_2\rangle = \langle a_{\psi_2, \psi_1}\phi_1|\phi_2\rangle,
\end{equation}
and with respect to which $(U(g)-1)|\phi]$ is a null vector for all $g\in \text{SO}_0(D,1)$ and all $|\phi]\in V$.

As a result, if we can find another inner product satisfying the above properties on the same domain $V$, it must be equivalent to \eqref{eq:innerproduct} up to an overall normalization (though we must then check that this normalization constant is finite and non-zero).  Let us therefore think of dS$_D$ as embedded in $D+1$ Minkowski space so that, choosing a standard set of inertial coordinates $X^0, \dots , X^{D}$ on that Minkowski space (with Minkowski metric $\eta_{\mu \nu} = {\rm diag}(-1,1,\dots, 1)$), we can write the generators of the de Sitter group as the generators $J_{\mu \nu} = - J_{\nu \mu}$ of Lorentz transformations on Minkowski space and define the operators 
\begin{equation}
\label{eq:JBdef}
J^2 : = \sum_{1\le i<j\le D} J_{ij}^2,  \ \ \ B^2: = \sum_{i=1}^{D} J_{0i}^2.
\end{equation}
For any real parameter $\alpha$ we may then consider the inner product
\begin{equation}
\label{eq:altIP}
(\phi_1|\phi_2) = \bigl[\phi_1\big|\, |B|^\alpha \delta(B^2) \Pi_{J^2=0}  \big|\phi_2\bigr] : = \frac{1}{2\pi}\int_{\lambda \in \mathbb{R}} d\lambda \bigl[\phi_1\big|\,|B|^\alpha e^{i\lambda B^2} \Pi_{J^2=0}  \big|\phi_2\bigr],
\end{equation}
where $|B|^\alpha : = (B^2)^{\alpha/2}$ and where $\Pi_{J^2=0}$ is the projection onto states with $J^2=0$; i.e., onto states that are invariant under the rotational subgroup SO$_0(D)\subset$ SO$_0(D,1)$.   This inner product is written with round brackets to distinguish it from the other inner products used in this work.  Expression \eqref{eq:altIP} satisfies the complex conjugation condition \eqref{eq:IPconj} since
\begin{eqnarray}
\label{eq:altIPconj}
(\phi_1|\phi_2)^*&=& \frac{1}{2\pi}\int_{\lambda \in \mathbb{R}} d\lambda \left(\bigl[\phi_1\big|\,|B|^\alpha e^{i\lambda B^2} \Pi_{J^2=0}  \big|\phi_2\bigr]\right)^* \cr &=& \frac{1}{2\pi}\int_{\lambda \in \mathbb{R}} d\lambda \left([\phi_2|\, |B|^\alpha e^{-i\lambda B^2} \Pi_{J^2=0}  \big|\phi_1\bigr]\right) = (\phi_2|\phi_1).
\end{eqnarray}

Furthermore, since $B^2$ commutes with all rotations, and since $\Pi_{J^2=0} = \int_{r\in \text{SO}_0(D)} U(r) dr$ (where $dr$ is the normalized Haar measure on SO$_0(D)$), we see that the factors $|B|^\alpha$,  $e^{i\lambda B^2}$, and  $\Pi_{J^2=0}$ (or equivalently $|B|^\alpha$,  $\delta(B^2)$, and  $\Pi_{J^2=0}$) all commute with each other.  In the same way we see that if the integral in \eqref{eq:altIP} converges, then we have
\begin{equation}
\label{eq:altIPJ}
(\phi_1|J_{\mu \nu} \phi_2) = 0
\end{equation}
for all SO$(D,1)$ generators $J_{\mu \nu}$.  For rotations this follows immediately from $\Pi_{J^2=0} J_{ij}=0$, while for boosts it follows from the fact that $\delta(B^2) =   \frac{1}{2\pi}\int_{\lambda \in \mathbb{R}} d\lambda  e^{i\lambda B^2}$ is the $\epsilon \rightarrow 0$ limit of projectors onto the part of the spectrum of $B^2$ in the range $[0,\epsilon]$.  Since $B^2$ is a positive-definite quadratic combination of the $J_{0i}$  (see \eqref{eq:JBdef}), this implies that at finite $\epsilon$ it also restricts the spectrum of any $J_{0i}$  to the interval $[0, \sqrt{\epsilon}]$.  For $(\mu, \nu)=(0,i)$, expressions like   \eqref{eq:altIPJ} then contain an extra factor of $\sqrt{\epsilon}$ and thus vanish as $\epsilon \rightarrow 0$.  As a result, for $\big|\phi_1\bigr], \big|\phi_2\bigr] \in V$ and any $g\in \text{SO}_0(D,1)$ we have
\begin{equation}
\label{eq:altNV}
(\phi_1|[U(g)-1]\phi_2) =0.
\end{equation}

We can also show that the alternative inner product \eqref{eq:altIP} satisfies the $*$-representation condition \eqref{eq:hermAPhi}.   To do so, we
simply note that $a_{\psi_1, \psi_2}$ commutes with all $U(g)$, and that it thus commutes with $\Pi_{J^2=0} $,  $|B|^\alpha$, and $e^{i\lambda B^2}$.  We may then use \eqref{eq:Herm} to write
\begin{equation}
(a_{\psi_2, \psi_1}\phi_1|\phi_2) = \bigl[\phi_1\big|  a_{\psi_1, \psi_2}   \int d\lambda |B|^\alpha e^{i \lambda B^2} \Pi_{J^2=0}  \big|\phi_2\bigr] =
(\phi_1|a_{\psi_1, \psi_2}\phi_2),
\end{equation}
as desired.

As a result, if we can find a value of $\alpha$ for which \eqref{eq:altIP} is finite on the domain $V$ spanned by global dS Fock basis states which contain enough particles for group averaging to converge absolutely, and if \eqref{eq:altIP} is finite non-zero for that $\alpha$ and any choice of $\big|\phi_1\bigr], \big|\phi_2\bigr] \in V$ then, up to an overall normalization,  for that $\alpha$ and all $\big|\psi_1\bigr], \big|\psi_2\bigr] \in V$ the alternative inner product $(\psi_1, \psi_2)$ must agree with the group averaging inner product \eqref{eq:innerproduct}.    Moreover, since $\delta(B^2)$ can be expressed as a limit of spectral projections, the alternative inner product \eqref{eq:altIP} is manifestly poistive-definite.  Finding the above $\alpha$ would then also establish positive semi-definiteness of group averaging on $V$.

The existence of such an $\alpha$ is certainly plausible, but the inner product \eqref{eq:altIP} appears difficult to compute.  We thus leave further investigation of this issue for future work. 

\subsection{One-particle states for free fields}
\label{subsec:ext}

In section \ref{subsec:positivity}, we suggested that the inner product \eqref{eq:altIP} (for some $\alpha$) provides an alternative way of writing the group-averaging inner product \eqref{eq:innerproduct} on the original domain $V$. However, it can happen that expression \eqref{eq:altIP} is well-defined for states where \eqref{eq:innerproduct} is not.  If the alternate and group-averaging inner products do in fact agree on $V$ for some $\alpha$, this would then suggest that \eqref{eq:altIP} is well-defined on a domain $V_{alt}$ that is strictly larger than the original domain $V$.  It would then be tempting to define the desired de Sitter invariant Hilbert space $\mathcal H_{LPG}$ using
\eqref{eq:altIP} on the full domain $V_{alt}$.  In this context, we could refer to \eqref{eq:altIP} as an extended group-averaging inner product.

Of course, doing so immediately raises the question of the extent to which this supposed extension would be unique.  We now will answer this question for the case of $N=1$ particle states for {\it any} free field. This is an interesting case since, when combined with the results reviewed in section \ref{subsec:summary} and with the discussion of section \ref{subsec:2ps} below, it provides a rather complete picture of group-averaging for both 3+1 gravitons and free scalar fields with $M>d/2\ell$.

The alternate inner product \eqref{eq:altIP} is well-defined for 1-particle states of  any free field and, in fact, defines any 1-particle state to be a null state.  To see this, note that the factor $\Pi_{J^2=0}$ in \eqref{eq:altIP} means that the only 1-particle state that could possibly have non-zero norm is the state $|\vec j=0\rangle$ associated with the zero angular-momentum mode of the field.  Yang-Mills fields and gravitons have no such modes, so this completes the analysis for such cases.  Similarly, for minimally-coupled massless scalars the zero angular-momentum mode is also a zero-frequency mode and so does not have particle excitations.   While it would be interesting to return to states of that zero mode in the future (in order to extend the 1+1 analysis of \cite{Marolf:2008it}), this again completes the analysis of 1-particle states for such fields.

It thus remains only to consider fields with minimally-coupled masses $M^2 >0$.  The 1-particle states of such fields are irreducible representations of SO$(D,1)$ with values of the quadratic Casimir $B^2 - J^2 = M^2\ell^2$. As a result, the rotationally-invariant one-particle state  $\big|\vec j=0\bigr]$ is an eigenstate of $B^2$ with eigenvalue $M^2\ell^2 >0$.  Thus $\delta(B^2)$ must annihilate $\big|\vec j=0\bigr]$, so that $\big|\vec j\bigr]$ is a null state under the inner product \eqref{eq:altIP}.

Furthermore, the above argument also shows these definitions to be unique.  Indeed, let us consider a basis of one-particle states that are eigenstates of $J^2$.  If $J^2\big|\phi\bigr] = \lambda\big|\phi\bigr]$ then, unless $\big|\phi\bigr]$ is proportional to the rotationally invariant state $\big|\vec j=0\bigr]$, the fact that $J^2$ is a sum of squares requires $\lambda >0$.  We may thus write
\begin{equation}
\label{eq:phinullJne0}
\big|\phi\bigr] = \frac{J^2}{\lambda}\big|\phi\bigr]= \sum_{i,j=1}^{D} J_{ij} \left( \frac{J_{ij}}{\lambda}\big|\phi\bigr] \right).
\end{equation}
But since $(U(g)-1)\big|\Psi\bigr]$ must be a null state for all $\psi$, the same must be true of $J_{ij} \big|\Psi\bigr]$.  Thus \eqref{eq:phinullJne0} must be null as well.  It then remains only to discuss the rotationally-invariant one-particle state  $\big|\vec j=0\bigr]$.  But it was shown above that this state is an eigenvector of $B^2$ with eigenvalue $M^2\ell^2 >0$.  It is thus of the form
\begin{equation}
\label{eq:phinullJ=0}
\frac{B^2}{M^2\ell^2} \big|\vec j=0\bigr] = \sum_{i=1}^{D} J_{0i} \left( \frac{J_{0i}}{M^2\ell^2}\big|\vec j=0\bigr] \right).
\end{equation}
The same argument used above then shows that $\big|\vec j=0\bigr]$ must be null in $\mathcal H_{LPG}$.

The statement that one-particle states map to null states in $\mathcal H_{LPG}$ is a direct analogue of the classical statement that any single particle in dS has at least one non-vanishing de Sitter charge.  In particular, for free particles with mass $M>0$, if one considers the static patch associated with the particle's geodesic, then the corresponding static-patch energy takes the value $M$.  There are thus no classical single-particle states for which all de Sitter charges vanish.

\subsection{The divergent group-averaging norm of 2-scalar-particle states}
\label{subsec:2ps}

We now briefly address the case of 2-particle states of scalar fields.  The asymptotic expansions of \cite{Marolf_2009} show that the group averaging norms of such states fail to converge absolutely and, in fact, that they in fact diverge linearly.  This is a slower divergence than the exponential divergence one finds for the group-averaging norm of the de Sitter-invariant vacuum $\big|0\bigr]$, but it is a divergence nonetheless.  In fact, it is precisely the degree of divergence one would expect if such states left unbroken a 1-dimensional subgroup of SO$(D,1)$ generated by some boost transformation.

\begin{figure}[h!]
    \centering
    \includegraphics[width=0.2\linewidth]{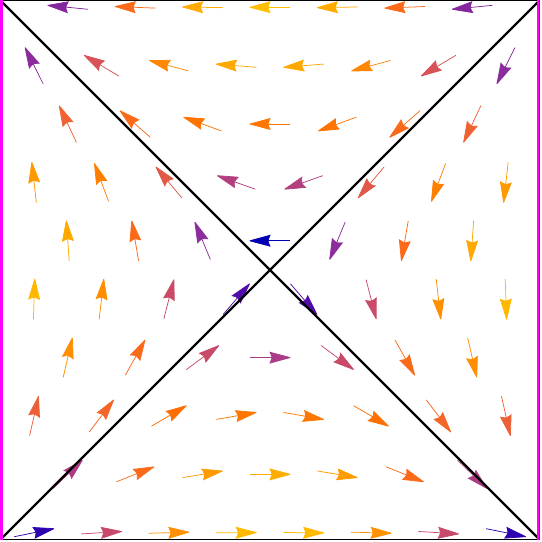}
    \caption{Conformal diagram of unperturbed dS with two marked antipodal geodesics (pink vertical lines).  Adding particles along these geodesics preserves a Killing field (arrows) with non-compact orbits.}
    \label{fig:antipodal_geos}
\end{figure}

While the 2-particle states $\big|\Psi\bigr]$ do not appear to leave such a group unbroken, the corresponding classical solutions {\it do} exhibit an unbroken such symmetry.  Indeed, setting the de Sitter charges to zero forces a pair of classical particles to travel along antipodally-related geodesics (as in the particle approximation to the Schwarzschild-de Sitter solution).  Such solutions clearly leave one boost symmetry intact; see figure \ref{fig:antipodal_geos}.   Some quantum version of this residual symmetry thus appears to be associated with the above divergence, though it would be useful to understand the relationship in more detail.  In particular, following similar discussions in \cite{Marolf:1994ae}, one might expect to be able to use an improved such understanding to argue that no well-defined de Sitter-invariant operator can cause such 2-particle states to mix with states having $N\ge 3$ particles.

\section{Generator moments in $\big|\psi_+\bigr]$ for general dimensions
}
\label{app:coeffs}

We now compute and analyze the coefficients in \eqref{eq:coeffeq} associated with the expectation values of $B_1, B_k^2, J_{ij}^2$ in the state $\big|\psi_+\bigr]$ for $D=d+1>2$.  Our goal is to extract useful expressions for these moments at leading order in large $j_{\mathsmaller \star}$.

Let us start by computing the expectation value $[B_1] = \bigl[\psi_+\big|B_1\big|\psi_+\bigr]$. On the states $\big|\psi_\pm\bigr]$, the action of $B_1$ is
\begin{equation}
\begin{split}
    B_1\big|\psi_\pm\bigr]=& N \sum_{j=0}^{j_{\mathsmaller \star}} c_j^\pm b_j^+\big|\Delta, j+1, \vec{0}\bigr]+ N \sum_{j=1}^{j_{\mathsmaller \star}} b_j^-\big|\Delta, j-1, \vec{0}\bigr],
\end{split}
\end{equation}
with
\begin{align}
    b_j^+=&\left[\frac{(j+d-1)(j+1)(j+d-\Delta)(j+\Delta)}{(2j+d-1)(2j+d+1)}\right]^{1/2},  \ \ \ {\rm and} \\
    b_j^-=&\left[\frac{(j+d-2)j(j-1+d-\Delta)(j-1+\Delta)}{(2j+d-3)(2j+d-1)}\right]^{1/2}. 
\end{align}
Utilizing $b_j^-=b_{j-1}^+$, this yields
\begin{equation}
    \bigl[\psi_\pm \big| B_1 \big| \psi_\pm \bigr] = \pm 2N^2 \sum_{j=0}^{{j_{\mathsmaller \star}}-1} c_j^+ c_{j+1}^+ b_j^+.
\end{equation}
As we will find with most of our other moments, a simple closed-form expression for this sum is not readily available. However, we are mostly interested in the asymptotic behavior at large ${j_{\mathsmaller \star}}$, since in this limit the group averaging kernel will approximate a Gaussian with shrinking width. In particular, given our results in dS$_{1+1}$, it is natural to take the ultrarelativistic limit $j_{\mathsmaller \star}\gg \mu$.  However, for now we will proceed with computing the remaining expectation values; we will then return later to the question of asymptotic behavior.

For $[B_1^2]$, we find the result
\begin{equation}
    \begin{split}
   \bigl[\psi_\pm\big| B_1^2\big|\psi_\pm\bigr]=&  N^2 \sum_{j=1}^{j_{\mathsmaller \star}} (c_j^+)^2(b_{j-1}^+)^2 +N^2 \sum_{j=0}^{j_{\mathsmaller \star}} (c_j^+)^2(b_j^+)^2 + 2N^2 \sum_{j=0}^{{j_{\mathsmaller \star}}-2} c_j^+ c_{j+2}^+ b_j^+b_{j+1}^+.
\end{split}
\end{equation}
Similarly, calculating $[J^2]$ gives
\begin{equation}
    \bigl[\psi_\pm\big| J^2 \big|\psi_\pm\bigr] = N^2 \sum_{j=0}^{j_{\mathsmaller \star}} j(j+d-1)c_j^2 = \frac{d}{d+2}j_{\mathsmaller \star}^2 + \frac{d^2}{d+2}j_{\mathsmaller \star}.
\end{equation}
Finally, to find the expectation value of $B_k^2$ for $k> 1$, we use the Casimir equation
\begin{equation}\label{eq:casimir}
    \mathcal{C}_2=B_1^2+B_\perp^2-J^2,
\end{equation}
where we have used the notation $B_\perp^2 := \sum_{n=2}^{d} B_n^2$. Since we have
\begin{equation}
    \bigl[\psi_\pm\big| \mathcal{C}_2 \big|\psi_\pm\bigr]  = \Delta (d-\Delta),
\end{equation}
we can easily compute $[B_\perp^2]$.

We now return to the question of asymptotic behavior. For $[B_1]$ and $[B_1^2]$, it is natural to assume that keeping only the leading terms in the summands will give the correct large ${j_{\mathsmaller \star}}$ behavior at leading order in $1/{j_{\mathsmaller \star}}$. We will then check this assumption. The leading order behavior of the normalization constant is, using Stirling's approximation, $N^2 = d{j_{\mathsmaller \star}}^{-d} [1+O(1/{j_{\mathsmaller \star}})]$. And, in the limit $j \gg d$ and $j \gg \Delta$, we have
\begin{align}\label{eq:bj_LO}
    b_j^+ =& \frac{j}{2}[1+O(1/j)], \\
    c_j^+ =& j^{(d-1)/2} [1+O(1/j)],\label{eq:cj_LO}
\end{align}
(which have the same asymptotic behavior for $j \to j+1$ or $j \to j+2$). While this large $j$ limit will not hold for the terms in the sum where $j$ is small, we will shortly show that the terms where it fails do not contribute significantly in the limit of large ${j_{\mathsmaller \star}}$. For convenience, we consider only $[\psi_{+}| B_1 \big|\psi_+\bigr]$, which differs from $[\psi_{-}| B_1 \big|\psi_-\bigr]$ only by an overall sign. For $\bigl[\psi_+\big|B_1\big|\psi_+\bigr]$, we have
\begin{equation}
    \bigl[\psi_+\big| B_1 \big|\psi_+\bigr] = \frac{d}{d+1} {j_{\mathsmaller \star}} + O(1),
\end{equation}
where we have used Faulhaber's formula for the sum: $\sum_{j=0}^{{j_{\mathsmaller \star}}-x} j^{d+y} = \frac{1}{d+y+1}{j_{\mathsmaller \star}}^{d+y+1} + O({j_{\mathsmaller \star}}^d)$ for the cases $x=0,1,2$ and $y=0,1$. Similarly, for $[ B_1^2 ]$ we find
\begin{equation}
    \bigl[\psi_{\pm}\big| B_1^2 \big|\psi_\pm\bigr] =  \frac{d}{d+2} {j_{\mathsmaller \star}}^2 + O({j_{\mathsmaller \star}}).
\end{equation}
Finally, the Casimir equation \eqref{eq:casimir} yields
\begin{equation}
    \bigl[\psi_{\pm}\big| B_\perp^2 \big|\psi_\pm\bigr] =  O({j_{\mathsmaller \star}}).
\end{equation}

We may now verify that the above guess indeed gives the correct large ${j_{\mathsmaller \star}}$ behavior of the desired moments (despite the fact that we used Equations \eqref{eq:bj_LO} and \eqref{eq:cj_LO} in regimes where they were not fully applicable).  For our summands $f(j)$, we have
\begin{equation}
    \int_{a-1}^b f(x) dx \leq \sum_{j=a}^b f(j) \leq \int_{a}^{b+1} f(x) dx,
\end{equation}
since $f(j)$ is increasing with $j$. In particular, for $[B_1 ]$ we find
\begin{equation}
    \lim_{{j_{\mathsmaller \star}}\to\infty }  \frac{2N^2 \int_{-1}^{{j_{\mathsmaller \star}}-1} f(x) dx}{\frac{d}{d+1}{j_{\mathsmaller \star}}} \leq \lim_{{j_{\mathsmaller \star}}\to\infty } \frac{\bigl[\psi_+\big|B_1\big|\psi_+\bigr]}{\frac{d}{d+1}{j_{\mathsmaller \star}}} \leq \lim_{{j_{\mathsmaller \star}}\to\infty } \frac{2N^2 \int_{0}^{{j_{\mathsmaller \star}}} f(x) dx}{\frac{d}{d+1}{j_{\mathsmaller \star}}}.
\end{equation}
The lower bound evaluates to
\begin{equation}
    \lim_{{j_{\mathsmaller \star}}\to\infty }  \frac{ 2\int_{-1}^{{j_{\mathsmaller \star}}-1} f(x) dx}{\frac{1}{d+1}{j_{\mathsmaller \star}}^{d+1}} = \lim_{{j_{\mathsmaller \star}}\to\infty }  \frac{ 2f({j_{\mathsmaller \star}}-1)}{{j_{\mathsmaller \star}}^d} = 1.
\end{equation}
We can see that the upper bound will also evaluate to $1$. Thus, we must have
\begin{equation}
    \lim_{{j_{\mathsmaller \star}}\to\infty } \frac{\bigl[\psi_{\pm}\big| B_1 \big|\psi_\pm\bigr]}{\pm \frac{d}{d+1}{j_{\mathsmaller \star}}} = 1.
\end{equation}
The same argument holds for $[B_1^2 ]$, proving we have the correct asymptotic behavior of our sums.

We will also need to find $[B_l^2]$ for $l \ge 2$ and $[J_{ij}^2]$. The symmetry of $\big|\psi_+\bigr]$ requires all $[B_l^2]$ to be equal for $l\ge 2$. Additionally, all $[J_{ij}^2]$ with $i=1$ (or $j=1$) must be equal, while all others must vanish. We thus find,
\begin{align}
\label{eq:leadingjcoeffs}
\left([B_1^2]-[B_1]^2\right) &=\left[\frac{d}{d+2}-\frac{d^2}{(d+1)^2}\right] {j_{\mathsmaller \star}}^2 + O(j_{\mathsmaller \star})=\frac{d}{(d+2)(d+1)^2}{j_{\mathsmaller \star}}^2 + O(j_{\mathsmaller \star}), \cr  [B_l^2] &= O(j_{\mathsmaller \star}), \cr 
 [J_{ij}^2] &= \left(\frac{1}{d+2} {j_{\mathsmaller \star}}^2 + \frac{d}{d+2} {j_{\mathsmaller \star}}\right) \delta_{i,1} \ \ \ {\rm for \ } j>i.
\end{align}

\addcontentsline{toc}{section}{References}
\bibliographystyle{JHEP}
\bibliography{draft}

\providecommand{\href}[2]{#2}\begingroup\raggedright\begin{thebibliography}{10}

\bibitem{DeWitt:1962}
B.~S. DeWitt, \emph{The quantization of geometry},  in \emph{Gravitation: An introduction to current research}, L.~Witten, ed., (New York, NY), Wiley, 1962.

\bibitem{DeWitt:1967yk}
B.~S. DeWitt, \emph{{Quantum Theory of Gravity. 1. The Canonical Theory}}, \href{https://doi.org/10.1103/PhysRev.160.1113}{\emph{Phys. Rev.} {\bfseries 160} (1967) 1113}.

\bibitem{Banks:1984cw}
T.~Banks, \emph{{T C P, Quantum Gravity, the Cosmological Constant and All That...}}, \href{https://doi.org/10.1016/0550-3213(85)90020-3}{\emph{Nucl. Phys. B} {\bfseries 249} (1985) 332}.

\bibitem{Hartle:1986gn}
J.~B. Hartle, \emph{{Prediction in Quantum Cosmology}}, \href{https://doi.org/10.1007/978-1-4613-1897-2_12}{\emph{NATO Sci. Ser. B} {\bfseries 156} (1987) 329}.

\bibitem{Rovelli:1990jm}
C.~Rovelli, \emph{{Quantum mechanics without time: a model}}, \href{https://doi.org/10.1103/PhysRevD.42.2638}{\emph{Phys. Rev. D} {\bfseries 42} (1990) 2638}.

\bibitem{Rovelli:1990pi}
C.~Rovelli, \emph{{Quantum reference systems}}, \href{https://doi.org/10.1088/0264-9381/8/2/012}{\emph{Class. Quant. Grav.} {\bfseries 8} (1991) 317}.

\bibitem{Kiefer:1993fg}
C.~Kiefer, \emph{{The Semiclassical approximation to quantum gravity}}, \href{https://doi.org/10.1007/3-540-58339-4_19}{\emph{Lect. Notes Phys.} {\bfseries 434} (1994) 170} [\href{https://arxiv.org/abs/gr-qc/9312015}{{\ttfamily gr-qc/9312015}}].

\bibitem{Marolf:1994nz}
D.~Marolf, \emph{{Almost ideal clocks in quantum cosmology: A Brief derivation of time}}, \href{https://doi.org/10.1088/0264-9381/12/10/007}{\emph{Class. Quant. Grav.} {\bfseries 12} (1995) 2469} [\href{https://arxiv.org/abs/gr-qc/9412016}{{\ttfamily gr-qc/9412016}}].

\bibitem{Giddings:2005id}
S.~B. Giddings, D.~Marolf and J.~B. Hartle, \emph{{Observables in effective gravity}}, \href{https://doi.org/10.1103/PhysRevD.74.064018}{\emph{Phys. Rev. D} {\bfseries 74} (2006) 064018} [\href{https://arxiv.org/abs/hep-th/0512200}{{\ttfamily hep-th/0512200}}].

\bibitem{Marolf:2015jha}
D.~Marolf, \emph{{Comments on Microcausality, Chaos, and Gravitational Observables}}, \href{https://doi.org/10.1088/0264-9381/32/24/245003}{\emph{Class. Quant. Grav.} {\bfseries 32} (2015) 245003} [\href{https://arxiv.org/abs/1508.00939}{{\ttfamily 1508.00939}}].

\bibitem{Witten:2021unn}
E.~Witten, \emph{{Gravity and the crossed product}}, \href{https://doi.org/10.1007/JHEP10(2022)008}{\emph{JHEP} {\bfseries 10} (2022) 008} [\href{https://arxiv.org/abs/2112.12828}{{\ttfamily 2112.12828}}].

\bibitem{Chandrasekaran:2022cip}
V.~Chandrasekaran, R.~Longo, G.~Penington and E.~Witten, \emph{{An algebra of observables for de Sitter space}}, \href{https://doi.org/10.1007/JHEP02(2023)082}{\emph{JHEP} {\bfseries 02} (2023) 082} [\href{https://arxiv.org/abs/2206.10780}{{\ttfamily 2206.10780}}].

\bibitem{Chandrasekaran:2022eqq}
V.~Chandrasekaran, G.~Penington and E.~Witten, \emph{{Large N algebras and generalized entropy}}, \href{https://doi.org/10.1007/JHEP04(2023)009}{\emph{JHEP} {\bfseries 04} (2023) 009} [\href{https://arxiv.org/abs/2209.10454}{{\ttfamily 2209.10454}}].

\bibitem{Jensen:2023yxy}
K.~Jensen, J.~Sorce and A.~J. Speranza, \emph{{Generalized entropy for general subregions in quantum gravity}}, \href{https://doi.org/10.1007/JHEP12(2023)020}{\emph{JHEP} {\bfseries 12} (2023) 020} [\href{https://arxiv.org/abs/2306.01837}{{\ttfamily 2306.01837}}].

\bibitem{Penington:2023dql}
G.~Penington and E.~Witten, \emph{{Algebras and States in JT Gravity}},  \href{https://arxiv.org/abs/2301.07257}{{\ttfamily 2301.07257}}.

\bibitem{Kudler-Flam:2024psh}
J.~Kudler-Flam, S.~Leutheusser and G.~Satishchandran, \emph{{Algebraic Observational Cosmology}},  \href{https://arxiv.org/abs/2406.01669}{{\ttfamily 2406.01669}}.

\bibitem{Chen:2024rpx}
C.-H. Chen and G.~Penington, \emph{{A clock is just a way to tell the time: gravitational algebras in cosmological spacetimes}},  \href{https://arxiv.org/abs/2406.02116}{{\ttfamily 2406.02116}}.

\bibitem{Giddings_2007}
S.~B. Giddings and D.~Marolf, \emph{A global picture of quantum de sitter space}, {\emph{Physical Review D} {\bfseries 76} (2007) } [\href{https://arxiv.org/abs/0705.1178}{{\ttfamily 0705.1178}}].

\bibitem{Kachru:2003aw}
S.~Kachru, R.~Kallosh, A.~D. Linde and S.~P. Trivedi, \emph{{De Sitter vacua in string theory}}, \href{https://doi.org/10.1103/PhysRevD.68.046005}{\emph{Phys. Rev. D} {\bfseries 68} (2003) 046005} [\href{https://arxiv.org/abs/hep-th/0301240}{{\ttfamily hep-th/0301240}}].

\bibitem{McAllister:2024lnt}
L.~McAllister, J.~Moritz, R.~Nally and A.~Schachner, \emph{{Candidate de Sitter Vacua}},  \href{https://arxiv.org/abs/2406.13751}{{\ttfamily 2406.13751}}.

\bibitem{Marolf:1994wh}
D.~Marolf, \emph{{Quantum observables and recollapsing dynamics}}, \href{https://doi.org/10.1088/0264-9381/12/5/011}{\emph{Class. Quant. Grav.} {\bfseries 12} (1995) 1199} [\href{https://arxiv.org/abs/gr-qc/9404053}{{\ttfamily gr-qc/9404053}}].

\bibitem{Marolf:1994ss}
D.~Marolf, \emph{{Observables and a Hilbert space for Bianchi IX}}, \href{https://doi.org/10.1088/0264-9381/12/6/010}{\emph{Class. Quant. Grav.} {\bfseries 12} (1995) 1441} [\href{https://arxiv.org/abs/gr-qc/9409049}{{\ttfamily gr-qc/9409049}}].

\bibitem{Aharonov:1967zza}
Y.~Aharonov and L.~Susskind, \emph{{Charge Superselection Rule}}, \href{https://doi.org/10.1103/PhysRev.155.1428}{\emph{Phys. Rev.} {\bfseries 155} (1967) 1428}.

\bibitem{Aharonov:1984zz}
Y.~Aharonov and T.~Kaufherr, \emph{{Quantum frames of reference}}, \href{https://doi.org/10.1103/PhysRevD.30.368}{\emph{Phys. Rev. D} {\bfseries 30} (1984) 368}.

\bibitem{Loveridge:2017pcv}
L.~Loveridge, T.~Miyadera and P.~Busch, \emph{{Symmetry, Reference Frames, and Relational Quantities in Quantum Mechanics}}, \href{https://doi.org/10.1007/s10701-018-0138-3}{\emph{Found. Phys.} {\bfseries 48} (2018) 135} [\href{https://arxiv.org/abs/1703.10434}{{\ttfamily 1703.10434}}].

\bibitem{Hoehn:2019fsy}
P.~A. Hoehn, A.~R.~H. Smith and M.~P.~E. Lock, \emph{{Trinity of relational quantum dynamics}}, \href{https://doi.org/10.1103/PhysRevD.104.066001}{\emph{Phys. Rev. D} {\bfseries 104} (2021) 066001} [\href{https://arxiv.org/abs/1912.00033}{{\ttfamily 1912.00033}}].

\bibitem{Fewster:2024pur}
C.~J. Fewster, D.~W. Janssen, L.~D. Loveridge, K.~Rejzner and J.~Waldron, \emph{{Quantum reference frames, measurement schemes and the type of local algebras in quantum field theory}},  \href{https://arxiv.org/abs/2403.11973}{{\ttfamily 2403.11973}}.

\bibitem{DeVuyst:2024pop}
J.~De~Vuyst, S.~Eccles, P.~A. Hoehn and J.~Kirklin, \emph{{Gravitational entropy is observer-dependent}},  \href{https://arxiv.org/abs/2405.00114}{{\ttfamily 2405.00114}}.

\bibitem{AliAhmad:2024wja}
S.~Ali~Ahmad, W.~Chemissany, M.~S. Klinger and R.~G. Leigh, \emph{{Quantum reference frames from top-down crossed products}}, \href{https://doi.org/10.1103/PhysRevD.110.065003}{\emph{Phys. Rev. D} {\bfseries 110} (2024) 065003} [\href{https://arxiv.org/abs/2405.13884}{{\ttfamily 2405.13884}}].

\bibitem{Klinger:2023auu}
M.~S. Klinger and R.~G. Leigh, \emph{{Crossed products, conditional expectations and constraint quantization}}, \href{https://doi.org/10.1016/j.nuclphysb.2024.116622}{\emph{Nucl. Phys. B} {\bfseries 1006} (2024) 116622} [\href{https://arxiv.org/abs/2312.16678}{{\ttfamily 2312.16678}}].

\bibitem{Higuchi_1991}
A.~Higuchi, \emph{{Quantum linearization instabilities of de Sitter space-time. 1}}, \href{https://doi.org/10.1088/0264-9381/8/11/009}{\emph{Class. Quant. Grav.} {\bfseries 8} (1991) 1961}.

\bibitem{Higuchi_1991_2}
A.~Higuchi, \emph{{Quantum linearization instabilities of de Sitter space-time. 2}}, \href{https://doi.org/10.1088/0264-9381/8/11/010}{\emph{Class. Quant. Grav.} {\bfseries 8} (1991) 1983}.

\bibitem{Deser:1973zza}
S.~Deser and D.~Brill, \emph{{Instability of Closed Spaces in General Relativity}}, \href{https://doi.org/10.1007/BF01645610}{\emph{Commun. Math. Phys.} {\bfseries 32} (1973) 291}.

\bibitem{Moncrief:1975}
V.~Moncrief, \emph{{Space-Time Symmetries and Linearization Stability of the Einstein Equations. 1.}}, \href{https://doi.org/10.1063/1.522572}{\emph{J. Math. Phys.} {\bfseries 16} (1975) 493}.

\bibitem{Moncrief:1976un}
V.~Moncrief, \emph{{Space-Time Symmetries and Linearization Stability of the Einstein Equations. 2.}}, \href{https://doi.org/10.1063/1.522814}{\emph{J. Math. Phys.} {\bfseries 17} (1976) 1893}.

\bibitem{Arms:1977}
J.~Arms, \emph{{Linearization stability of the Einstein–Maxwell system}}, \href{https://doi.org/10.1063/1.523312}{\emph{J. Math. Phys.} {\bfseries 18} (1977) 830}.

\bibitem{Arms:1979au}
J.~M. Arms, \emph{{Linearization Stability of Gravitational and Gauge Fields}}, \href{https://doi.org/10.1063/1.524094}{\emph{J. Math. Phys.} {\bfseries 20} (1979) 443}.

\bibitem{Marolf:1994ae}
D.~Marolf, \emph{{The Spectral analysis inner product for quantum gravity}},  in \emph{{7th Marcel Grossmann Meeting on General Relativity (MG 7)}}, pp.~851--853, 9, 1994, \href{https://arxiv.org/abs/gr-qc/9409036}{{\ttfamily gr-qc/9409036}}.

\bibitem{Ashtekar:1995zh}
A.~Ashtekar, J.~Lewandowski, D.~Marolf, J.~Mourao and T.~Thiemann, \emph{{Quantization of diffeomorphism invariant theories of connections with local degrees of freedom}}, \href{https://doi.org/10.1063/1.531252}{\emph{J. Math. Phys.} {\bfseries 36} (1995) 6456} [\href{https://arxiv.org/abs/gr-qc/9504018}{{\ttfamily gr-qc/9504018}}].

\bibitem{Giulini:1998rk}
D.~Giulini and D.~Marolf, \emph{{On the generality of refined algebraic quantization}}, \href{https://doi.org/10.1088/0264-9381/16/7/321}{\emph{Class. Quant. Grav.} {\bfseries 16} (1999) 2479} [\href{https://arxiv.org/abs/gr-qc/9812024}{{\ttfamily gr-qc/9812024}}].

\bibitem{Marolf:2000iq}
D.~Marolf, \emph{{Group averaging and refined algebraic quantization: Where are we now?}},  in \emph{{9th Marcel Grossmann Meeting on Recent Developments in Theoretical and Experimental General Relativity, Gravitation and Relativistic Field Theories (MG 9)}}, 7, 2000, \href{https://arxiv.org/abs/gr-qc/0011112}{{\ttfamily gr-qc/0011112}}.

\bibitem{Chakraborty:2023los}
T.~Chakraborty, J.~Chakravarty, V.~Godet, P.~Paul and S.~Raju, \emph{{Holography of information in de Sitter space}}, \href{https://doi.org/10.1007/JHEP12(2023)120}{\emph{JHEP} {\bfseries 12} (2023) 120} [\href{https://arxiv.org/abs/2303.16316}{{\ttfamily 2303.16316}}].

\bibitem{Marolf:1996gb}
D.~Marolf, \emph{{Path integrals and instantons in quantum gravity: Minisuperspace models}}, \href{https://doi.org/10.1103/PhysRevD.53.6979}{\emph{Phys. Rev. D} {\bfseries 53} (1996) 6979} [\href{https://arxiv.org/abs/gr-qc/9602019}{{\ttfamily gr-qc/9602019}}].

\bibitem{Streater:1989vi}
R.~F. Streater and A.~S. Wightman, \emph{{PCT, spin and statistics, and all that}}. 1989.

\bibitem{VK}
N.~Y. Vilenken and A.~U. Klimyk, \emph{{Representations of Lie Groups and Special Functions}}, vol.~1-3. Dordrecht: Kluwer Acad. Publ, 1991-1993.

\bibitem{Wong:1974cv}
M.~K.~F. Wong, \emph{{Unitary representations of so(n,1)}}, \href{https://doi.org/10.1063/1.1666496}{\emph{J. Math. Phys.} {\bfseries 15} (1974) 25}.

\bibitem{Sun:2021thf}
Z.~Sun, \emph{{A note on the representations of $\text{SO}(1,d+1)$}},  \href{https://arxiv.org/abs/2111.04591}{{\ttfamily 2111.04591}}.

\bibitem{Marolf_2009}
D.~Marolf and I.~A. Morrison, \emph{Group averaging for de sitter free fields}, {\emph{Classical and Quantum Gravity} {\bfseries 26} (2009) 235003} [\href{https://arxiv.org/abs/0810.5163}{{\ttfamily 0810.5163}}].

\bibitem{Gao:2000ga}
S.~Gao and R.~M. Wald, \emph{{Theorems on gravitational time delay and related issues}}, \href{https://doi.org/10.1088/0264-9381/17/24/305}{\emph{Class. Quant. Grav.} {\bfseries 17} (2000) 4999} [\href{https://arxiv.org/abs/gr-qc/0007021}{{\ttfamily gr-qc/0007021}}].

\bibitem{Aichelburg:1970dh}
P.~C. Aichelburg and R.~U. Sexl, \emph{{On the Gravitational field of a massless particle}}, \href{https://doi.org/10.1007/BF00758149}{\emph{Gen. Rel. Grav.} {\bfseries 2} (1971) 303}.

\bibitem{Halliwell:1990qr}
J.~J. Halliwell and J.~B. Hartle, \emph{{Wave functions constructed from an invariant sum over histories satisfy constraints}}, \href{https://doi.org/10.1103/PhysRevD.43.1170}{\emph{Phys. Rev. D} {\bfseries 43} (1991) 1170}.

\bibitem{Reisenberger:1996pu}
M.~P. Reisenberger and C.~Rovelli, \emph{{'Sum over surfaces' form of loop quantum gravity}}, \href{https://doi.org/10.1103/PhysRevD.56.3490}{\emph{Phys. Rev. D} {\bfseries 56} (1997) 3490} [\href{https://arxiv.org/abs/gr-qc/9612035}{{\ttfamily gr-qc/9612035}}].

\bibitem{Marolf:2020xie}
D.~Marolf and H.~Maxfield, \emph{{Transcending the ensemble: baby universes, spacetime wormholes, and the order and disorder of black hole information}}, \href{https://doi.org/10.1007/JHEP08(2020)044}{\emph{JHEP} {\bfseries 08} (2020) 044} [\href{https://arxiv.org/abs/2002.08950}{{\ttfamily 2002.08950}}].

\bibitem{Giulini_1999}
D.~Giulini and D.~Marolf, \emph{A uniqueness theorem for constraint quantization}, {\emph{Classical and Quantum Gravity} {\bfseries 16} (1999) 2489–2505} [\href{https://arxiv.org/abs/9902045}{{\ttfamily 9902045}}].

\bibitem{Higuchi:2011vw}
A.~Higuchi, D.~Marolf and I.~A. Morrison, \emph{{de Sitter invariance of the dS graviton vacuum}}, \href{https://doi.org/10.1088/0264-9381/28/24/245012}{\emph{Class. Quant. Grav.} {\bfseries 28} (2011) 245012} [\href{https://arxiv.org/abs/1107.2712}{{\ttfamily 1107.2712}}].

\bibitem{Seery_2010}
D.~Seery, \emph{Infrared effects in inflationary correlation functions}, \href{https://doi.org/10.1088/0264-9381/27/12/124005}{\emph{Classical and Quantum Gravity} {\bfseries 27} (2010) 124005} [\href{https://arxiv.org/abs/1005.1649}{{\ttfamily 1005.1649}}].

\bibitem{Linde_2005}
A.~Linde, \emph{Particle physics and inflationary cosmology},  2005.

\bibitem{Giddings_2011}
S.~B. Giddings and M.~S. Sloth, \emph{Cosmological observables, infrared growth of fluctuations, and scale-dependent anisotropies}, \href{https://doi.org/10.1103/physrevd.84.063528}{\emph{Physical Review D} {\bfseries 84} (2011) } [\href{https://arxiv.org/abs/1104.0002}{{\ttfamily 1104.0002}}].

\bibitem{Marolf:2010zp}
D.~Marolf and I.~A. Morrison, \emph{{The IR stability of de Sitter: Loop corrections to scalar propagators}}, \href{https://doi.org/10.1103/PhysRevD.82.105032}{\emph{Phys. Rev. D} {\bfseries 82} (2010) 105032} [\href{https://arxiv.org/abs/1006.0035}{{\ttfamily 1006.0035}}].

\bibitem{Marolf:2008it}
D.~Marolf and I.~A. Morrison, \emph{{Group Averaging of massless scalar fields in 1+1 de Sitter}}, \href{https://doi.org/10.1088/0264-9381/26/3/035001}{\emph{Class. Quant. Grav.} {\bfseries 26} (2009) 035001} [\href{https://arxiv.org/abs/0808.2174}{{\ttfamily 0808.2174}}].

\end{thebibliography}\endgroup

\end{document}